\documentclass[a4paper,fleqn,usenatbib]{mnras}

\usepackage{url}
\usepackage{hyperref}

\usepackage{amsmath}    
\usepackage{amssymb}    
\usepackage{times,txfonts}

\usepackage[T1]{fontenc}
\usepackage{ae,aecompl}

\usepackage{graphicx}   

\usepackage{wasysym}

\title[Distance bias affecting Hi-GAL source properties]{Distance biases in the estimation of the physical properties of Hi-GAL compact sources-I.
  Clump properties and the identification of high-mass star forming candidates.
}

\author[Baldeschi et al.]{
Adriano Baldeschi$^{1,2} $ \thanks{E-mail:adriano.baldeschi@iaps.inaf.it},
D.~Elia $^{1},$
S.~Molinari $^{1},$
S. Pezzuto $^{1},$
E. Schisano $^{1},$ \newauthor
M. Gatti $^{3},$ 
A. Serra $^{4},$ 
M. Merello $^{1},$
M. Benedettini $^{1},$
A. M. Di Giorgio $^{1},$
J. S. Liu $^{1}$  
\\
$^{1}$Istituto di Astrofisica e Planetologia Spaziali - INAF,
Via Fosso del Cavaliere 100, I-00133 Roma, Italy \\
$^{2}$Dipartimento di Fisica, ‘Sapienza’ Universit\`a di Roma, Piazzale Aldo Moro 5, I-00185 Roma, Italy \\
$^{3}$Institut de Fisica d'Altes Energies (IFAE). Edifici Cn, Universitat Autonoma de Barcelona (UAB), E-08193 Bellaterra (Barcelona), Spain \\
$^{4}$IBM Italia - Via Sciangai 53, I-00144 Roma, Italy 
}

\begin{document}

\date{\today}


\maketitle

\label{firstpage}

\begin{abstract}
The degradation of spatial resolution in star-forming regions observed at large distances  ($d\gtrsim1$ kpc) 
with \emph{Herschel},
 can lead to estimates of the physical parameters 
 of the detected compact sources 
(clumps) which do not necessarily mirror the properties of the original population
of cores. This paper aims at quantifying  the 
bias introduced in the estimation of these parameters by the distance effect.
To do so, we consider \emph{Herschel} maps of nearby star-forming 
regions taken from the \emph{Herschel}-Gould-Belt survey, and simulate the effect of increased distance to understand
what amount of information is lost  when a distant 
star-forming region is observed with \emph{Herschel} resolution.
In the  maps displaced  to different distances 
we extract compact sources, and
we derive their physical parameters 
as if they were original Hi-GAL maps of the extracted source samples.
In this way, we are able to discuss how the main physical properties 
 change with distance. 
In particular, we discuss  the ability of clumps to form massive stars:
we estimate the fraction of distant sources that are classified as  high-mass stars-forming objects due to their position
in the mass vs radius diagram, that are  only ``false positives''.
We give also a threshold for  high-mass star-formation  $M>1282 \ \left(\frac{r}{ [\mathrm{pc}]}\right)^{1.42} M_{\odot}$.
In conclusion, this paper provides the astronomer dealing with \emph{Herschel} 
maps of distant star-forming regions with a set of  prescriptions
to partially recover the  character of the core population in 
unresolved clumps.

\end{abstract}

\begin{keywords}
ISM: clouds, stars: formation, infrared: ISM, methods: statistical.
\end{keywords}


\begin{figure*}
\includegraphics[scale=0.4]{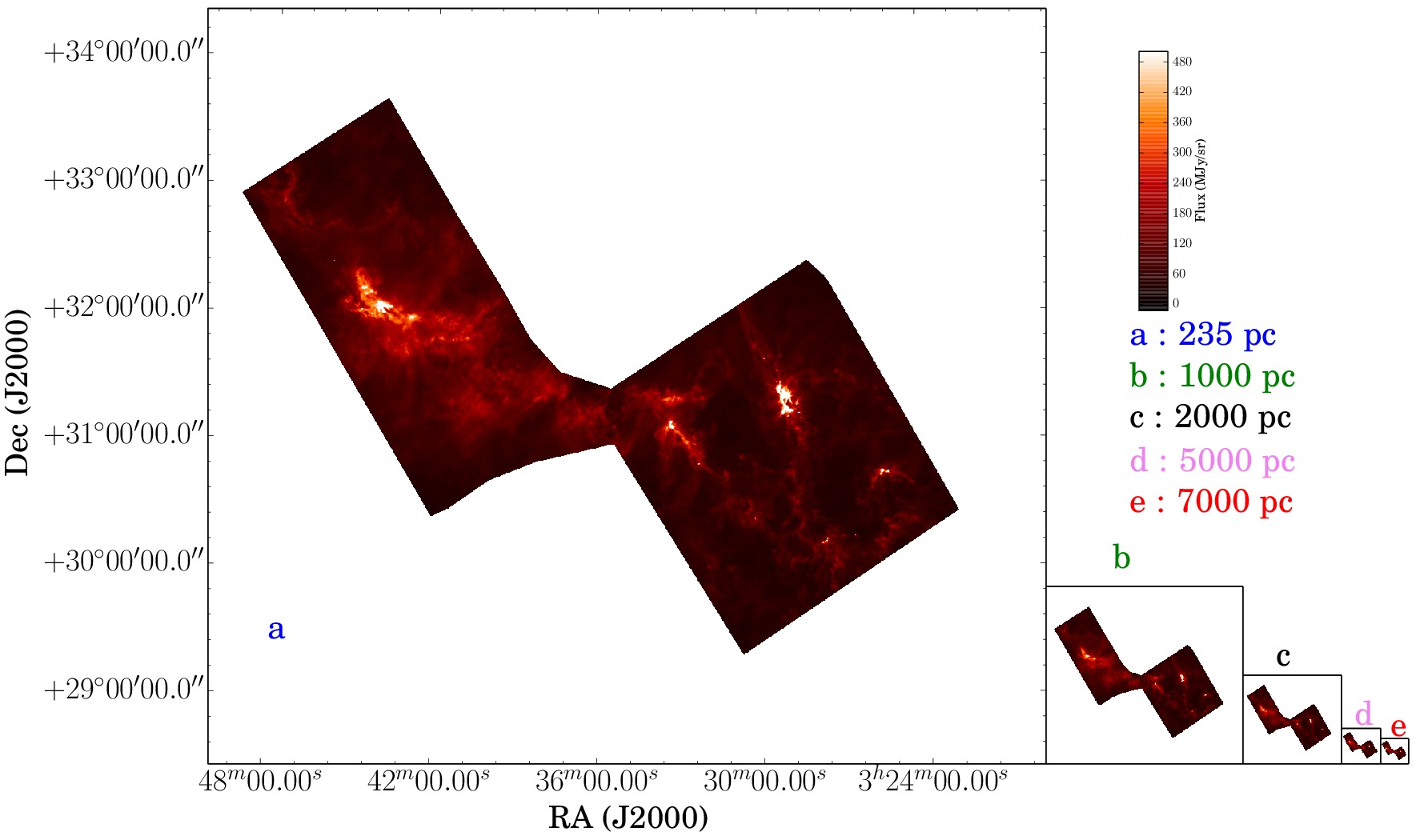}
\caption{Original (a) and moved (b-e) maps of Perseus  at $250 \ \mu \mathrm{m}$. Here  only a few of the simulated distances are shown.
}
\label{fig:PesreusPSW}
\end{figure*}

\begin{figure*}
\includegraphics[scale=0.53]{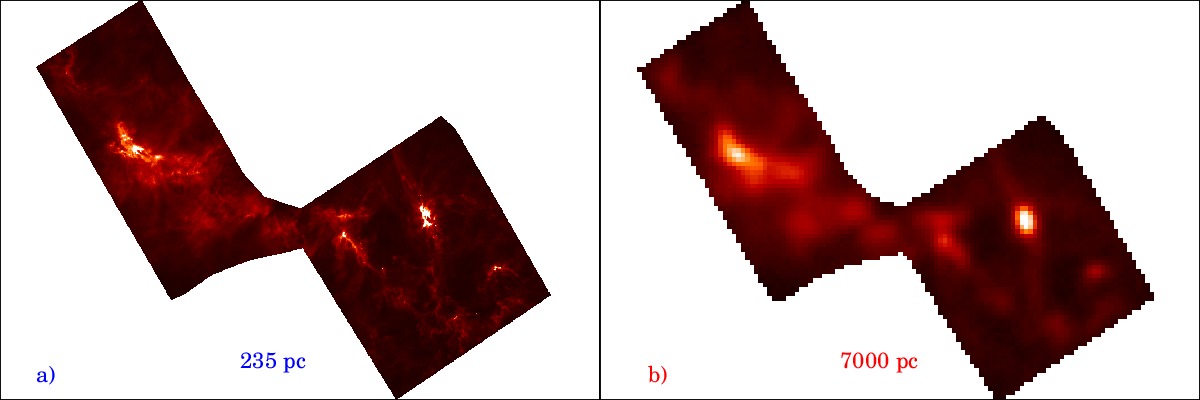}
\caption{ Perseus map at $\mathrm{250 \ \mu m}$ as it appears at the original distance (panel a) is the same as panel a) of Fig.~\ref{fig:PesreusPSW}
  and how it would look
 like at the distance of 7 kpc (panel b).
 This latter panel has been enlarged to make the figure have the same size of panel a). The loss of spatial resolution at 7 kpc is evident.         
}
\label{fig:PerseusPSW_deg}
\end{figure*}

\section{Introduction}
\label{sec:intro}
The impact of massive stars in the Milky Way is predominant with respect to low-mass stars: they
produce most of the heavy elements and energize the interstellar medium (ISM) through the emission
of ultraviolet photons. Therefore, understanding massive star formation is one of the most important
goals of modern Astrophysics \citep[see, e.g.,][]{Tan2014}.
The \emph{Herschel} infrared Galactic Plane Survey \citep[Hi-GAL,][]{Molinari2010}, based on photometric observations in five bands between
70 and 500~$\mu$m, was designed to study the early phases of star formation across the Galactic plane,
with particular interest in the high-mass regime \citep{Elia2010,Veneziani2013,Beltran2013,Olmi2013,Molinari2014}.

Star forming regions observed in Hi-GAL span a wide range of heliocentric distances
\citep[1~kpc~$\lesssim d \lesssim$~15~kpc,][]{Russeil2011,Elia2016}, therefore the physical size of the detected
compact sources (i.e. not resolved or poorly resolved) could correspond to quite different types of structures,
depending on the combination of their angular size and distance.

The smallest and  densest  structures in the ISM, considered as the last product of cloud fragmentation, then
progenitors of single stars or multiple systems, are called dense cores \citep[$D \lesssim$~0.2~pc][]{Bergin2007},
while larger unresolved overdensities in Giant Molecular Clouds  which host these cores are called clumps (0.2~pc$ \ \lesssim D \lesssim$~3~pc).
In few cases, very distant unresolved Hi-GAL sources may have a diameter $D>3$~pc \citep{Elia2016}, even fulfilling
the definition of cloud. Correspondingly, other distance-dependent source properties, as mass and luminosity,
are found to span a wide range of values, from typical conditions of a core to those of an entire cloud. \\ 
Unfortunately, for distant sources \emph{Herschel} is not able to resolve the internal structure  and the contained
 population of
cores \citep[e.g.][]{Elia2013}, therefore only global and/or averaged parameters can be quoted to describe the physical conditions
of the source and the characteristics of possible star formation ongoing in its interior.

 Observations at higher resolution would be needed (for example
by means of sub-mm interferometry) to fully resolve any individual clump identifying all its single components,
 but this would require very large observing programs with different
facilities, to cover the entire Galactic plane. \\
To overcome empirically the lack of spatial resolution, we pursued a completely different approach, namely
to consider \emph{Herschel} maps of nearby star forming regions and degrade their spatial resolution to simulate the view of the same region
if located at a larger heliocentric distance. We implemented this idea using some nearby ($d < 0.5$~kpc) molecular
clouds observed in the \emph{Herschel} Gould Belt survey \citep[HGBS,][]{Andre2010}, where the  compact sources correspond
to dense cores. 
Obviously, these maps do not represent the ``reality'', in turn being those regions located at a heliocentric distance of some hundred pc.
However they constitute the closest, then best resolved, available view of star-forming regions on which to base our analysis.
``Moving away'' these regions to farther distances, we aim to understand not only how a region
would appear if it were placed at a distance farther than the actual, but mainly at linking the physical properties
of the compact sources detected in the new maps with those of the underlying source populations present in the
original maps. In this way we can evaluate the degree of information lost as a function of distance, in other words the distance
bias affecting the estimation of Hi-GAL compact source physical properties. To reach this goal, we probed a set of
different simulated distances for each considered region, and at each distance we
repeated the typical procedures applied to  Hi-GAL maps for extracting the compact sources \citep[e.g.][]{Elia2013}, treating the
simulated maps as a completely new data set, with no reminiscence either to the original map or to those ``moved''
at other distances. \\
The first paper is organized as follows:
in section~\ref{sec:data} we present the regions of the Gould Belt survey that we use in this paper,
in section~\ref{sec:MAM} we describe the procedure of ``moving away'' the regions, and in
section~\ref{sec:Sextract} we report how the detection and the photometry of the sources in all the produced
maps has been carried out.
In section~\ref{sec:Number} and \ref{sec:protopre} we describe how the number of detected sources and
the fraction starless-to-protostellar changes with distance.
In section~\ref{sec:size} we present the distribution of the size of the detected objects as a function of distance.
Section~\ref{sec:assosiation} shows a procedure to associate the sources detected at different distances.
Section~\ref{sec:temp} describe describe uncertainties in the average temperature for protostellar and prestellar 
objects due to distance effect
Section~\ref{sec:mass}  describes the distance bias in the mass vs radius relation.  
A summary of main conclusions is reported in
 section~\ref{sec:conc}.
In a second paper we will discuss the effects of the distance bias 
on the  the luminosity vs mass diagram and on the derived star-formation rate.

\section{Observations and methodology}
\label{sec:obsand_meth}

\subsection{Observations and data reduction}
\label{sec:data}

The observations used in this paper were taken from the \emph{Herschel} \citep{Pilbratt2010}
 Gould Belt survey \citep[][]{Andre2010} for the study of nearby star forming regions.  A full 
description of the HGBS is given by \citet{Konyves2015}.
For this paper we concentrated on a few regions among those observed in the HGBS, namely: Orion~A, Perseus, Serpens and  Lupus~III 
and IV. 

We selected these regions both because they are close, so that we can reasonably assume 
that most of the cores we detect are not blended, and because they cover a range of different core masses: 
Orion A is a high-mass star forming region, Perseus is an intermediate-to-low mass star-forming region, while 
Lupus and Serpens are forming low-mass stars. Other similar \emph{Herschel}  programs, such as HOBYS \citep{Motte2010} and Hi-GAL itself,
 observed farther regions, for which confusion is an issue: for example, in the Hi-GAL clump catalogue of the
 inner Galaxy \citep{Elia2016} only 166 sources 
out of 36644 having a distance estimate are located at $d<500$ pc.
Furthermore, each HGBS region has the advantage 
of being self-consistent in distance, which is a basic requirement for our kind of analysis. Instead the aforementioned nearby Hi-GAL sources
are generally found to be mixed,
in the same maps, with sources belonging to other distance components, which would represent an irreparable contamination of our data set.
On the other hand the disadvantage of using HGBS data consists of a limited statistics due to limited map size
 at the furthest simulated distances.

HGBS observations were taken at $60\arcsec/\mathrm{s}$ in parallel mode with the two cameras PACS \citep{Poglitsch2010} and SPIRE \citep{Griffin2010}: the observed wavelengths were 70 $\mu \mathrm{m}$ and 160 $\mu \mathrm{m}$ for PACS, and 250 $\mu \mathrm{m}$, 350 $\mu \mathrm{m}$ and 500 $\mu \mathrm{m}$ for SPIRE.

Maps were generated using the Unimap \citep{Piazzo2015} map-maker for both instruments. 
The area considered for our work is that common to the  PACS and SPIRE fields of view.

We assumed the following  distances to the selected regions: 150 pc and 200 pc \citep{Comeron2008} for Lupus IV and III,
respectively; 235 pc \citep{Hirota2008} 
for Perseus; 
230 pc for Serpens \citep{Eiroa2008}, 
 415 pc for Orion A \citep{Menten2007}.
The distance of the Serpens molecular cloud has been a matter of controversy: some authors place the Serpens molecular cloud 
at approximately 400 pc.
In appendix~\ref{sec:serp} we discuss, briefly, how the physical properties of the Serpens molecular cloud change if we assume a distance of 
436 pc \citep{Leon2016}. 
These regions will be the subject of dedicated papers to be published by the HGBS consortium: in this respect we stress 
 that in this paper we are not interested in deriving the physical properties of the sources at the nominal
 distances, but we want to derive how the intrinsic properties
 of the sources change with the distance. For this reason we do not provide any catalogue of cores.


\subsection{The simulation of increased distance}
\label{sec:MAM}

The methodology adopted in this paper  simulates the view, through \emph{Herschel}, of a region of the Gould-Belt  as if it were placed at
a distance larger than the actual one. Of course,
this procedure implies a loss of spatial resolution and detail since the angular size of the moved maps (MM) decreases.
The pipeline adopted to obtain a MM is the following:
\begin{enumerate}
\item Rescaling/rebinning the original map.
\item Convolving the new rescaled map with the PSF of the instrument at the given wavelength.
\item Adding white Gaussian noise to the map.
\end{enumerate}
In detail, \\
(i) A structure of spatial size $L$ placed at distance $d_0$ subtends an angle  $\varphi_0=\frac{L}{d_0}$, 
while if the same object is moved to  a distance $d_1>d_0$ its angular dimension
  becomes $\varphi_1=\frac{L}{d_1}<\varphi_0$. Therefore $\varphi_1=\varphi_0 \frac{d_0}{d_1}$, 
so an image  rebinned by a factor $\frac{d_0}{d_1}$ mimics the movement of the region from $d_{0}$ to $d_{1}$. \\
(ii) To reproduce more realistically the effect of a region  observed with \emph{Herschel}, the  rescaled map must be 
 re-convolved  with the PSF of the instrument. However, one must take  into account the fact that 
the original map  already results from a convolution of the sky with the PSF, i.e. a kernel which can be approximated with a Gaussian of width,
  $\theta_{\mathrm{beam}}$, which is equal to 8.4, 13.5, 18.2, 24.9, 36.3 arcsec at 70, 160, 250, 350, 500 $\mu \mathrm{m}$, respectively
\citep{Molinari2016}. So the width of the kernel we use to
re-convolve the map is:
\begin{equation}\label{eq:moved_PSF}
\theta_{\mathrm{conv}}=\sqrt{\theta_{\mathrm{beam}}^2\left(1-\left(\frac{d_0}{d_1}\right)^2\right)}.
\end{equation} 
 (iii) The noise in the maps can be well modeled  as a combination of correlated noise 
\citep[which is strongly attenuated by the map making algorithm, see][]{Piazzo2015}
and white noise.
The map rescaling, of course, reduces white noise with respect to the original map by a factor $\sqrt{d_0/d_1}$.
 The sample standard deviation of the noise in the original and rescaled map are  
 $s_N$ and $s_N(\sqrt{d_0/d_1})$, respectively, then to restore the noise level of the original map one has to 
add a white noise image to the rebinned map.
 In this noise image, each pixel is the realization of a Gaussian process with 0 mean and a standard deviation of $s_N\sqrt{1-d_0/d_1}$
  (to
keep in all the simulated maps the same white noise level of the original one).
The $s_N$ was estimated in a box of the original map where no sources and quite low diffuse emission are found, and therefore where the signal 
is essentially due to statistical fluctuation. 

 This procedure is applied to every map at each band.
We decide to  ``move'' the maps of each region for each band to the following virtual distances: 0.75, 1, 1.5, 2, 3, 5 and 7 kpc.
Fig.~\ref{fig:PesreusPSW} shows the original and MMs of the Perseus nebula at $250 \hspace{1 pt} \mu \mathrm{m}$ while in the 
Appendix~\ref{sec:app1} the maps of the remaining  regions are shown.
 Fig.~\ref{fig:PerseusPSW_deg}  displays how the MM of Perseus lose detail at increasing distance, and correspondingly, 
how sources resolved at the original distance become unresolved at large distances.

\subsection{Source extraction and catalogue compilation}
\label{sec:Sextract}


 The detection and photometry of compact sources on the original and moved maps is carried out with CuTEx 
\citep[Curvature Thresholding Extractor, presented in][]{Molinari2011}.
 This algorithm detects the sources as local maxima in the second derivative images and then fits
 an elliptical Gaussian to the source brightness profile to estimate the integrated flux.
 The main output parameters  of the fit are: the peak position, the minimum and maximum FWHM ($\phi_{min}$, $\phi_{max}$) 
 of the fitting ellipse, the peak flux and the integrated flux.
 Since in this paper we intend to treat the moved region at each simulated distance as an independent data set, to be analysed according 
to typical procedure applied to Hi-GAL fields \citep[e.g][]{Elia2016}, we run CuTEx on all the maps at  each simulated distance for each region.
Depending on the brightness level of the region, we adopted a CuTEx set-up more suitable for the inner Galaxy
 \citep[][]{Molinari2016}
or for the outer Galaxy (Merello et al., in prep).
Regions in the outer Galaxy have in general a lower median flux per map with respect to the inner part,
 due to  much lower background emission.
Moreover the detection of sources in the outer Galaxy is more sensitive to pixel-to-pixel noise.
For maps in the outer Galaxy, the CuTEx set-up  also includes the smoothing  of the PACS image and  a lower detection threshold  for   SPIRE and PACS
$160 \, \mu m$.
We therefore made a check on our set of HGBS maps to ascertain which ones require a CuTEx set-up similar to the Hi-GAL one for the outer Galaxy.
Orion A and Perseus have a larger median flux ($\sim70$ MJy/sr) respect to the others.
Furhermore Serpens, Lupus III and IV have 
a median absolute deviation comparable ($\sim11$ MJy/sr) with the outer  Galaxy regions such as the $2^\circ \times 2^\circ$ field centred at
$l\sim160^\circ$,
showing the lack of an extended background.

Therefore  we used the inner Galaxy set up of CuTEx for Orion~A and Perseus while we used the outer Galaxy set up for Serpens, Lupus~III and Lupus~IV.
The fluxes measured with CuTEx are then corrected with the procedure 
described in \citet{Pezzuto2012} to take into account for the fact that
  the instrumental PSF  is not a Gaussian, while CuTEx performs a Gaussian fit.

After the source detection and flux measurement in all  five \emph{Herschel} bands,
 we select the good compact source candidates (band-merging) by applying the procedure described in 
  \citet[][]{Elia2010,Elia2013} as well as \citet{Elia2016}.
The band merging makes it  possible to build the  Spectral Energy Distribution (SED) of the sources.
A SED eligible for a grey body fit must satisfy the following criteria:
1) at least three consecutive fluxes between 160 and 500 $\mu \mathrm{m},$ 
2) showing no dips (negative second derivative),
3) not peaking at $500 \hspace{1 pt}\mu \mathrm{m}$.

 An additional issue in our case is that the absolute astrometry of the  MMs  has no physical meaning because the rescaling of
 the image shrinks their size.
 In the band merging procedure we take  
 care of this issue, to ensure that the coordinates of the detected sources, at different wavelengths, are consistent with each other. 

We address this issue as follows:
the function that we used to execute the band merging works  with  Equatorial coordinates.
For this reason we had to perform the band merging considering the pixel coordinates and the angular extent of
the objects in the MMs, then  rescaled them to the
original 
map and then finally convert them
into the equatorial system. With these quantities
it is finally possible to perform the band merging.

This procedure is repeated for all maps at each wavelength. 
Once the SEDs are built, it is possible to perform a  modified  black body (hereafter grey-body) fit  \citep[e.g.][]{Elia2010} 
described by the equation:
\begin{equation}
F_{\nu}=\frac{M}{d^2}k_0 \left(\frac{\nu}{\nu_0}\right)^{\beta} B_{\nu }(T),
\end{equation}
where $ F_{\nu}$ is the flux at frequency $\nu$; $M$ is the mass of the source located at the distance 
$d$, $k_0$ is the opacity  at the frequency $\nu_0$. We adopt
 $k_0 = 0.1 \ \mathrm{cm^2  \ g^{-1}}$ at $\nu_0= 1000 \ \mathrm{GHz}$ \citep[i.e. $\lambda_{0}= 300 \, \mu \mathrm{m},$][]{Beckwith90}; 
$B_{\nu}(T)$ is the Planck function 
at the temperature $T$. We fixed $\beta$  to 2 as in \citet{Elia2013}.
The flux at 70 $\mu \mathrm{m}$, where present, is not considered for the fit since it is 
mostly due to the protostellar content of a clump, rather than its large-scale envelope emitting as a grey-body \citep{Elia2013}.
 In this way we obtain estimates of temperature and mass for each  source in both the original and moved maps.


\section{Discussion of distance bias}
\label{sec:dist_bias}

\subsection{Number of sources as a function of distance }
\label{sec:Number}

The amount of objects detected by CuTEx in \emph{Herschel} maps is expected to decrease whith increasing simulated distance 
(for each band).
In this section we analyse this effect from a quantitative point of view.
Notice that, since in this section  we study each band separately,
  not all the  sources discussed here constitute a regular SED as those considered in the following sections. \\

\begin{figure}
\includegraphics[scale=0.22]{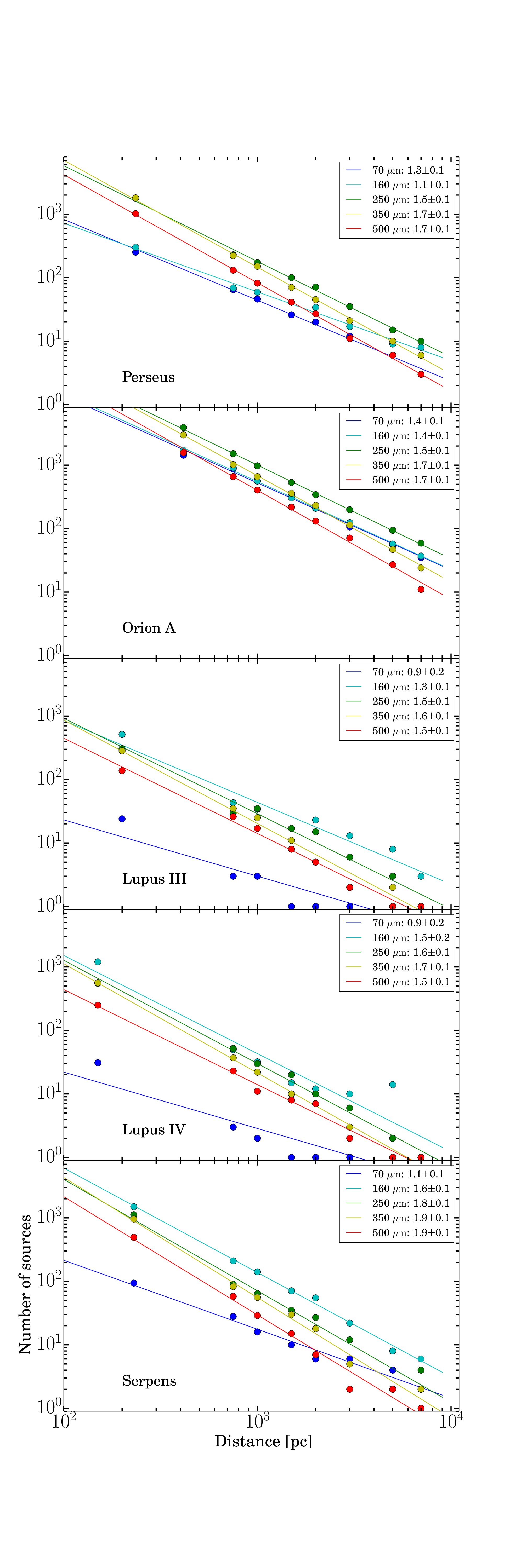}
\caption{Number of sources detected by CuTEx at various bands for each region as a function of the virtual distance at which the map is moved.
The power-law exponents $\delta$ estimated through the best fit are reported in the upper right corner of each panel.
}
\label{fig:number_source_region}
\end{figure}

The decrease of the number of objects with distance (see Fig.~\ref{fig:number_source_region}) 
is due to two main combining effects. 
The first one is related to the decrease of the flux with distance ($F_{\lambda}\propto d^{-2}$): 
if the flux goes below the sensitivity 
limit at a given band the source is not detected any more.
The second one is due to blending of sources that are close to each other in the original map and hence are not resolved any more
at larger
distances. The blending effect may also  prevent  losing some sources (since the flux is an additive quantity) that would be undetected 
 when 
their flux would be below the sensitivity limit.  

In principle since the flux decreases quadratically with distance  and
since in a log-log plot, data shows a remarkable linear correlation, we fit a power law to the data:
\begin{equation}
N_{s}(d)\propto d^{-\delta},
\end{equation}
where $N_{\mathrm{s}}$ is the number of sources at distance $d$.
 The best-fitting power-laws with the
 corresponding values of the $\delta$ exponent are shown in Fig.~\ref{fig:number_source_region} and reported in 
 Table~\ref{tab:table_number}. This exponent is smaller for the $70 \, \mu \mathrm{m}$ and $160 \ \mu \mathrm{m}$ bands 
and is found generally to be between 1 and 1.9.
If we assume that the decreasing of sources is  due to the decrease 
of the flux with distance,  and  
assuming a power-law relation between the number of detected sources and the flux ($N_{\mathrm{s}}  \propto F_{\lambda}^{-\eta}$)
it turns out that $N_{\mathrm{s}}(d) \propto d^{2\eta-2}$. To get the value of $\eta$,
we fit the distribution of the fluxes, for the different regions  at $70 \ \mu m$.
We find that $2\eta-2$ is equal to 1.6, 1.0 and 1.0 for Orion A, Serpens and Perseus respectively. 
There sample is too small for the Lupus III and IV.
We did this analysis only at $70 \mu m$ because in the other bands the effects of the blending are more prominent.
These values are in good agreement with the values of $\delta$ in table \ref{tab:table_number}, considering that we make the simple assumption that 
the distribution of the fluxes is a power-law.

Knowing the slope $\delta$ can turn out to be very interesting for practical purposes, as shown by the following example:
suppose we detect  $N$ objects at a certain wavelength in a  Hi-GAL region located at a distance $d > 1 \, \mathrm{kpc}$; it is possible 
to obtain an estimate of the number of ``real''
sources (at a specific band) that one would expect to observe if the cloud was located at a distance $d_{0}<d$,
using the formula:
\begin{equation}\label{eq:N_d}
N_{0}=N  \left(\frac{d}{d_{0}}\right)^{\delta}.
\end{equation}

A critical point is represented by the choice of $d_0$, which can lead to quite different values of $N_0$.
We suggest using values in the range $200 \, \mathrm{pc} <d_{0}< 400 \, \mathrm{pc}$ 
since this corresponds to compact source scales typical of cores.
For example, let us consider a star-forming region located at $d=5000$ pc, whose Hi-GAL map at $250 \, \mu m$ contains 20 sources:  
how many sources we would see if the region was located at $d_{0}=300$ pc?
Using equation \ref{eq:N_d} we can give a rough estimate of this number assuming $\delta=1.6$ (see Table \ref{tab:table_number})
namely $N_{0}=1802$. 
Applying instead equation \ref{eq:N_d} to a larger $d_{0}$ (e.g. $d_{0}>1000$ pc) would imply to deal 
with clumps also at the original distance (i.e. with still unresolved structures).

Certainly Equation \ref{eq:N_d} requires to confirmation by independent observational evidence, 
such as interferometric measurements aimed at exploring the 
real degree of fragmentation of clumps.

\begin{table}
  \centering
  \caption{Values of the slope $\delta$ of the power law relation between the number of sources and the distance.
 Uncertainty on the values of $\delta$ is almost always 0.1 (see Fig.~\ref{fig:number_source_region}). 
The values of $\delta$ for
 the $70 \ \mu \mathrm{m}$ band for Lupus III and Lupus IV are not very reliable since the number of detected sources is very low.     }
  \label{tab:table_number}
  \begin{tabular}{l*{6}{c}r}
\hline
                     & 70 $\mu \mathrm{m}$ &        160 $\mu \mathrm{m}$             & 250 $\mu \mathrm{m}$      & 350 $\mu \mathrm{m}$ &       500 $\mu \mathrm{m}$   \\
\hline       
Perseus              & $1.3 $     & $1.1 $    & $1.5 $   & $1.7$     & $1.7 $       \\
Orion                & $1.4 $     & $1.4 $    & $1.5 $   & $1.7 $    & $1.7 $           \\
Lupus III             & $0.9 $     & $1.3 $    & $1.5 $   & $1.6 $    & $1.5 $            \\
Lupus IV             & $0.9 $     & $1.5 $    & $1.6$    & $1.7$     & $1.5 $        \\
Serpens              & $1.1 $     & $1.6 $    & $1.8 $   & $1.9 $    & $1.9 $       \\
\hline
  \end{tabular} 
\end{table}

\subsection{Starless and protostellar fraction vs distance}
\label{sec:protopre}

After assembling the SEDs the sources  are classified as follows: if  a $70 \: \mu \mathrm{m} $ counterpart is present,
the source is classified as 
protostellar candidate \citep[e.g.][]{Giannini2012}, hereafter protostellar, otherwise it is considered a 
starless object. The starless sources  can be further subdivided into bound (objects that can form  stars and which are usually called prestellar)
or unbound (transient objects that will not form stars) whether their mass is larger or smaller, respectively, 
than the mass given by the relation of  \citet{Larson1981}
\begin{equation}\label{eq:Larson}
M_{\mathrm{Lars}}=460\left(\frac{r}{ [\mathrm{pc}]}\right)^{1.9}M_{\odot}.
\end{equation}

\begin{figure}
\includegraphics[scale=0.4]{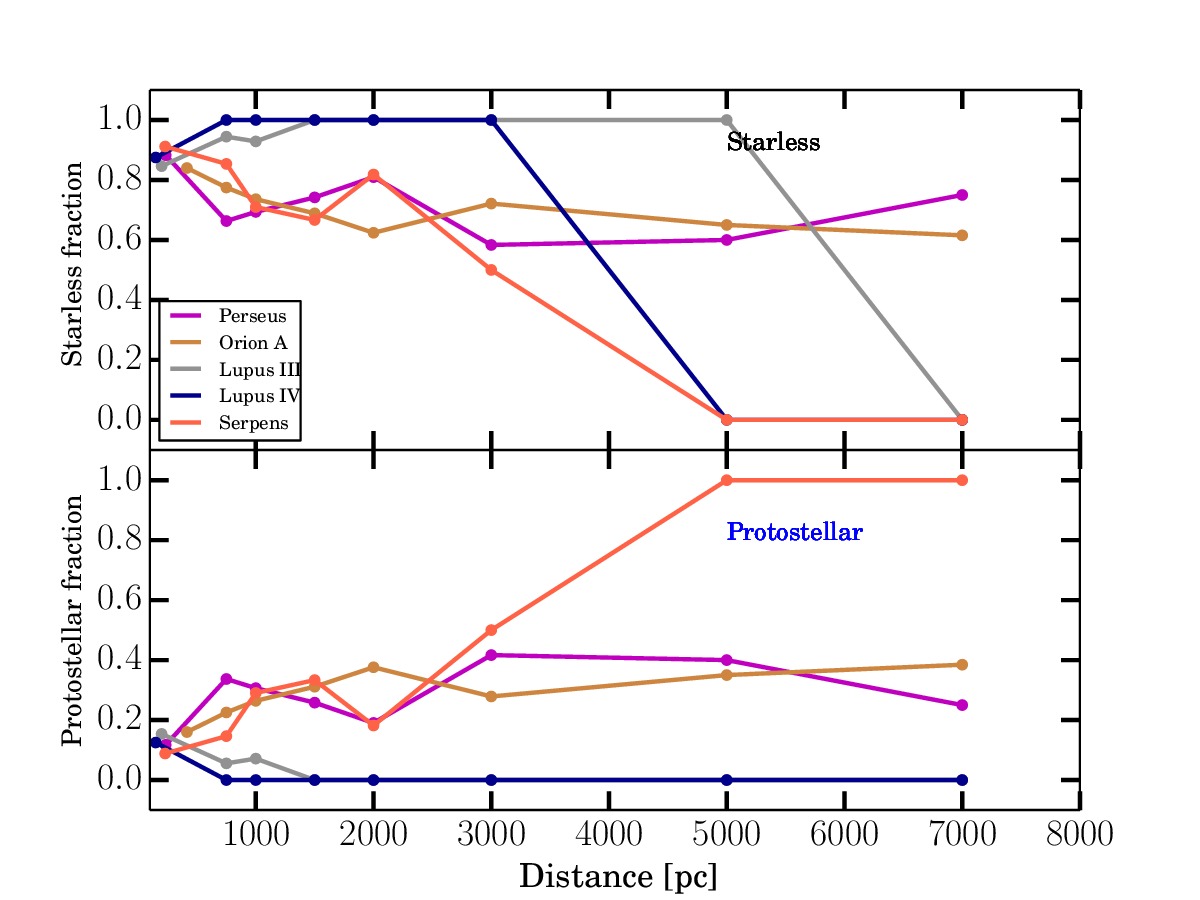}
\caption{Fraction of starless and protostellar objects ($n_{\mathrm{sl}}/n$ and $n_{\mathrm{pro}}/n$) as a function of
 the distance for the five considered regions 
respectively.
}
\label{fig:fraction_pre_pro}
\end{figure}

\begin{figure}
\includegraphics[scale=0.4]{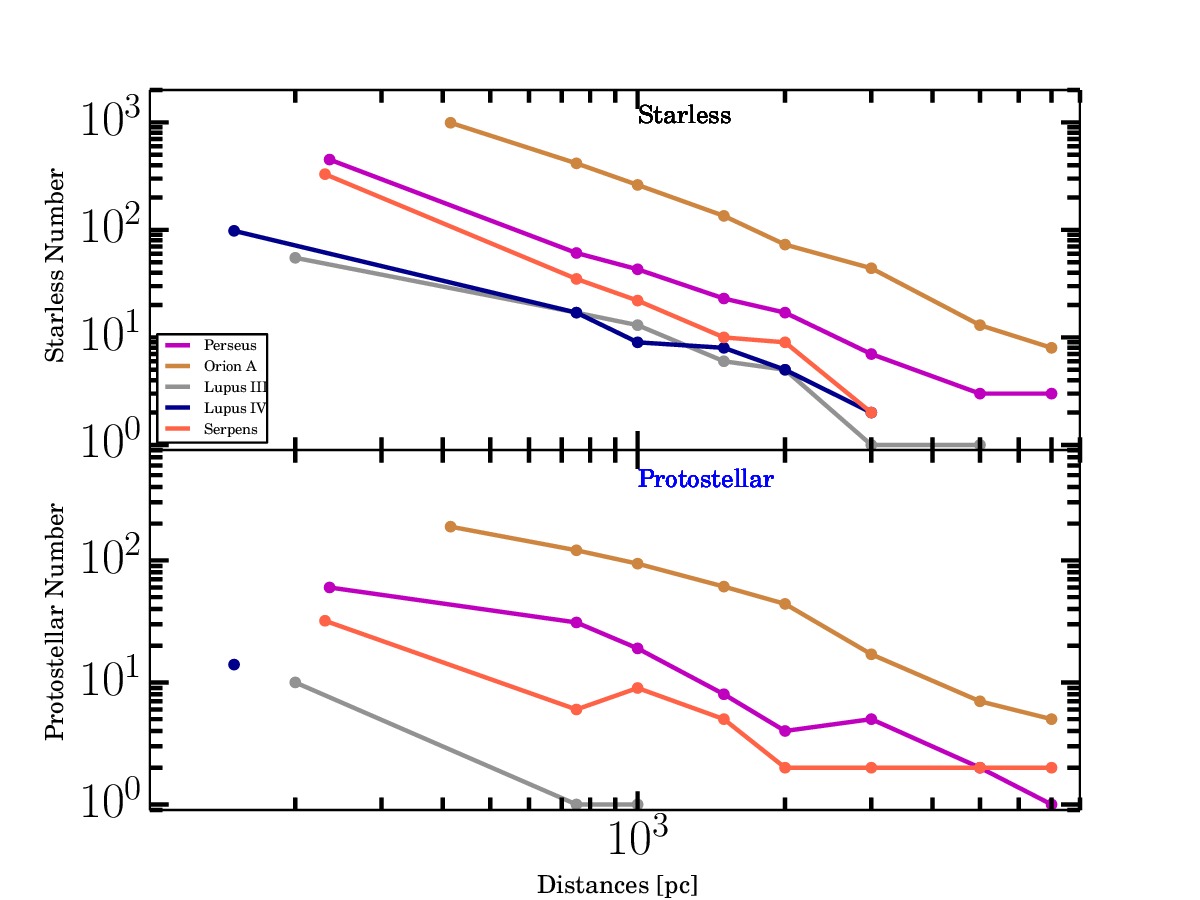}
\caption{Number of starless and protostellar objects ($n_{\mathrm{sl}}$ and $n_{\mathrm{pro}}$) as a function of the distance.
}
\label{fig:number_pre_pro}
\end{figure}

We define the starless and protostellar fraction as $n_{\mathrm{sl}}/n$ and $n_{\mathrm{pro}}/n$, respectively, 
where $n_{\mathrm{sl}}$ and $n_{\mathrm{pro}}$ 
are  the number of starless  and  of protostellar sources respectively and
$n=n_{\mathrm{sl}}+n_{\mathrm{pro}}$ 
is the total number of sources.
In Fig.~\ref{fig:fraction_pre_pro} we show $n_{\mathrm{sl}}/n$  and  $n_{\mathrm{pro}}/n$ while in Fig.~\ref{fig:number_pre_pro} we show
$n_{\mathrm{sl}}$ and
 $n_{\mathrm{pro}}$ as a function of distance, respectively.
The general trend for $n_{\mathrm{pro}}/n$  in  Orion A, Perseus and Serpens is  to increase with distance until it reaches a plateau.
A different behaviour is found in Lupus III, where $n_{\mathrm{pro}}/n$ decreases and in Lupus IV where $n_{\mathrm{pro}}= 0$ to all moved distances
due to the complete lack of detected protostellar sources.

The increase of $n_{\mathrm{pro}}/n$ with distance is explained with  the fact that a source detected at a larger distance is an unresolved object
that probably  contains multiple cores (protostellar but also starless).
Suppose there were an unresolved source, classified as protostellar, detected at a simulated distance $d_{1}$, and actually containing
a certain amount of sources 
 at the original distance $d_{0}$, and suppose that one of them is a protostellar and the remaining ones are starless: 
in that case one would assign a protostellar character to a source which actually contains also a certain number of prestellar cores.
This simple example suggest to us that  in principle $n_{\mathrm{pro}}/n$  is expected to increase  with distance, as long as
$n_{\mathrm{pro}}/n$ reaches a plateau  where the number of detected objects is very low due to of the sensitivity effect.  
The decrease of $n_{\mathrm{pro}}/n$ with distance found for Lupus III and IV may, instead be due to a weak emission at 70 $\mu \mathrm{m}$
at the original distance, which leads to a decrease in the number
of protostellar at the 
larger virtual distances (Fig.~\ref{fig:number_pre_pro}).

\citet{Ragan2016} found that the behaviour of $n_{\mathrm{pro}}/n$ as a function
of the heliocentric distance, in the Hi-GAL catalogue \citep{Molinari2016}, is in agreement with that we derived from 
Fig.~\ref{fig:fraction_pre_pro}. Indeed they found that $n_{\mathrm{pro}}/n$ grows up to 2 kpc and then $n_{\mathrm{pro}}/n$ reaches a plateau.
In their Fig. 4 they found that $n_{\mathrm{pro}}/n \lesssim 0.2$ up to 2 kpc but the order of 0.2 for larger distances.
This  significantly supports  the result we obtained here.

\subsection{Size distribution vs distance}
\label{sec:size}

\begin{figure}
\includegraphics[scale=0.2]{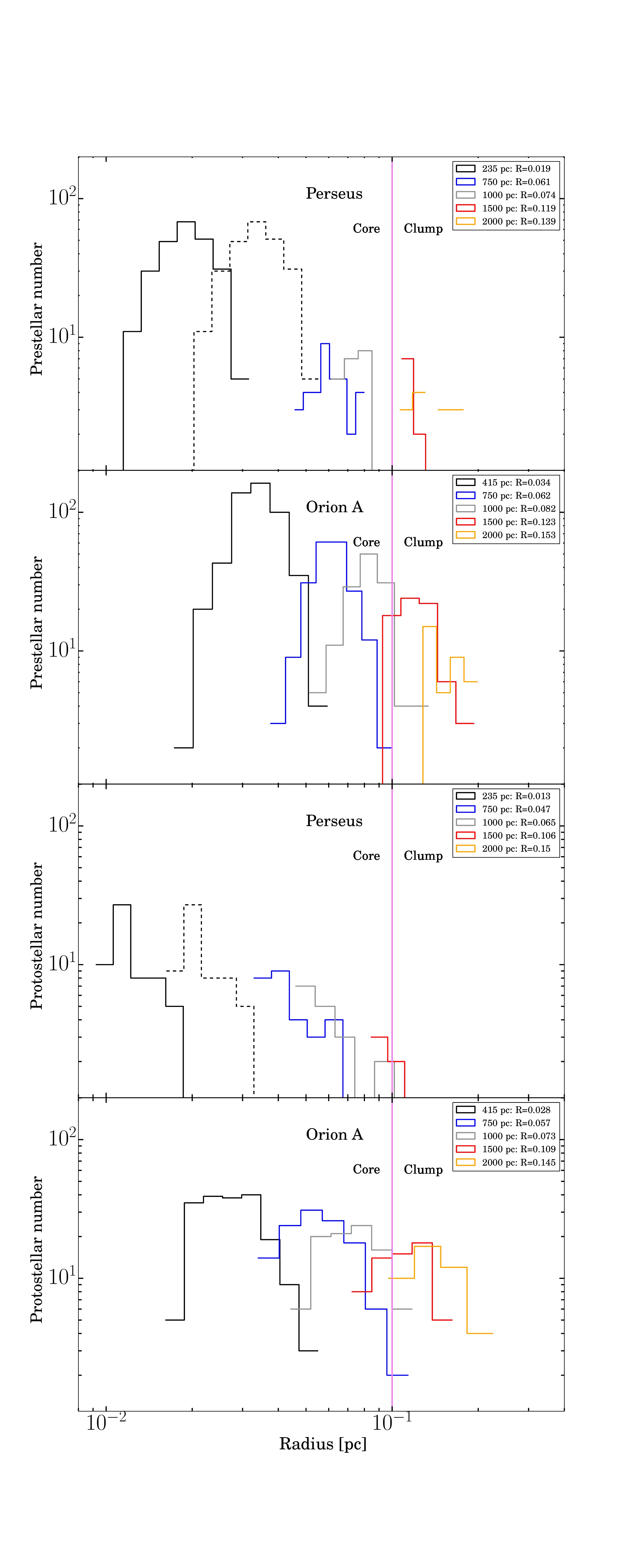}
\caption{Radii distribution of the protostellar and prestellar sources for different distances for the Orion A and Perseus  regions. 
The vertical magenta line represents the separation between 
the cores/clumps according to the classification of \citet[][]{Bergin2007}, namely $R_{\mathrm{sep}}=0.1$ pc.
The values of the mean radius for all regions at each simulated distance are also reported.
Dashed histogram represents the distribution of the radii, in the original map, for Perseus if it was located at an
original distance of 415 pc (same as Orion A) instead of 235 pc (see appendix \ref{sec:serp}). 
}
\label{radius_distribution_all_regions}
\end{figure}

A common feature of molecular clouds is  the hierarchical structure they show, containing bright agglomerates called clumps, that
in turn are formed
by smaller condensations called dense cores (see section~\ref{sec:intro}). 
In Fig.~\ref{radius_distribution_all_regions} the distribution of the radii (at $250 \ \mu m$) of the
detected sources is shown, for the Perseus and Orion A maps, at different distances.
We do not show the same for the Lupus III, Lupus IV and Serpens  due to poor statistics.
As one can see from these figures, the physical radius increases on average, and hence the fraction of cores in the overall population
of detected compact sources
decreases  with distance.
As it emerges  from this figure, up to $d=1000$ $\mathrm{pc}$ most of the detected sources are classified as cores, while at larger
distances they are generally classified as clumps.
Of course  the 
threshold $r\lesssim$~0.1~pc is not so strict and the transition between core and clump definition is not so sharp. 
Furthermore, a difference is found between the  prestellar
and protostellar source distribution:
the former is characterized on average by larger radii than  the latter, as found, e.g., by \citet{Giannini2012}.
In Fig~\ref{radius_distribution_all_regions} 
the behaviour of the average radius 
for each of the two populations is also reported (top right corner): that of prestellar 
objects $\left<R_{\mathrm{pre}}\right>$ is larger than $\left<R_{\mathrm{pro}}\right>$ for
protostellar sources, but this gap appears to get smaller at
larger distances.
If we define $q$ such that $\left<R_{\mathrm{pro}}\right> = q \left<R_{\mathrm{pre}}\right>$, $q$ is found to be
  0.68, 0.77, 0.88, 0.89 and 1.08  for the Perseus region at 235, 750, 1000, 1500 and 2000 pc, respectively, while $q$ is 
  0.82, 0.92, 0.89, 0.89 and 0.95   for the Orion A region at 415, 750, 1000, 1500 and 2000 pc, respectively.



\subsection{Association of the ``moved'' sources with the original ones}
\label{sec:assosiation}

\begin{figure}
\includegraphics[scale=0.33]{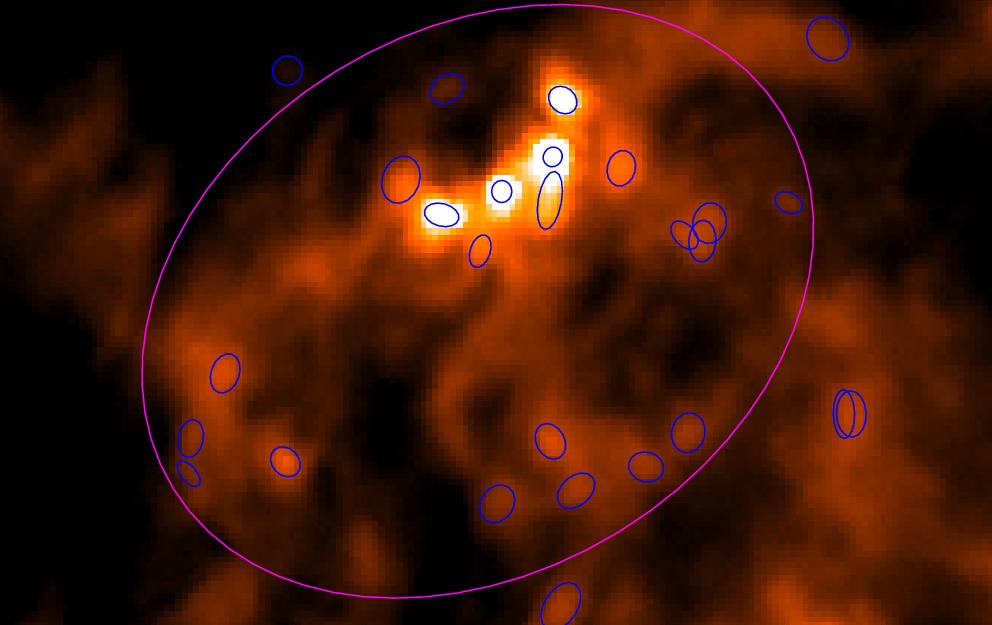}
\caption{Example of association between a source detected in a moved map and the population of cores present in the original map.
A portion of the original Perseus map at 250 $\mu m$ is shown, with blue ellipses representing the sources detected with CuTEx.
The magenta ellipse represents a source detected in the MM at 5~kpc and projected on to the original map. 
}
\label{fig:associazione}
\end{figure}

In section~\ref{sec:size} we have seen that for $d\gtrsim1.5$ kpc the source sizes are such
that the objects are 
 classified as clumps. For further analysis it becomes 
important to identify and count the  cores present in the original map that are contained in  clumps found in the MM.
This information can be used, for example, to estimate which 
fraction of mass  of such clumps comes 
from the contained cores, and what comes from the diffuse ``inter-core'' material.
To associate the sources detected at a given distance $d$ with the original population of  
objects, 
we projected the ellipse, found at 250 $\mu m$ at distance $d$, back to the original distance $d_{0}$, and consider the sources falling
 within such ellipse:
an example  is shown in Fig.~\ref{fig:associazione}.
At this point it is possible to associate the sources of the moved map and those of the original one in two ways: 

1)  by doing band-specific associations, including those without regular SEDs.
For example, suppose to consider an object, detected at $250 \, \mu \mathrm{m}$ at 5 kpc and 
then count the sources that, in the map corresponding to the original distance and to the same band, lie inside the area occupied by this object. 
2)Association using only sources with regular SEDs.
As an example, let us  suppose there is a clump with a regular SED at 5 kpc and then count the number of cores
 with a regular SED
that are contained in the clump, in the original map.
This association between a clump and the contained cores is very useful because one can  decompose an unresolved object (clump) 
into its smaller components (cores).
Therefore, from a practical point of view, this corresponds to observing an unresolved object with a higher resolution and hence 
revealing its internal structure.

The former approach  will be used in section~\ref{sec:diffuse} where we discuss the contribution of the diffuse emission separately 
for the various different wavelengths, while the latter will be used both in the following and in section~\ref{sec:association_mass_l}, 
where we will discuss the relation between 
the physical properties of the moved sources and the original core population. 

\begin{figure}
\includegraphics[scale=0.61]{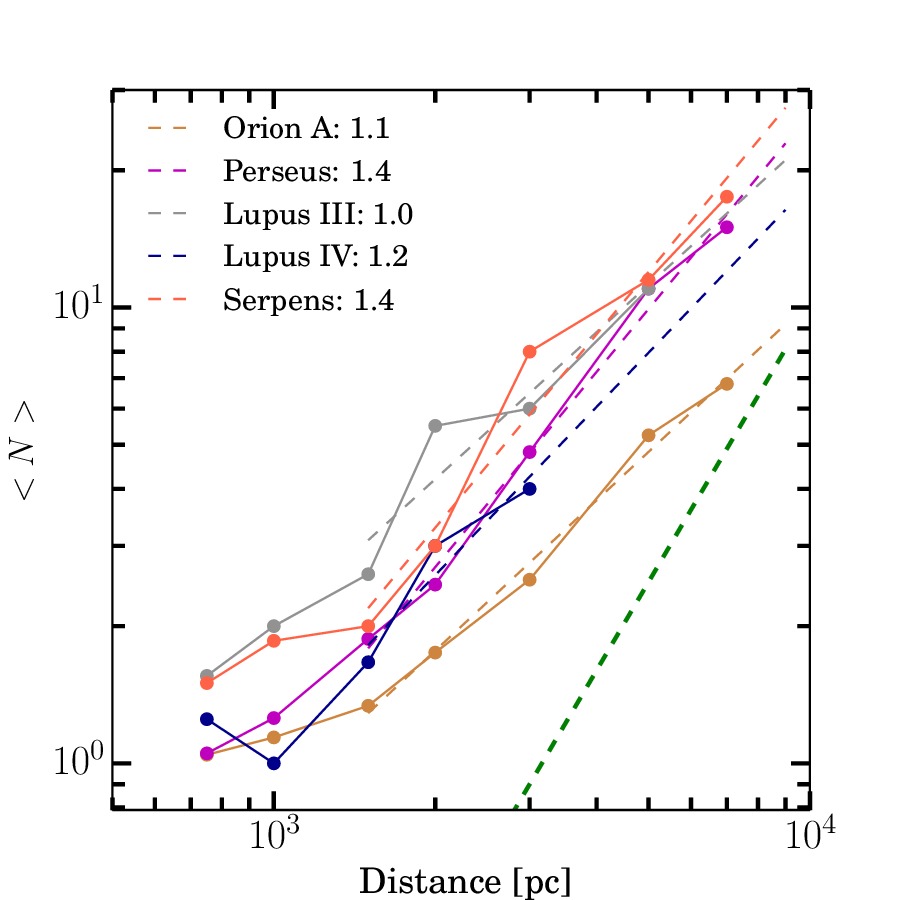}
\caption{Average number of cores that are contained within a clump at the moved distances for each region.
  The best power-law fit is plotted as a dashed line, and the corresponding exponent, estimated from $d \geq 1500$~pc, is also reported.
Green dashed thick line: a reference power law with exponent 2.
}
\label{fig:association_average_number}
\end{figure}

Fig.~\ref{fig:association_average_number} shows the average number of cores $\left<N\right>$, at the original distance, 
that are merged into one single  source
with a regular SED, at the moved distances.
As one can see from this figure, $\left<N \right>$ increases slowly, with distance, up to 1500 pc and then tends to increase faster.
We fit the data with a power law $\left<N\right> \propto d^{\zeta}$, starting from 1500 pc and we find values for $\zeta$ 
between 1 and 1.5 for the different regions.
The slower increase of $\left<N \right>$  at smaller distances
 is simply due to the fact that the sizes of the 
objects in the MMs are more similar to the sizes of the original cores.

Moreover, the values of $\zeta$ are always smaller than 2, that is  the expected value if the distribution of cores in 
the maps was uniform.
Since, on the other hand, the actual core distribution  is far from being uniform, but rather  it is clustered,  it is reasonable
 to find values of $\zeta<2$. 

 The values of  $\left<N \right>$ depend both  on the original average  surface density $\Sigma$ (measured as the number of sources per
 $\mathrm{pc}^2$)
of the core population, and on the
minimum spatial scale present in the original maps.
The latter effect explains, for example,
 why the  values of  $\left<N \right>$  in the Orion A are smaller than in the other regions: the spatial 
detail corresponding to the \emph{Herschel} resolution
in Orion A is coarser than  in Lupus, Serpens and Perseus due to the larger distance.
 The former effect, is instead, related to $\Sigma$ which is  93.0, 56.4, 50.4, 39.7, 129.8 $\mathrm{pc^{-2}}$ for
Orion A, Perseus, Lupus III, Lupus IV and Serpens, respectively.
Considering two regions with comparable original distance, namely Serpens and Perseus, this effect can be appreciated, since 
$\Sigma$ is larger in Serpens and, correspondingly, larger $\left<N\right>$ are systematically 
found for Serpens than for  Perseus.


\subsubsection{Diffuse emission}
\label{sec:diffuse}
\begin{figure*}
\includegraphics[scale=0.687]{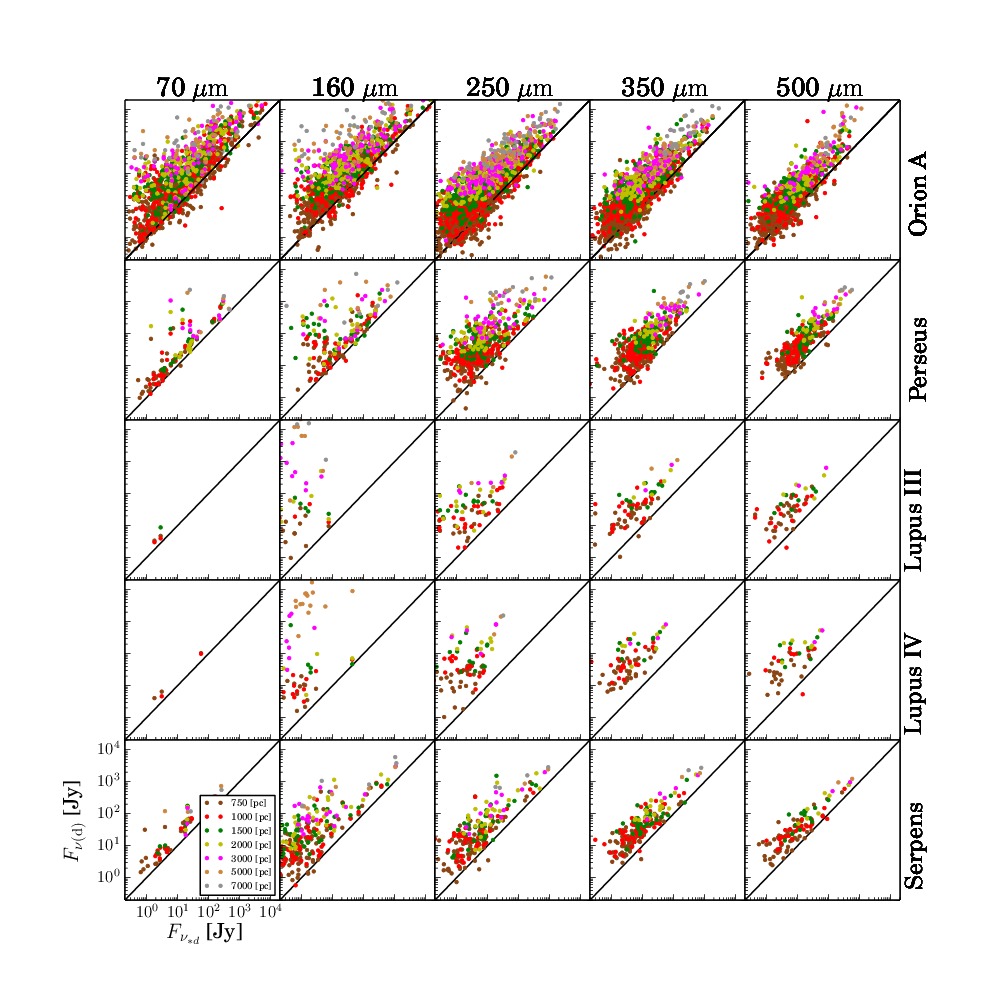}
\caption{Flux of a source in the MM ($F_{\nu(\mathrm{d})}$) vs the sum of the flux  $F_{\nu_{*\mathrm{d}}}$ of the sources in the original map that are within 
the moved source. The columns correspond to five different Herschel bands, while the rows correspond to the five considered regions.
}
\label{fig:associazione_flux}
\end{figure*}

A first comparison between the properties  of the sources  detected in the MM and those  of the 
sources at the original distance that are within the rescaled object can be carried out looking at the total fluxes.
We perform such analysis separately 
 at each wavelength,
therefore we also include detections that do not contribute to build regular SEDs.
Therefore we associate sources through  the method 1) described in section~\ref{sec:assosiation}.
Let $F_{\lambda(\mathrm{d})}$ the flux of a source detected in a MM at a wavelength $\lambda$ and at distance $d$ rescaled to the original map,
and $F_{\lambda_{*\mathrm{d}}}=\sum_{\mathrm{i=1}}^{n} f_{\lambda_{\mathrm{i}}} $ the sum of the flux of the sources that are within the moved source at the original distance.
Fig.~\ref{fig:associazione_flux} shows $F_{\lambda(\mathrm{d})}$ vs $F_{\lambda_{*\mathrm{d}}}$ for different regions at each  wavelength:
it appears that, as expected, the contribution of the diffuse emission $F_{\lambda(\mathrm{d})}$-$F_{\lambda_{*\mathrm{d}}}$ increases
 with distance.
This is not surprising since the objects in the map  are patchily distributed (see Fig.\ref{fig:associazione}) and since the  size of 
 the sources increases with $d$ (see section~\ref{sec:size}).
 As the radius becomes larger, the contribution of the diffuse inter-core emission, which is expected to be more
uniformly distributed, increases quite regularly with the increasing area of the source (in addition to this,
an increasing amount of background emission is included in such flux estimate, as shown at the end of this section
through Fig.~\ref{fig:bck}). 
On the contrary, the contribution of emission from cores increases with distance depending on the degree
of clustering.
 
\begin{figure*}
\includegraphics[scale=0.46]{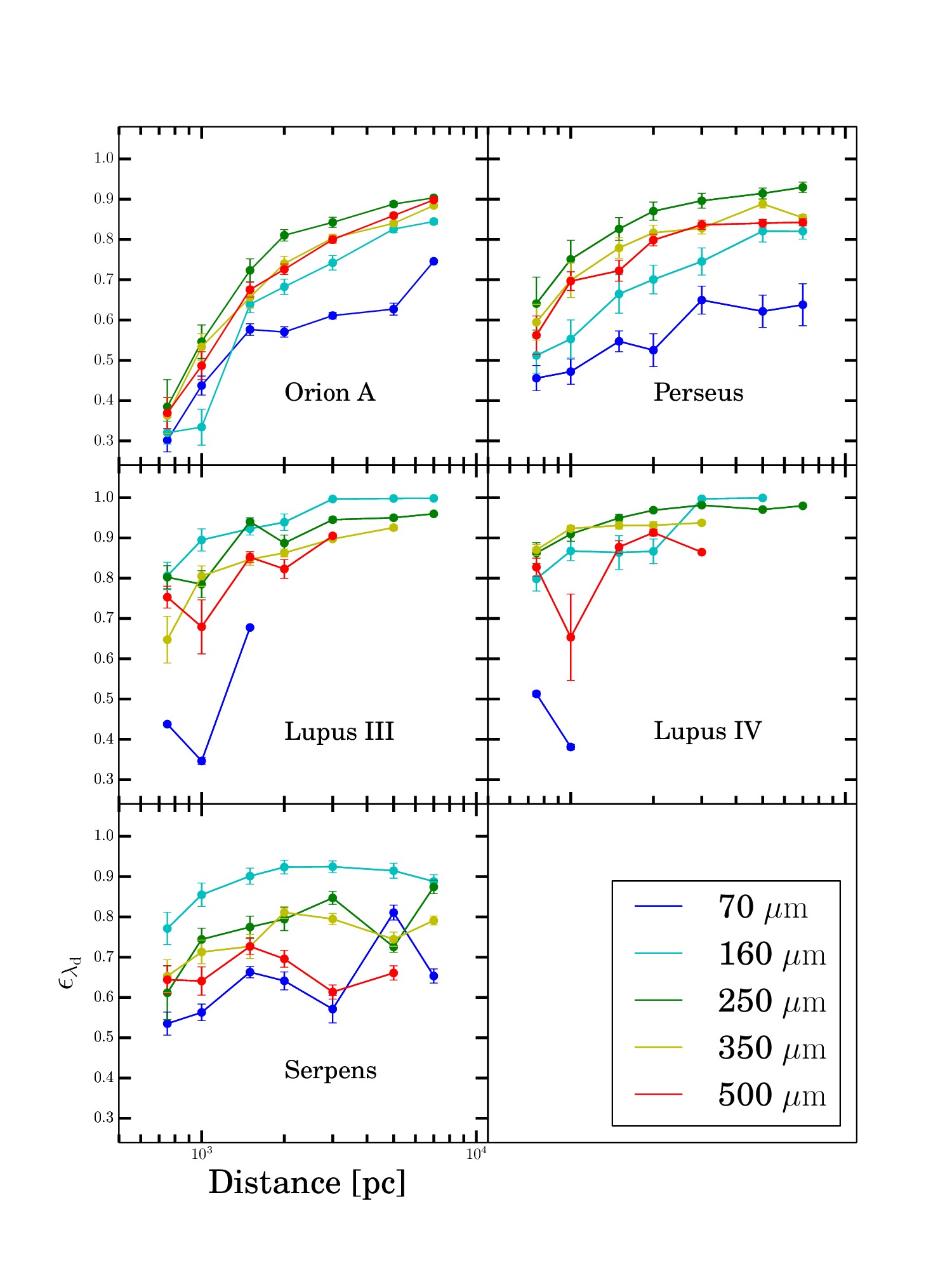}
\caption{Mean fraction of diffuse emission 
 $\epsilon_{\lambda(\mathrm{d})}=  \left<( F_{\lambda(\mathrm{d})}-F_{\lambda_{*\mathrm{d}}})/F_{\lambda(\mathrm{d})}\right>$ vs virtual distance.
}
\label{fig:associazione_flux_efficiency}
\end{figure*}

\begin{figure}
\includegraphics[scale=0.46]{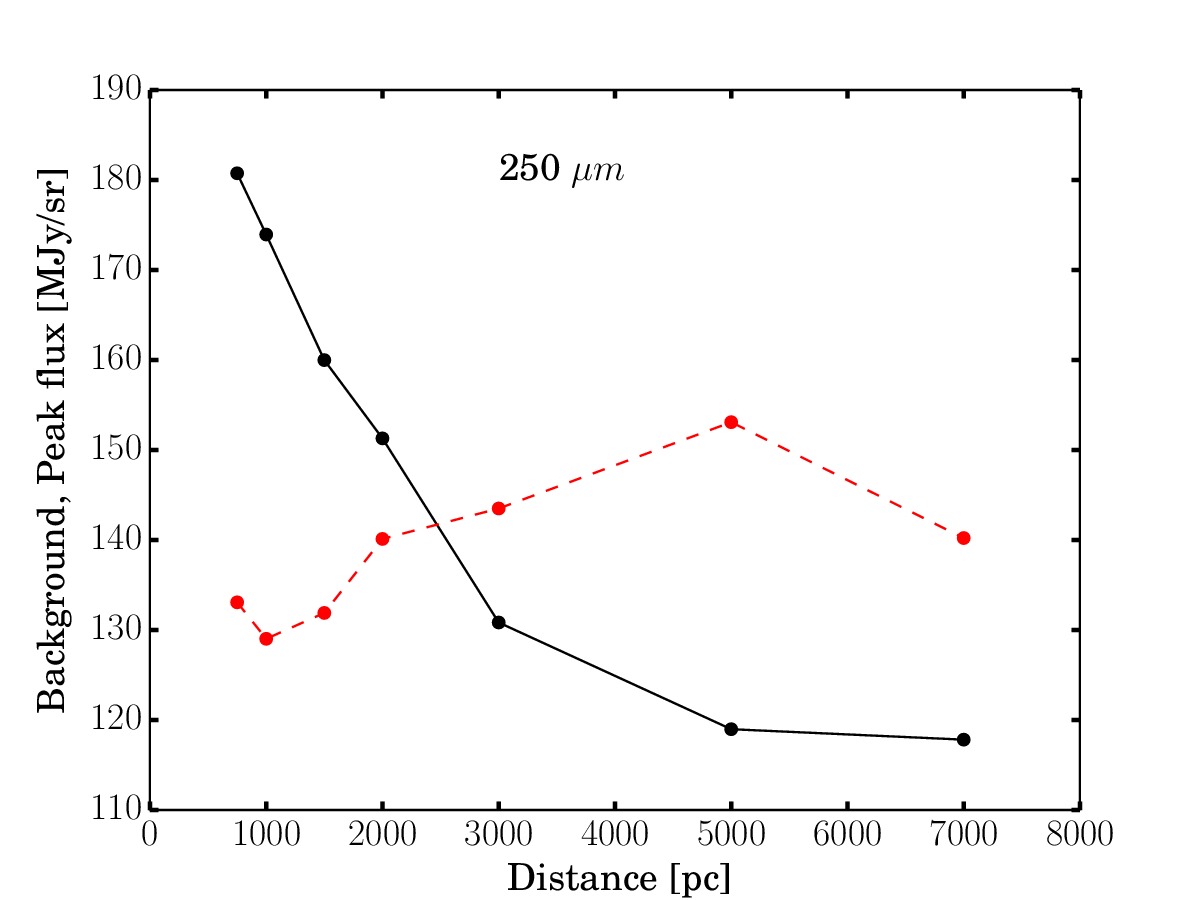}
\caption{Red dashed: mean peak flux $F_{\mathrm{p}}$ at 250 $\mu \mathrm{m}$ for all the regions merged as a function of distance.
Black solid: same as red dashed but for the background emission.
}
\label{fig:bck}
\end{figure}

Figure.~\ref{fig:associazione_flux_efficiency} shows the average fraction of the diffuse emission 
  $\epsilon_{\lambda(\mathrm{d})}=  \left<( F_{\lambda(\mathrm{d})}-F_{\lambda_{*\mathrm{d}}})/F_{\lambda(\mathrm{d})}\right>$ vs distance.
As one can see from this figure,  the contribution of the diffuse emission at $\mathrm{70 \ \mu m}$ is lower than at larger wavelengths
for all the considered regions; this  is  due to the fact that the $70 \, \mu \mathrm{m}$ emission is typically
associated with  protostellar activity 
 \citep[e.g.][]{Dunham2008,Elia2013} and is therefore more concentrated in compact structures
 than the emission  in the other bands, 
which appears instead arranged in a diffuse network of cold filaments.
Furthermore, $\epsilon_{\lambda(\mathrm{d})}$ increases with distance up to a certain point, which depends on the region, after which
it tends to reach a plateau.
In conclusion this analysis suggests that the contribution of the diffuse emission for large distances ($d\gtrsim1$ kpc), i.e. a typical
case for Hi-GAL sources,
goes from  $50\%$ up to $95\%$ (depending  on the region) 
 except for the 70 $\mu m$ band where it goes from $50\%$ up to $80\%$ (see Fig.~\ref{fig:associazione_flux_efficiency}). 
 These  large values of  $\epsilon_{\lambda(\mathrm{d})}$ suggest that most of the clump emission is due 
to the diffuse inter-core emission when the clump is located far away ($d\ge 1$ kpc). This has to be taken into account 
to distinguish the whole clump mass from the fraction of it contained in denser substructures, possibly involved in star formation processes.
Such behaviour can not be simply explained with the increasing physical size of the source, but it must be take 
into account how the background level 
estimate changes with distance.
Indeed we expect that the background level for the sources detected in the original map is higher than that generally found 
 in the moved maps, since the background, in the original map, is due to the inter-core emission, while in the MM
it is the weaker cirrus emission on which the entire clump lies.

Since CuTEx derives the flux of a source by fitting a 2-D Gaussian with the formula $F_{\lambda} = 4.53 \, a \, b \, F_{\mathrm{p}}$,
where $a$ and $b$ are the semi-axes at half maximum of the elliptical Gaussian and $F_{\mathrm{p}}$ is the peak flux measured 
in MJy/sr,
with increasing distance, $a$ and $b$ are constrained to have physical plausible size as in general is done with the extraction tool,
so the physical area we use to scale $F_{\lambda}$ increases quadratically, on average.
Moreover the background level is generally expected to be estimated lower and lower (see above), producing an increase of $F_{\mathrm{p}}$.
To show that,
we computed the mean value of the background and  of  $F_{\mathrm{p}}$ for all the regions merged together 
 as a function of distance:  in
 Fig.~\ref{fig:bck} 
the background is
 found  on average to become smaller with $d$, and 
correspondingly $F_{\mathrm{p}}$ becomes larger. 

Therefore, the derived values of $\epsilon_{\lambda(\mathrm{d})}$ may be explained roughly with the following considerations: 
the value of $\epsilon_{\lambda(\mathrm{d})}$ for one source can be modeled by the formula
\begin{equation}
\epsilon_{\lambda(\mathrm{d})}=1-\frac{\sum_{i=1}^{N} a_{i} b_{i} F_{\mathrm{p_{i}}}}{a b F_{\mathrm{p}}}
\end{equation}
where $a_{i}$, $b_{i}$,  $F_{\mathrm{p_{i}}}$ are the parameters  of the objects in the original map that are contained within a 
larger object at the moved distance $d$ having $a$, $b$, $F_{\mathrm{p}}$ parameters in turn.
The product $ab$ is proportional to the square of the distance and  we roughly assume that the peak flux of the sources at the original distance 
is the same for all sources. The peak flux   
$F_{\mathrm{p}}$ is weakly dependent on distance (Fig.~\ref{fig:bck}): on one hand, the dramatic drop of the background emission
 at increasing distance should be expected to produce correspondingly, 
by subtraction, a strong increase of $F_\mathrm{p}$. However this is not observed mostly due to the fact that the
distance increase implies the averaging of boxes of pixels implied by the map rebinning we impose to
simulate the distance effect. 
 In particular we find that $F_{\mathrm{p}}$ at the largest
probed distance is at most 
1.5 times the average peak flux for smaller distances, therefore adopting this value we get
that $\epsilon_{\lambda(\mathrm{d})}=1-\left(\frac{d_{0}}{d}\right)^2 N(d)/1.5$ where $N$ is the number of contained sources
 within the moved source.
For example, if we consider the Perseus region observed at 250 $\mu \mathrm{m}$ at 5000 pc  assuming $N(d)=22$ (as in Fig.~\ref{fig:associazione}),
 we get 
$\epsilon_{\lambda(\mathrm{d})}=0.97$, which is in good agreement, despite the naivety of the model, with
 the result of Fig.~\ref{fig:associazione_flux_efficiency},
namely $\epsilon_{\lambda(\mathrm{d})}=0.9$.

\subsubsection{Physical properties of the ``moved'' clump and those of the original core population}
\label{sec:association_mass_l}
\begin{figure}
\includegraphics[scale=0.59]{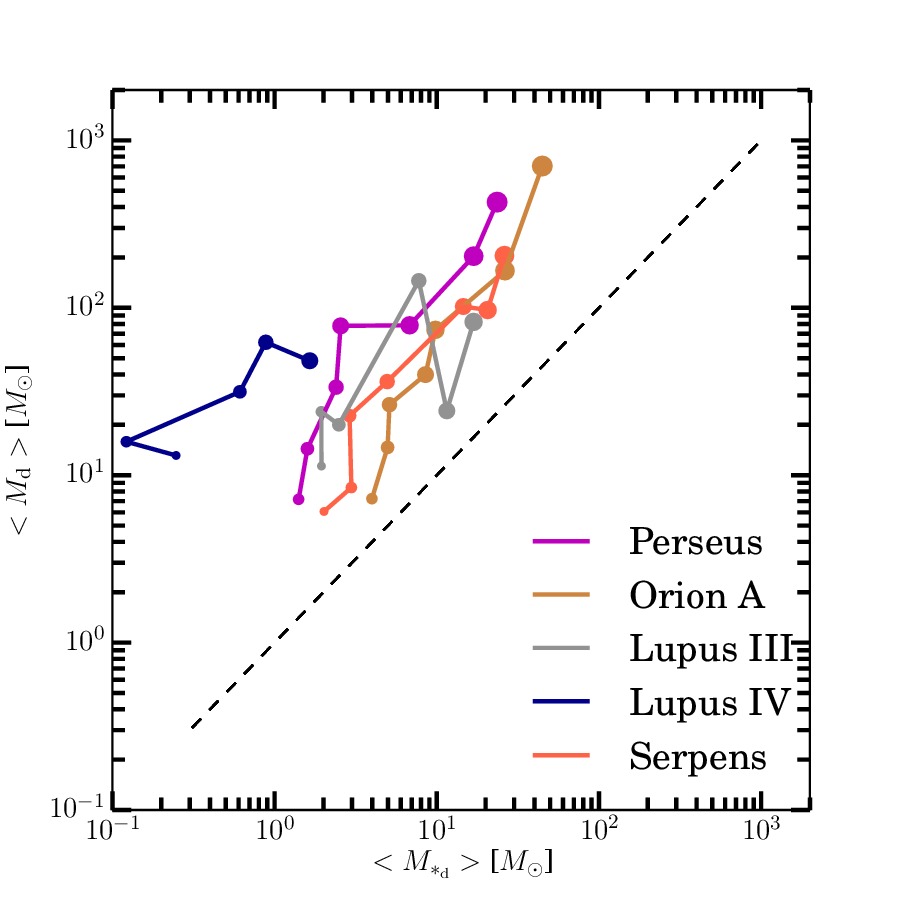}
\caption{Average mass $M_{\mathrm{d}}$ of the moved sources for different distances, vs the average of the sum of the 
masses of the sources in the original 
map that are inside the rescaled sources.
The size of the filled circles
 increases with distance. Black dashed line is the bisector  of the plot.
}
\label{fig:associazione_mass_lum}
\end{figure}

In this section we consider the mass of those sources of the MM for which we performed the grey-body fit,
to understand how these quantities for a moved 
 source mirror those of  
the original core population rather than those of the diffuse material.
We define  with $M_{\mathrm{d}}$  the mass  of the
moved detected sources (protostellar and prestellar) at distance $d$, and with $M_{*_{\mathrm{d}}}=\sum_{\mathrm{i}=1}^{n} m_{\mathrm{i_{d}}}$ 
 the sum of the masses   of all 
the original sources (protostellar and starless) 
that lie inside 
the  source when it is reported in the original map.
Figure~\ref{fig:associazione_mass_lum}  shows the average values of $M_{\mathrm{d}}$ vs $M_{*_{\mathrm{d}}}$ at distance $d$:
we find that $\left<M_{\mathrm{d}}\right>$ is always larger than $\left<M_{*_{\mathrm{d}}}\right>$. 
A similar conclusion can be obtained if one uses the median statistics, instead of the mean: the median mass of the clumps 
at large distances is by far larger than the
sum of the masses of the contained cores, as weel.

\begin{figure}
\includegraphics[scale=0.41]{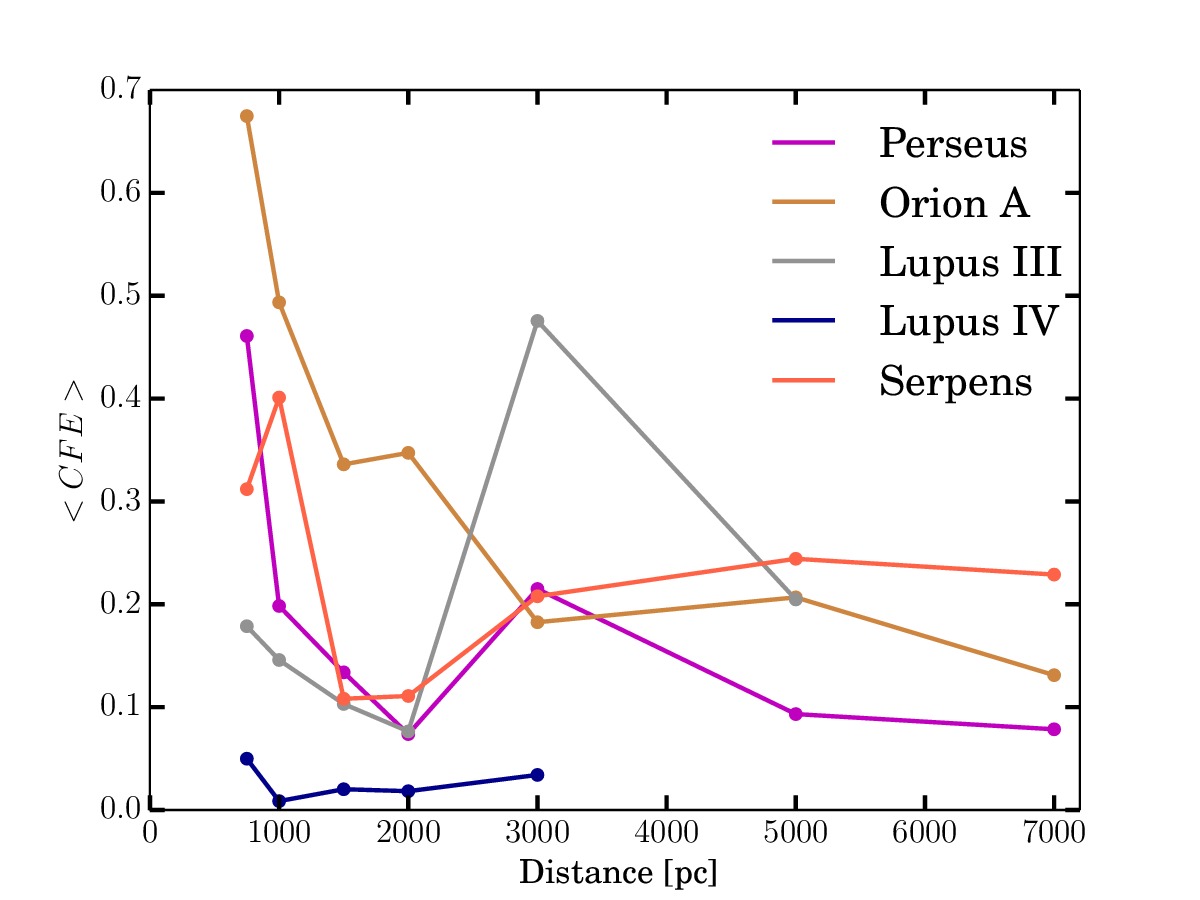}
\caption{Average core formation efficiency   vs distance for each of the considered regions.
}
\label{fig:CFE}
\end{figure}

We want to know the effect of distance on the determination of the core formation efficiency.
The average
core formation efficiency, that we define as $\left<CFE\right> =\left<M_{*_{\mathrm{d}}}/M_{\mathrm{d}}\right>$, 
can be derived  from the  values plotted in  Fig.~\ref{fig:associazione_mass_lum} (panel a).
In Fig.~\ref{fig:CFE} we show  $\left<CFE\right>$ as a function of distance for each of the considered regions.
It can be seen that
 the $\left<CFE\right>$  changes from region to region
and  tends to decrease 
with $d$ for distances below 1500 pc while,
at large distances, it becomes  independent of $d$.
The decrease of the $\left<CFE\right>$ with $d$ below 1500 pc is due to the fact that the size of the clumps
and the contained cores are quite similar and hence the fraction of diffuse material remains low. In fact, this effect is particularly prominent
  in Orion A, because it is located at 415 pc while we do not see this effect in the Lupus IV which is located at 150 pc.
The values of the $\left<CFE\right>$ at large distances are between  $1\%$ and $20\%$, depending on the region.

\subsection{Mean temperature vs Distance}
\label{sec:temp}

\begin{figure}
\includegraphics[scale=0.4]{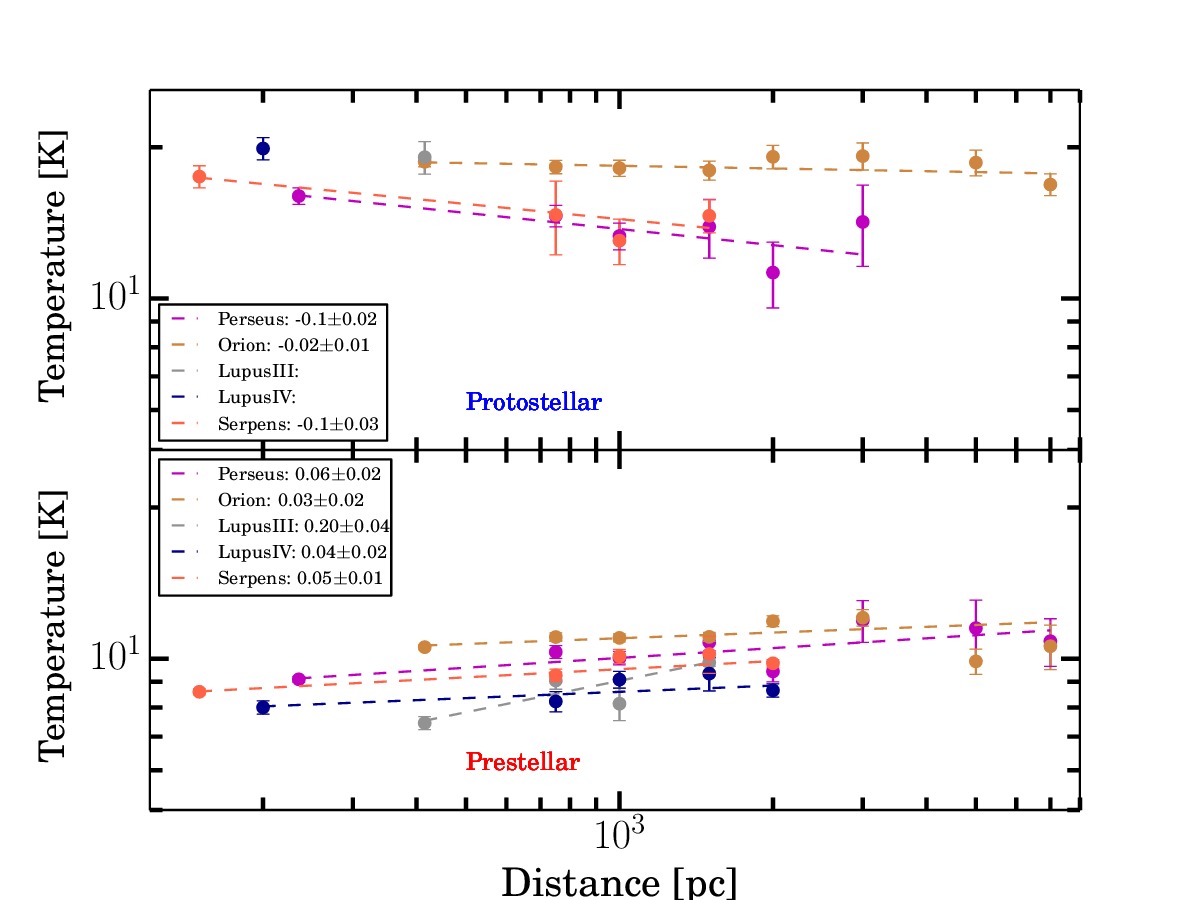}
\caption{Mean temperature of the prestellar (bottom) and protostellar (top) sources as a function of distance. The best fit is also plotted with
the relative power-law exponent.
}
\label{fig:temperature_mean}
\end{figure}

Let us consider now the effects of the distance on the mean temperature of the sources  
detected in each region at all virtual distances, comparing it with the temperature of the 
original population of cores that fall within each clump.
We  fit the average temperature with a  power-law  \\
$\left<T\right>_{\mathrm{d}} \propto d^\xi$  .
The uncertainties of $\left<T_{\mathrm{d}}\right>$ are  given by $\frac{S_{\mathrm{T_{d}}}}{\sqrt{n}}$, where $S_{\mathrm{T_{d}}}$ is the sample
standard deviation of the temperature at distance $d$,
and $n$ is the sample size.
Fig.~\ref{fig:temperature_mean} shows  $\left<T_{\mathrm{d}}\right>$ for each $d$ together with the best power-law fit and the values of $\xi$.
We find that
$\left<T\right>_{\mathrm{d}}$ increases slowly with distance for the prestellar objects while it decreases 
slowly for the protostellar ones.
The values of $\xi$ for the prestellar objects (see Fig.~\ref{fig:temperature_mean}, lower panel) are 0.06, 0.03, 0.2, 0.04, 0.05 for 
Perseus, Orion A, Lupus III, Lupus IV and Serpens respectively. These slopes are very shallow but are  positive anyway.
An opposite trend (i.e. temperature weakly decreasing with distance) is found for the protostellar objects: the values of $\xi$ 
 are  -0.1, -0.02 and -0.1 for Perseus, Orion A, and Serpens
respectively. The statistics are not sufficient at large distance to perform the fit for the Lupus III and IV regions.
It appears unlikely that both of those behaviours are due by chance, since we systematically find these trends
in all the considered regions.

We are quite confident that the decreasing behaviour of temperature for the protostellar sources and the increasing one  for the 
prestellar ones  is  due to the effect of confusion  and can be possibly explained by taking into account two main factors likely to be concomitant
in most cases.

\begin{figure}
\includegraphics[scale=0.5]{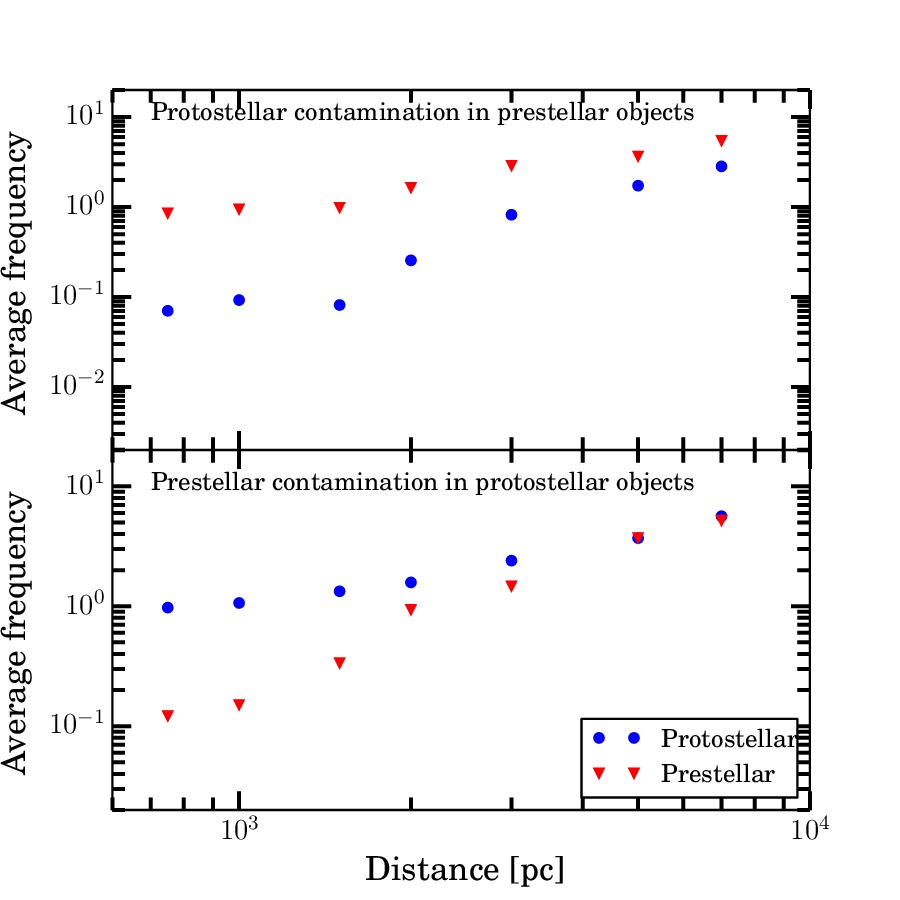}
\caption{Upper panel: average protostellar  contamination in prestellar clumps.
Lower panel: average prestellar contamination in protostellar clumps.
At each distance all the regions are merged  together.
}
\label{fig:proto_pre_contamination}
\end{figure}

The first is related to the fact that the protostellar objects are typically warmer than the prestellar ones \citep[e.g.][]{Elia2013}.
A  protostellar clump detected at a large distance is probably an unresolved object  containing in turn some smaller objects 
that can be starless or protostellar. If inside the unresolved  protostellar clump there are some prestellar  ones
  this can decrease 
the global average temperature.
 In the lower panel of Fig.~\ref{fig:proto_pre_contamination}
we show the prestellar contamination of sources, namely the average number of prestellar cores contained in protostellar clumps (for
 all the regions merged together).
The prestellar contamination, as expected, is close to 0  up to 1000 pc and then starts to get larger, in particular between 5000
and  7000~pc the average number of original prestellar cores contained in the protostellar moved clumps 
becomes the same as the original protostellar objects.
 On the other hand, if we take an unresolved  prestellar clump at large distance this can contain some protostellar cores, whose flux  at
70 $\mu \mathrm{m}$ goes below the sensitivity threshold (and hence not detected), 
 that can contribute anyway to increase the average temperature.
Indeed, although the signature of the presence of a protostellar component would remain undetected, in such case the effect 
of this component on the remaining wavelengths ($\lambda>160 \ \mu m$) would remain still observable as an unnaturally high temperature for a 
prestellar source.
In the upper panel of  Fig.~\ref{fig:proto_pre_contamination} we show the average protostellar  contamination in  
the prestellar clumps:
 the protostellar contamination is close to 0 at low distances while it becomes larger at increasing distances. 
The second effect  responsible for the temperature decrease for the protostellar sources  and  increase  
for the prestellar ones is related to the presence of the diffuse inter-core material, which is known for being typically
 warmer than the prestellar sources and colder 
than the protostellar ones \citep{Elia2013}.  This  might
homogenize the temperature of the protostellar and prestellar as the physical radius of the clumps increases with distance.

\section{Mass-Radius relation }
\label{sec:mass}

The mass vs radius relation (MR) for cores/clumps is a useful tool for checking the  conditions  for  massive star
formation (MSF). 
Indeed several authors \citep[e.g.][]{Krumholz2008,Kauffmann2010} use such diagram to identify 
thresholds in column density supposedly required for the
formation of stars with $M>10M_{\odot}$. 

\begin{figure}
\includegraphics[scale=0.5]{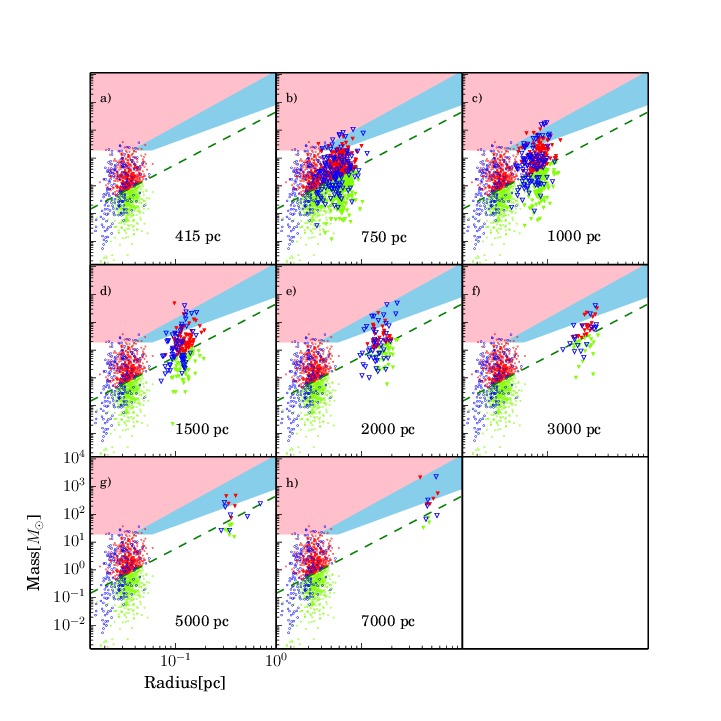}
\caption{Mass vs radius relation for the Orion A region. Red, blue and green circles are prestellar, protostellar and starless unbound sources
at the original distance and are shown in all panel for reference. 
Larger triangles represent the objects detected in the moved maps at different distances (panels b-i).
Dashed green line is the Larson's mass (see equation~\ref{eq:Larson}).
Pink and sky-blue zones correspond to thresholds for compatibility with MSF provided by \citet{Krumholz2008} and \citet{Kauffmann2010}, respectively.
}
\label{fig:mass_radius_Orione_unito}
\end{figure}

\begin{figure}
\includegraphics[scale=0.5]{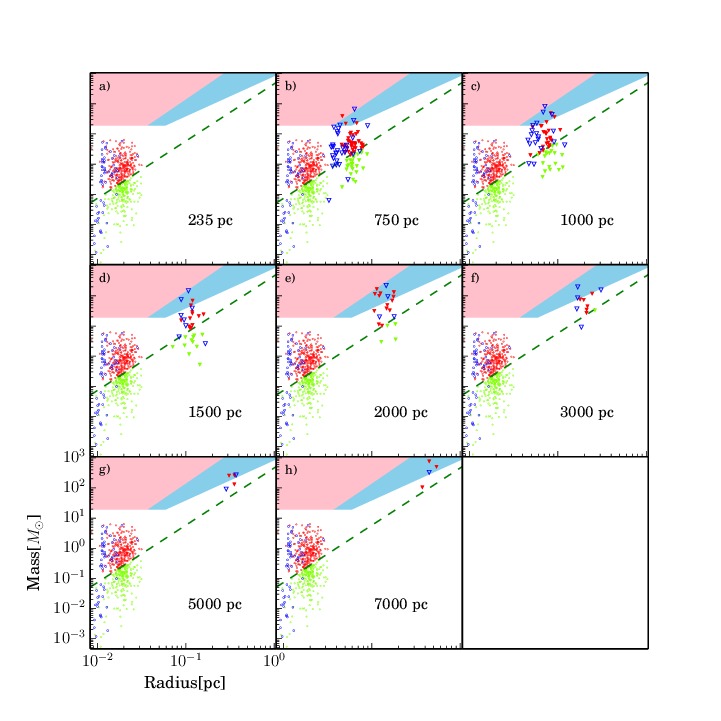}
\caption{The same as Fig.\ref{fig:mass_radius_Orione_unito} but for the Perseus region.
}
\label{fig:mass_radius_perseus}
\end{figure}

\begin{figure}
\includegraphics[scale=0.5]{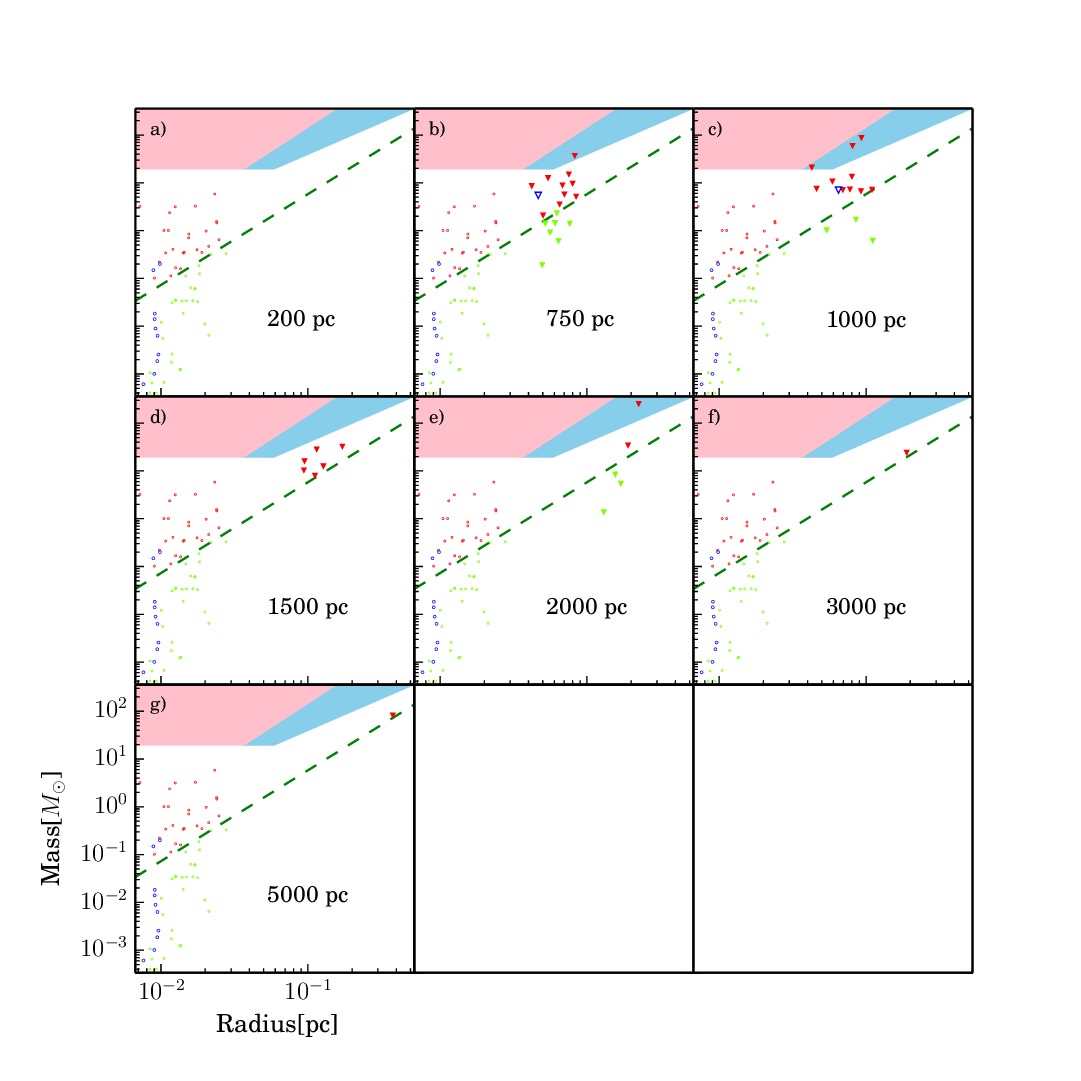}
\caption{ The same as Fig.\ref{fig:mass_radius_Orione_unito} but for the Lupus III region.
The panel for 7 kpc is missing since no sources are detected at this virtual distance.
}
\label{fig:mass_radius_lupusIII}
\end{figure}

\begin{figure}
\includegraphics[scale=0.5]{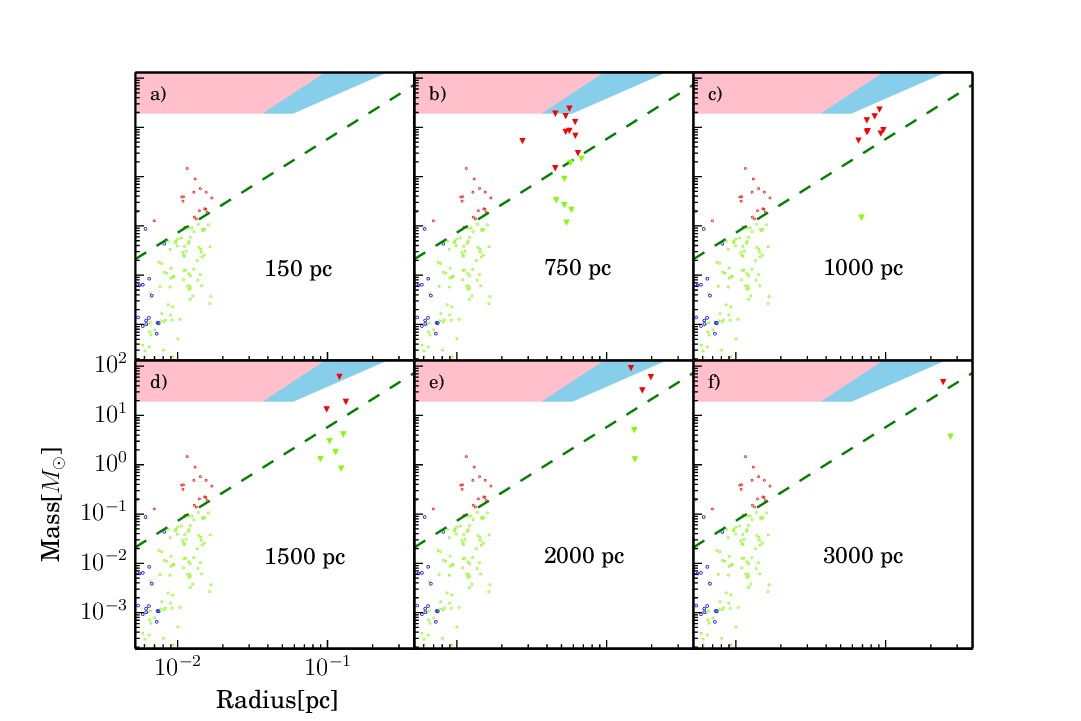}
\caption{ The same as Fig.\ref{fig:mass_radius_Orione_unito} but for the Lupus IV region.
The panels for 5 and 7 kpc are missing since no  sources are detected at these virtual distances.
}
\label{fig:mass_radius_lupusIV}
\end{figure}

\begin{figure}
\includegraphics[scale=0.5]{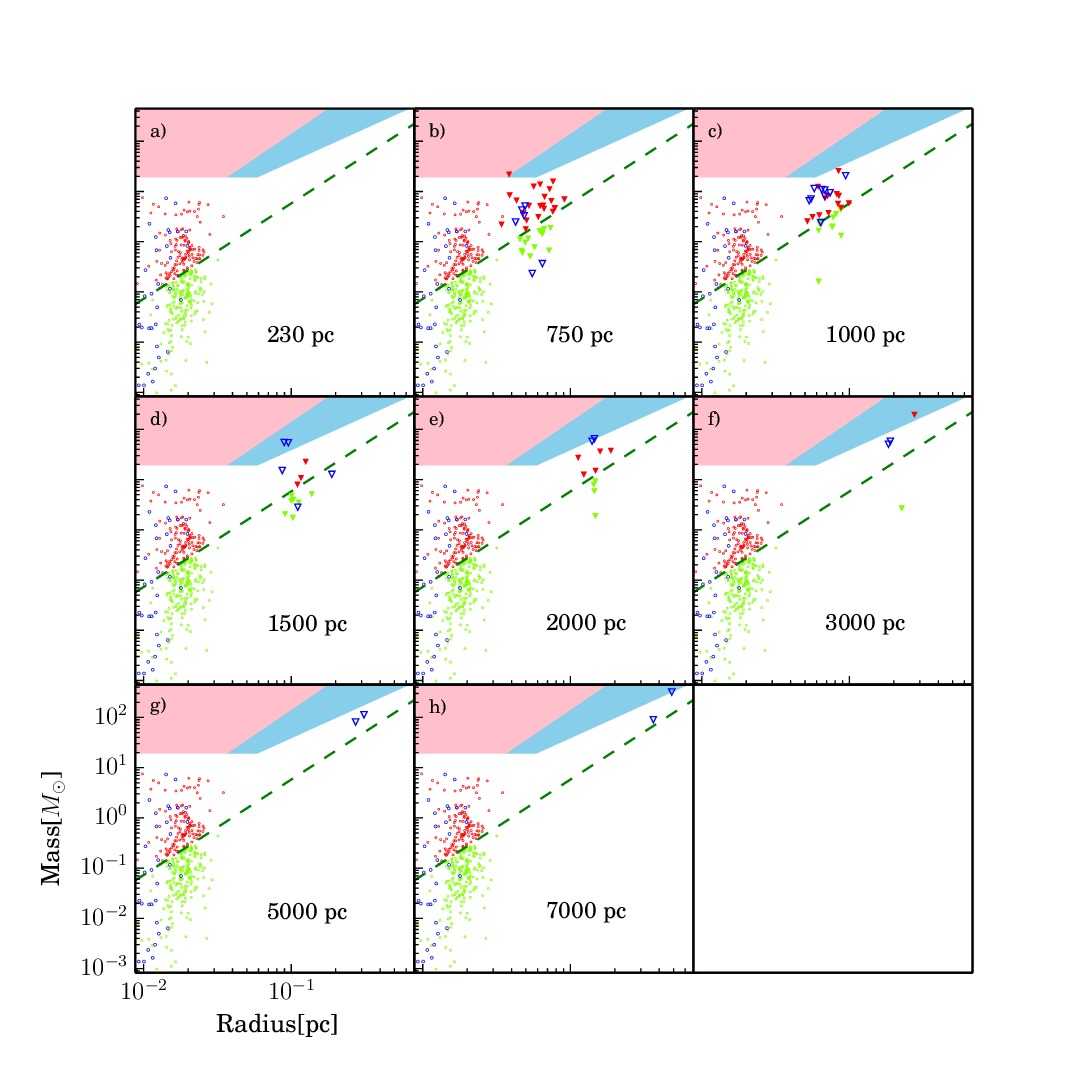}
\caption{ The same as Fig.\ref{fig:mass_radius_Orione_unito} but for the Serpens region.
}
\label{fig:mass_radius_Serpens00}
\end{figure}

Here we want to investigate how the predictive ability of the MR plot is affected by the distance effects.  
In Figs.~\ref{fig:mass_radius_Orione_unito} - \ref{fig:mass_radius_Serpens00} the MR plot is shown for the  different regions.
The MR diagram for sources found in the original map is reported in the panel $a$ of each of these figures.
 With open red and green  circles we indicate  bound and unbound starless objects, respectively, 
while with open blue circles we indicate a protostellar
objects. The  green dashed line represents the Larson's relation reported in equation~\ref{eq:Larson}, while the filled sky-blue and pink
 area of the diagram correspond to the conditions compatible with
  possible MSF, according to two different prescriptions respectively:
the first (which includes the second) corresponds to the criterion of 
 \citet[][hereafter KP]{Kauffmann2010}, namely $M(r) \geq 870\left(\frac{r}{ [\mathrm{pc}]}\right)^{1.33}M_{\odot}$, 
 which is an empirical limit for MSF based on observations of Infrared Dark Clouds; the latter corresponds to the criterion of 
  \citet[][hereafter KM]{Krumholz2008}, which is a more demanding threshold on column density of $ 1 \ \mathrm{g \ cm^{-2}}$ for MSF, based on 
theoretical calculations.
This theoretical limit can be also expressed as $M(r) \geq 15042\left(\frac{r}{ [\mathrm{pc}]}\right)^{2}M_{\odot}$.
We assume that these MSF thresholds break  at masses lower than $M=20 M_{\odot}$,  since the typical values of the core-to-star conversion efficiency are $0.5-0.33$ 
\citep{Alves2007,Kauffmann2010} and hence it is not reasonable that a core with a mass lower than 20 $M_{\odot}$ 
can form  a $10M_{\odot}$ star.
 
In the panels from b) to i) of Fig.~\ref{fig:mass_radius_Orione_unito}-\ref{fig:mass_radius_Serpens00} the  diagram is built
for the objects 
 detected in the MM of different regions for various distances, and compared with the case of the original map (panel a).
Fig.~\ref{fig:mass_radius_Orione_unito}, in particular, displays the MR diagram for the Orion~A region;
this nebula is known to be   
a MSF molecular cloud \citep[see e.g.][]{Genzel1989,Polychroni2013}, and this is consistent with the fact that 
several sources lie inside the MSF zone of the plot. If the Orion A map is moved to larger
distances (panels from b to i),
this region can still be considered a MSF nebula, based on this plot. This implies  that apparently the intrinsic character of a MSF region
 is still conserved if this region is observed at larger distances.
This remark is based on the case of a single region. However, from panel a) of Fig.~\ref{fig:mass_radius_Orione_unito}  one can deduce that 
for another MSF region containing cores denser and more massive than those found in Orion A, clumps extracted in the MMs would likely continue 
to populate even more so the KP region of the plot.
In Figs.~\ref{fig:mass_radius_perseus}-\ref{fig:mass_radius_Serpens00}  
 the same analysis is repeated for the other regions. These regions are known for not being  MSF, and this is consistent with the fact that
 there are no sources
 inside the blue and pink zone in the panel ``a'' of  all of these figures. However, 
at larger distances (panels b-i) sources are found inside the  zone of the plot compatible with both MSF prescriptions, 
and in particular in the KP zone;
 therefore, clearly, the distance is biasing  the 
character one assigns to a region regarding its ability to form massive stars. 

The Perseus region (Fig.~\ref{fig:mass_radius_perseus}) at the virtual distances (panels b-i) would be always classified as MSF according
the KP prescription. The Serpens nebula (Fig.~\ref{fig:mass_radius_Serpens00}) at the moved distances would be classified as MSF,
except at  1000 and 5000 pc.
The Lupus III and IV regions (Figs.~\ref{fig:mass_radius_lupusIII}-\ref{fig:mass_radius_lupusIV}) are
characterized by a regime of somewhat lower masses  compared with other regions \citep{Benedettini2015}.  
The Lupus III, at the virtual distances, would be classified as MSF at 750, 1000 and 2000 pc while it is not classified as MSF at 1500-3000 and 5000 pc. 
The Lupus IV nebula is classified as MSF at 750, 1500 and 2000 pc while it would be classified as non-MSF  at 1000 and 3000 pc.

\begin{figure}
\includegraphics[scale=0.4]{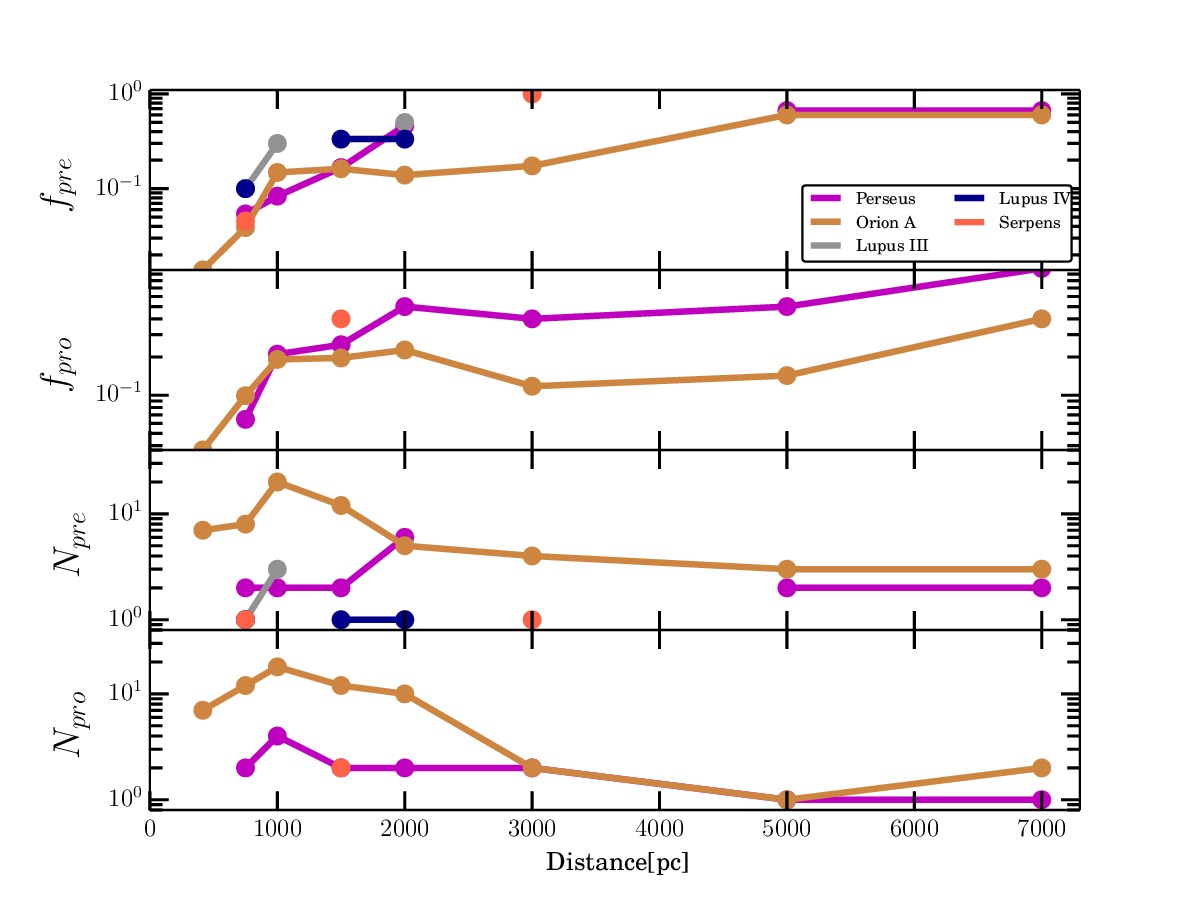}
\caption{Fraction of prestellar $f_{\mathrm{pre}}$ and protostellar $f_{\mathrm{pro}}$ objects above the KP zone in the mass vs 
radius plot
as a function of distance, with also the absolute number of prestellar ($N_{\mathrm{pre}}$)
 and protostellar ($N_{\mathrm{pro}}$) .
}
\label{fig:fraction_massive_pillai}
\end{figure}

At this point we want to
quantify   how the fraction of supposed MSF objects varies with distance:
we define, at each distance, the  fraction of prestellar and protostellar objects inside the MSF region (one of the two zones of the plot where MSF is possible
 according to KP and KM prescriptions), 
as $f_{\mathrm{pre}}=N_{\mathrm{pre}}/N_{\mathrm{PRE}}$, $f_{\mathrm{pro}}=N_{\mathrm{pro}}/N_{\mathrm{PRO}}$  where
$N_{\mathrm{pre}}$ and $N_{\mathrm{pro}}$ are the number of prestellar and protostellar inside the MSF region while $N_{\mathrm{PRE}}$ and
 $N_{\mathrm{PRO}}$                                   
are the total number of prestellar and protostellar.                       
Looking at Figs.~\ref{fig:mass_radius_Orione_unito} - \ref{fig:mass_radius_Serpens00} 
it  appears that $f_{\mathrm{pre}}$, $f_{\mathrm{pro}}$ increase with distances for the KP zone, while they seem  to decrease for 
the KM one.
This statement can be made more quantitative through  Fig.~\ref{fig:fraction_massive_pillai}, 
that shows  $f_{\mathrm{pre}}$ and  $f_{\mathrm{pro}}$ as a function of distance for the KP zone.
This figure highlights a clear trend of $f_{\mathrm{pre}}$ and $f_{\mathrm{pro}}$ to increase in the KP zone 
up to  $\sim1000-2000$ pc (depending on the region)  and then to reach a plateau (Fig.~\ref{fig:fraction_massive_pillai}).
The former of these trends is quite steep for both $f_{\mathrm{pre}}$ and $f_{\mathrm{pro}}$.
The gap $f_{\mathrm{pre}}(d)-f_{\mathrm{pre}}(d_{0})$ (and similarly for $f_{\mathrm{pro}}$) is found to be larger
 than the largest error bar associated with these points,
 indicating that the increase of this fraction 
is statistically significant\footnote{Errorbars are estimated by assuming a counting statistics}.
For example, for Orion A, $f_{\mathrm{pro}}=0.03$ at $d_{0}$ and $f_{\mathrm{pro}}=0.2$ at $d= 1000$ pc, while the largest uncertainty
among  $f_{\mathrm{pro}}$ values at $d=d_{0}$ is 0.06.
As for the plateau behaviour 
for the two regions with enough statistics, namely Orion~A and Perseus, the average value of these  fractions is $\sim 37\%$  and
 $\sim 60\%$ for the prestellar sources, respectively and $\sim 22\%$  and
 $\sim 60\%$ for the protostellar ones, respectively.
We do not show the trend of $f_{\mathrm{pre}}$ and $f_{\mathrm{pro}}$ for the KM zone because of  the poor statistics.
With respect to the KP criterion, the ratios $f_{\mathrm{pre}}$ and $f_{\mathrm{pro}}$ are found to increase for distances up to 1000 pc.
This is mostly due to the sharp break of the KP relation we impose at $M<20 \ M_{\odot}$ (corresponding to $r_{\mathrm{break}}= 0.06$ pc):
indeed, cores in the original map at the smallest probed distances typically have radius smaller than $r_{\mathrm{break}}$.
At the virtual distance of 1000 pc, all detected sources have, instead, $r>r_{\mathrm{break}}$. Therefore some of them, classified as
not-MSF in the original map, can be more easly found inside the KP area of the diagram (see
 Figs.~\ref{fig:mass_radius_Orione_unito} - \ref{fig:mass_radius_Serpens00}).
This turns out in the behaviour of Fig.~\ref{fig:fraction_massive_pillai} (first two panels starting from the top), where a plateau  in the
$f_{\mathrm{pre}}$ and $f_{\mathrm{pro}}$ vs $d$ relation is reached at $d=1000$ pc.

\begin{figure}
\includegraphics[scale=0.4]{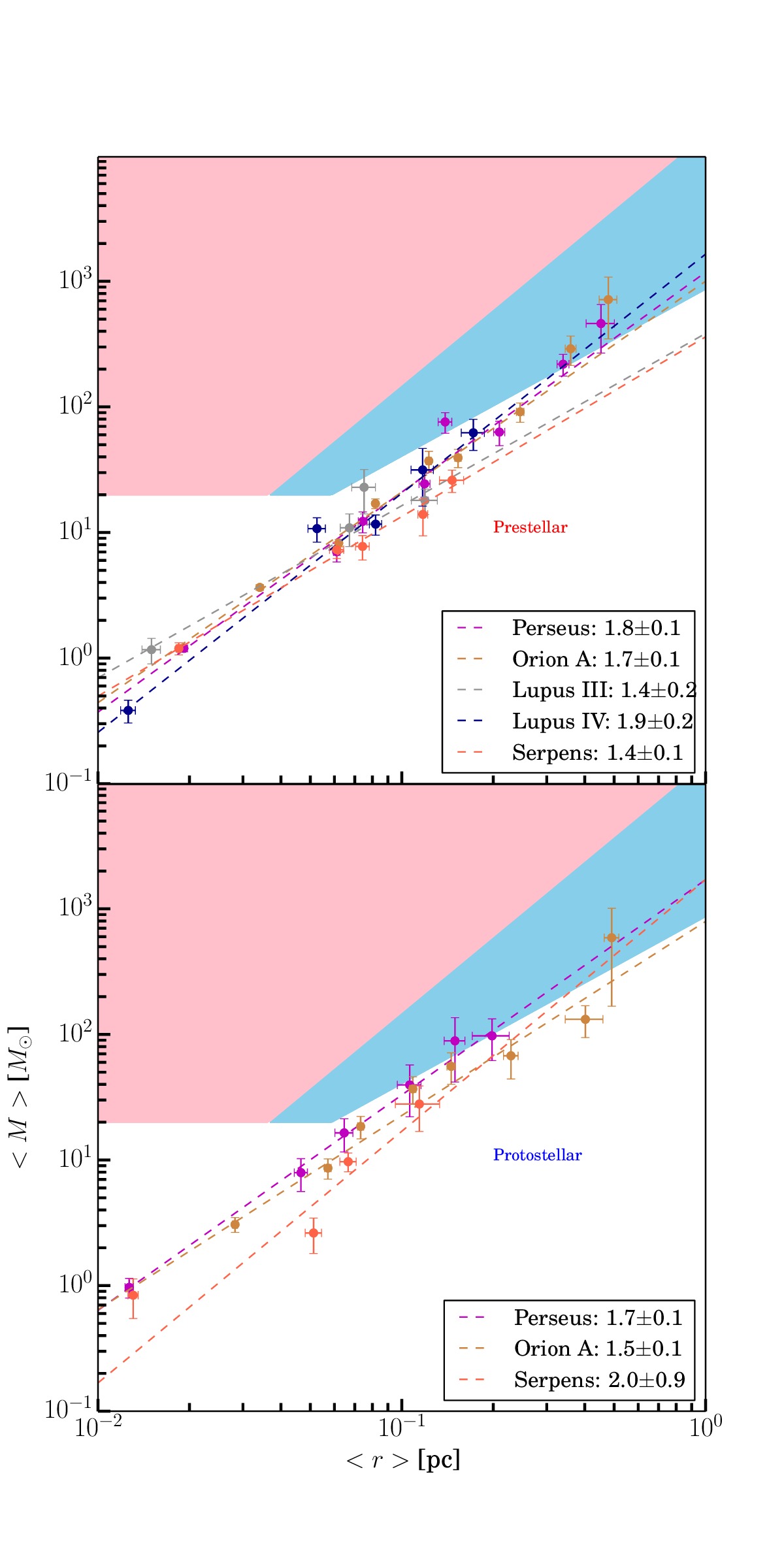}
\caption{ Mass vs radius plot  for
  the mean mass and the mean radius of the sources found at each simulated distance. The best power-law fit
of these points is shown,
and the corresponding  power-law exponent, estimated through best-fit, is reported.
}
\label{fig:fit_mass_radius}
\end{figure}

To understand the effects of distance on the MR relation we fit the data
 for prestellar and protostellar objects, with a power-law  $M\propto r^{\alpha}$ in the following way: 
for each distance $d$ we calculate the average values of the mass $\left<M\right>_{\mathrm{d}}$ and radius $\left<r\right>_{\mathrm{d}}$ of the 
sources in the sample and 
the corresponding standard error. 
The fit is shown in Fig.~\ref{fig:fit_mass_radius} with the corresponding $\alpha$ exponent evaluated separately
for prestellar and protostellar 
objects. They range mostly between  the exponent of the KP relation, 1.33,  and the one
of the KM prescription, 2.0;
this means that  if one decides to use the KM criterion 
for inferring conditions for MSF, one probably loses sources with  increasing  distance, while on the contrary one  tends to get some false positives if the 
KP criterion is adopted. \\

\begin{figure}
\includegraphics[scale=0.3]{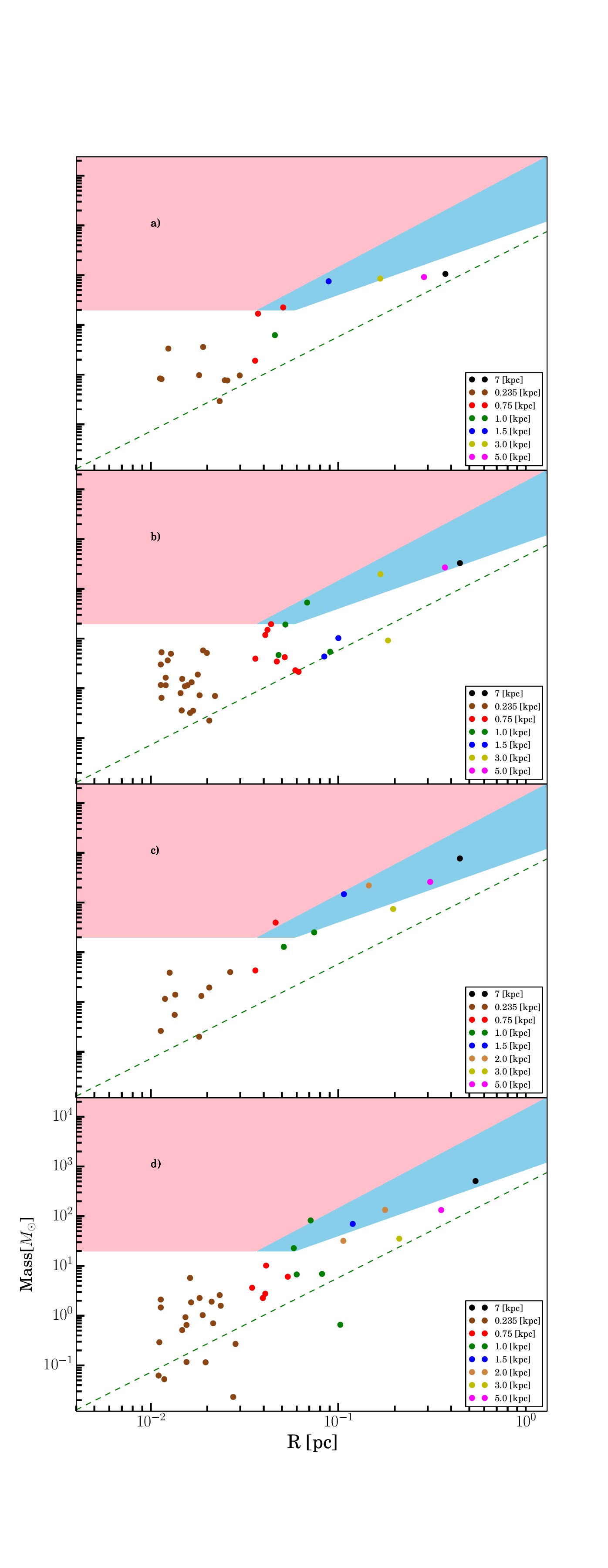}
\caption{Mass vs radius plots of sources found in the Perseus maps moved to various distances.
Each plot is built starting from one of the four  sources found at the largest distance (7 kpc), and contains all the sources found
at shorter distances and contained in it.
}
\label{fig:ass_perseus_M_vs_R}
\end{figure}

\begin{figure}
\includegraphics[scale=0.3]{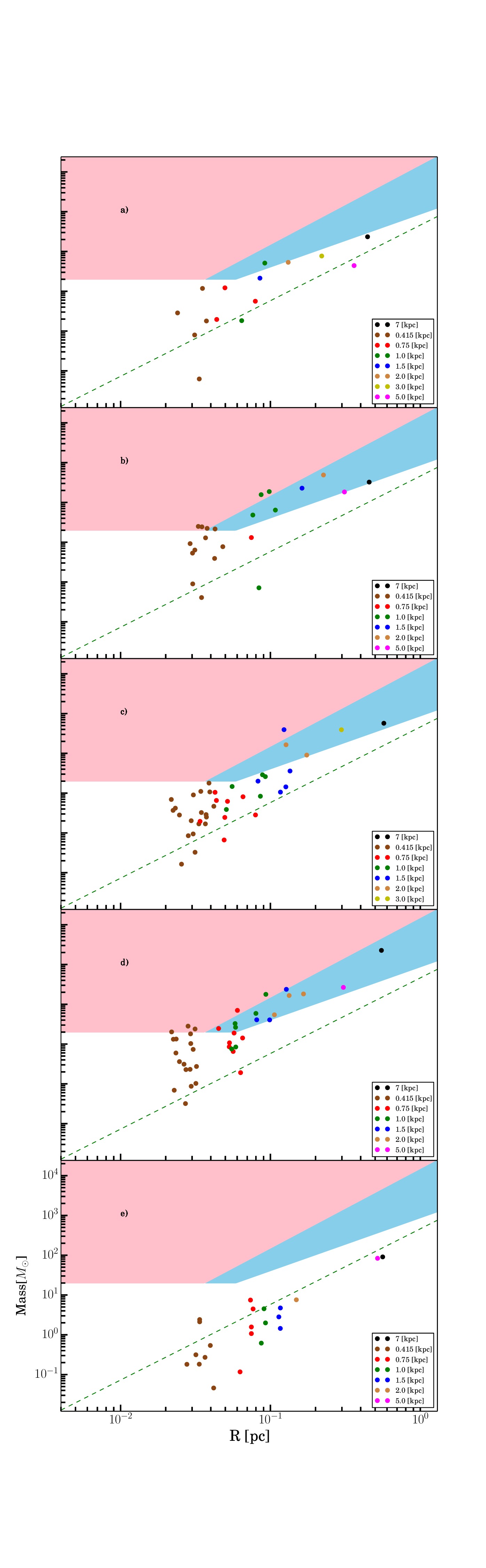}
\caption{Same as Fig.\ref{fig:ass_perseus_M_vs_R} but for the Orion A. Here we show only five out of ten cases of associations between 
the sources  found
at shorter distances.
}
\label{fig:ass_OrionN_M_vs_R}
\end{figure}

\begin{figure}
\includegraphics[scale=0.4]{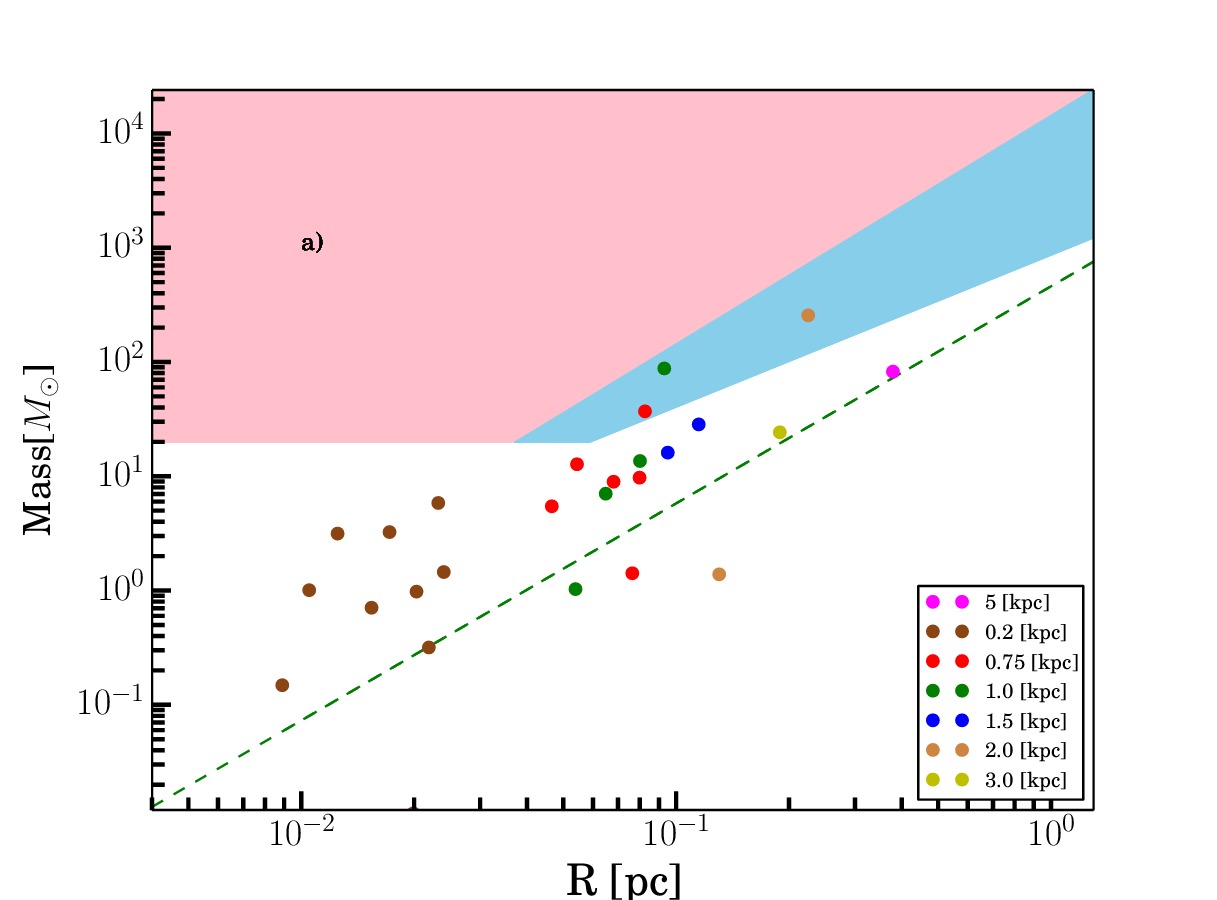}
\caption{Same as Fig.\ref{fig:ass_perseus_M_vs_R} but for the Lupus III region.
}
\label{fig:ass_LupusIII_M_vs_R}
\end{figure}

\begin{figure}
\includegraphics[scale=0.4]{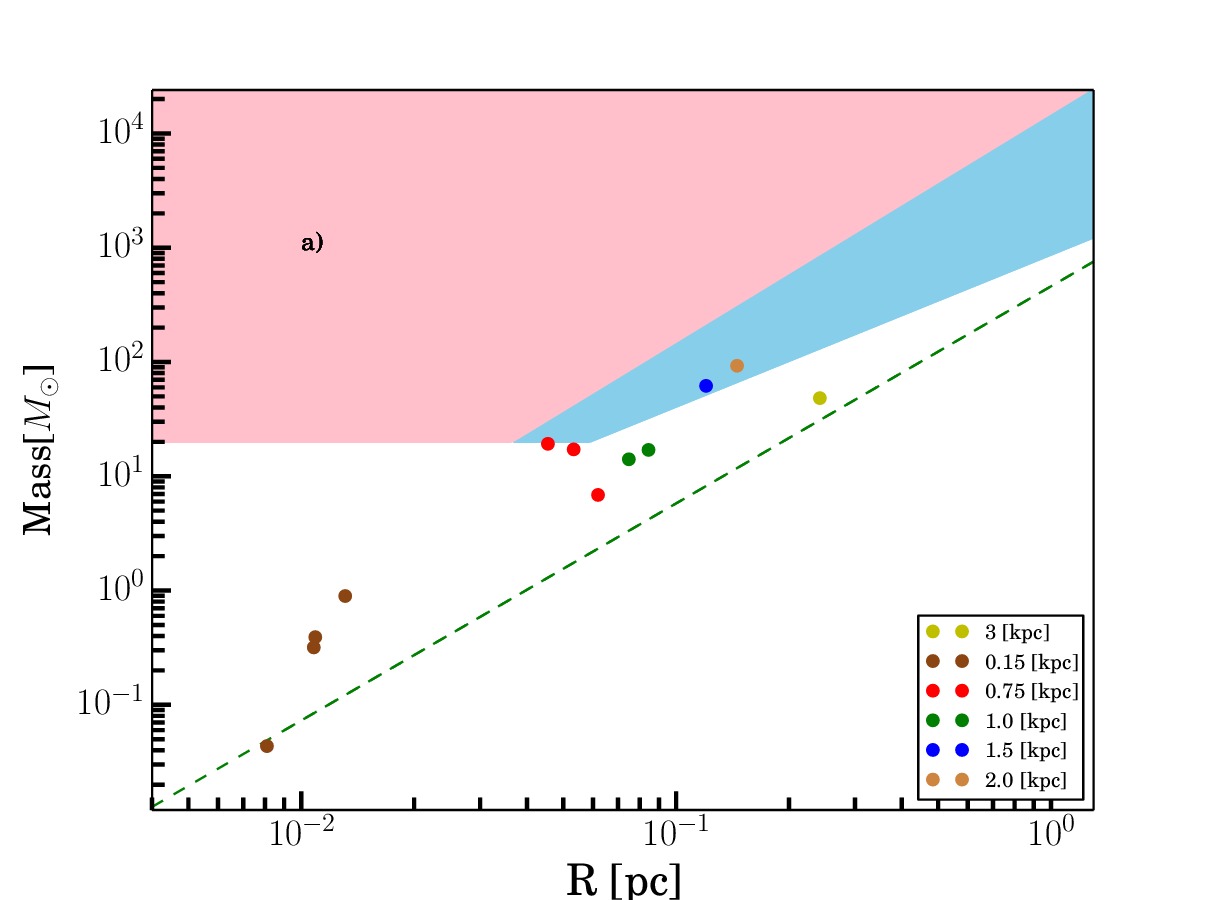}
\caption{Same as Fig.\ref{fig:ass_perseus_M_vs_R} but for the Lupus IV region.
}
\label{fig:ass_LupusIV_M_vs_R}
\end{figure}

\begin{figure}
\includegraphics[scale=0.4]{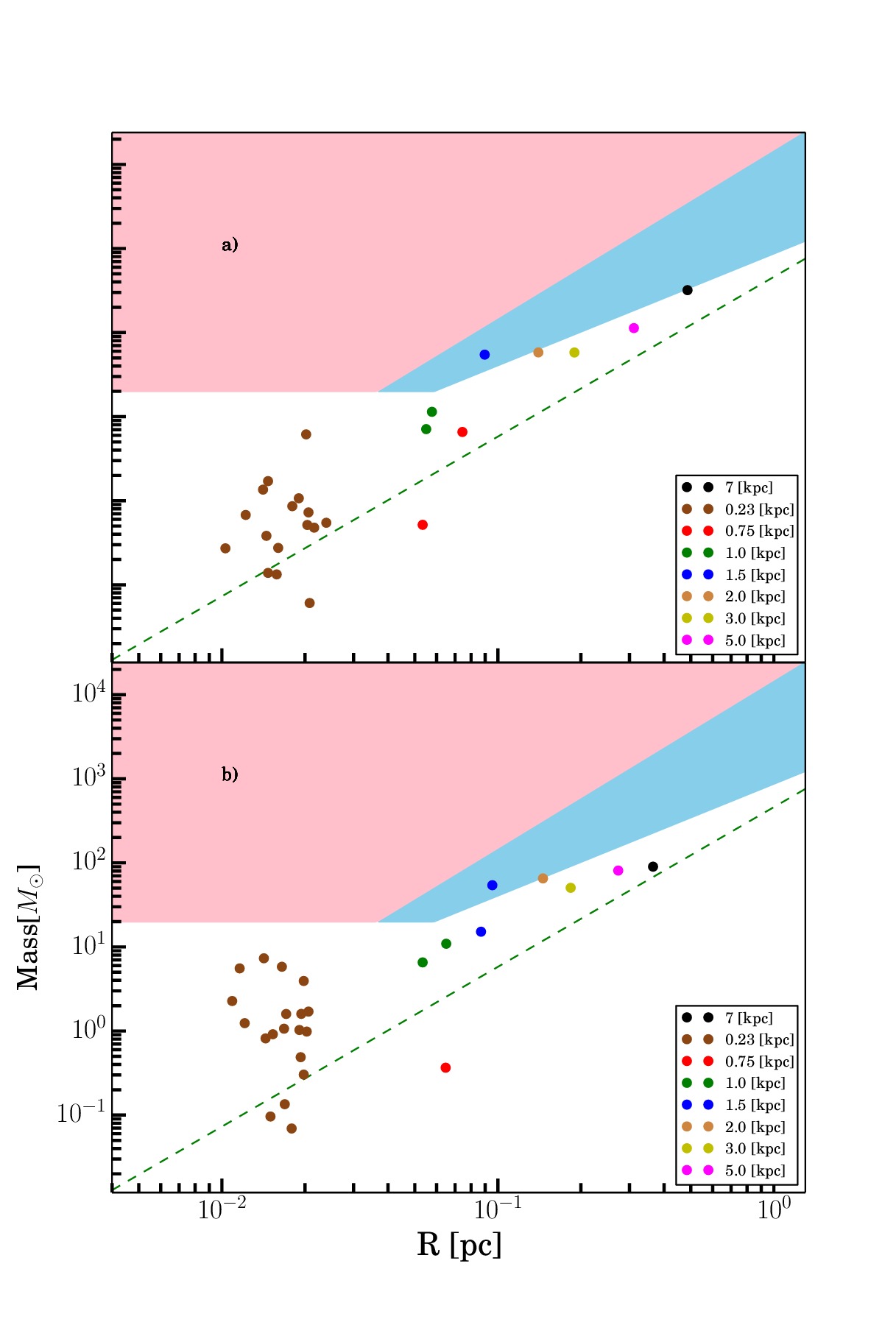}
\caption{Same as Fig.\ref{fig:ass_perseus_M_vs_R} but for the Serpens region.
}
\label{fig:ass_Serpens_M_vs_R}
\end{figure}

To examine in more detail how the properties of a far source mirror those of the contained core population, now let us consider
the relation between the clumps detected at the largest probed distance and the  ``contained'' sources at all smaller distances,
adopting the procedure described in Section \ref{sec:assosiation}.
Let $O_{\mathrm{d_{max}}}$ a source detected at the largest distance $d_{\mathrm{max}}$ and $O_{\mathrm{d}}$ the sources at distance 
$d<d_{\mathrm{max}}$ that
fall within $O_{\mathrm{d_{max}}}$.
Figs.~\ref{fig:ass_perseus_M_vs_R}-\ref{fig:ass_Serpens_M_vs_R} contain  a MR diagram for each
$O_{\mathrm{d_{max}}}$
and the corresponding $O_{\mathrm{d}}$.
In  detail, in Fig.~\ref{fig:ass_perseus_M_vs_R} we display  four $O_{\mathrm{d_{max}}}$ cases found in the Perseus nebula:
 three of them are classified as MSF  according to KP criterion 
(panels $b$, $c$, and $d$) and one of them as low-mass star 
forming (panel $a$). The corresponding $O_{\mathrm{d}}$  are classified in some cases as MSF, while in other cases   as low-mass star forming;
in particular, looking at the sources detected at the original distance, none of them is found to be  MSF. 

Fig.~\ref{fig:ass_OrionN_M_vs_R} contains the cases found in Orion~A.
Again, at the moved distances, many sources are classified as MSF according to the KP criterion.
In panels $b$ and $d$, also some $O_{\mathrm{d}}$ found at the original distance are classified as MSF.
This suggests that the character of an intrinsically MSF clump  (or of an entire MSF region as  Orion~A)
 is preserved under the moving procedure we apply, while
for a low-mass star forming region like Perseus a spurious classification as MSF can be introduced as an effect of distance bias. 

This can be observed also in the behaviour of $O_{\mathrm{d}}$ in the other low-mass star forming regions we considered, namely
Lupus III, IV and Serpens (Figs.~\ref{fig:ass_LupusIII_M_vs_R}, \ref{fig:ass_LupusIV_M_vs_R} and \ref{fig:ass_Serpens_M_vs_R}):
also in these cases some $O_{\mathrm{d}}$ at intermediate distances are found to lie in the KP zone of the 
MR diagram, although they do not contain sources in the original map with this property.
In Figs.~\ref{fig:ass_LupusIII_M_vs_R} and \ref{fig:ass_LupusIV_M_vs_R} we notice an apparently unnatural behaviour:
a  $O_{\mathrm{d}}$ source has a mass larger than that of the corresponding $O_{\mathrm{d_{max}}}$ containing it. 

Many factors can contribute to such cases occurring: multiplicities in source detection which appear/disappear at different distances 
and intrinsic fluctuations in CuTEx flux extraction can change the shape of the SEDs of a source observed at two different distances.
A shift of the SED peak results in a different temperature estimate, and in turn in a change of the mass which can be very 
relevant at very small temperatures, a shift of few K may lead to an order of magnitude change in mass \citep[see, e.g.,][]{Elia2016B}.

In conclusion the analysis of the MR diagram, carried out by means of 
Figs.~\ref{fig:mass_radius_Orione_unito}-\ref{fig:ass_Serpens_M_vs_R} suggests that:
1) the region  that is recognized as MSF at the nominal distance (Orion A) is still classified as MSF at each of the moved distances;
2) the regions that are not recognized as MSF at the original distance (Perseus, Serpens, Lupus III and IV) are  classified as MSF
for most of the virtual distances;
3) the fraction of objects fulfilling the KP relation increases  for each region up to 1000 pc and then reaches 
a plateau;
4) the mean value of the clump mass at a certain distance is generally related to the mean value of the radius with 
a power law with an index larger than 1.33 (KP criterion)
 and smaller than 2 (KM criterion) for all regions.
Therefore we find a trend to introduce  MSF objects, on average, if one adopts the KP prescription and  to lose MSF objects if one chooses
 the KM one.


For these reasons it is important to estimate the fraction of misclassified objects according to the KP prescription. 
To increase statistics, all regions are merged together in this analysis.
We define as false positive (FP) the objects, detected at the moved distance, that are classified as MSF although they do not
contain MSF cores at the original distance; with true negative (TN) the objects that are not classified as MSF 
at the moved distance and  do not have MSF association
with the objects at the original distance. The true positive (TP) are objects that are classified as MSF at the moved distances and have associations
with MSF objects at the true distance. Finally the false negative (FN) are the objects that are not classified as MSF at the moved distances and  are 
associated with MSF at the original distance.
Given these definitions, our attention has to be focused on the fraction of FP with respect to the 
population from which they originate (and to which they should ideally belong), namely that of TN.
Therefore we define
$f_{\mathrm{kauff}}=n_{\mathrm{FP}}/(n_{\mathrm{FP}}+n_{\mathrm{TN}})$, where $n_{\mathrm{FP}}$ is the number FP at distance $d$
and $n_{\mathrm{TN}}$ is the number of TN at distance $d$;
the $f_{\mathrm{kauff}}$ fraction as a function of distance gives an indication of the rate of misclassification of the objects introduced by distance.

\begin{figure}
\includegraphics[scale=0.43]{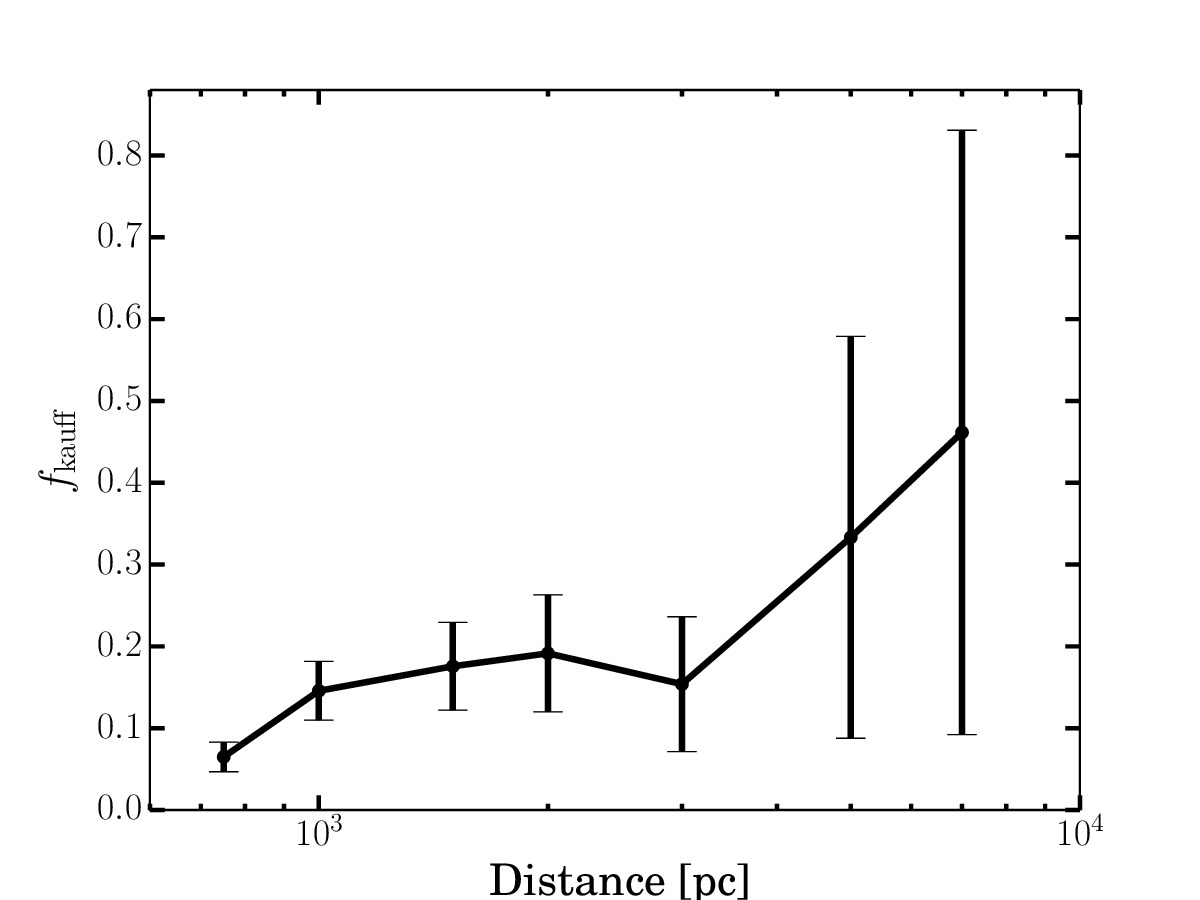}
\caption{Black thick: fraction of objects $f_{\mathrm{kauff}}=n_{\mathrm{FP}}/(n_{\mathrm{FP}}+n_{\mathrm{TN}})$ that are classified as MSF
at the moved distances,
 according to the KP relation,
while in  reality  they are only false positive (FP). 
}
\label{fig:fraction_above_kauff_dist}
\end{figure}

\begin{figure*}
\includegraphics[scale=0.85]{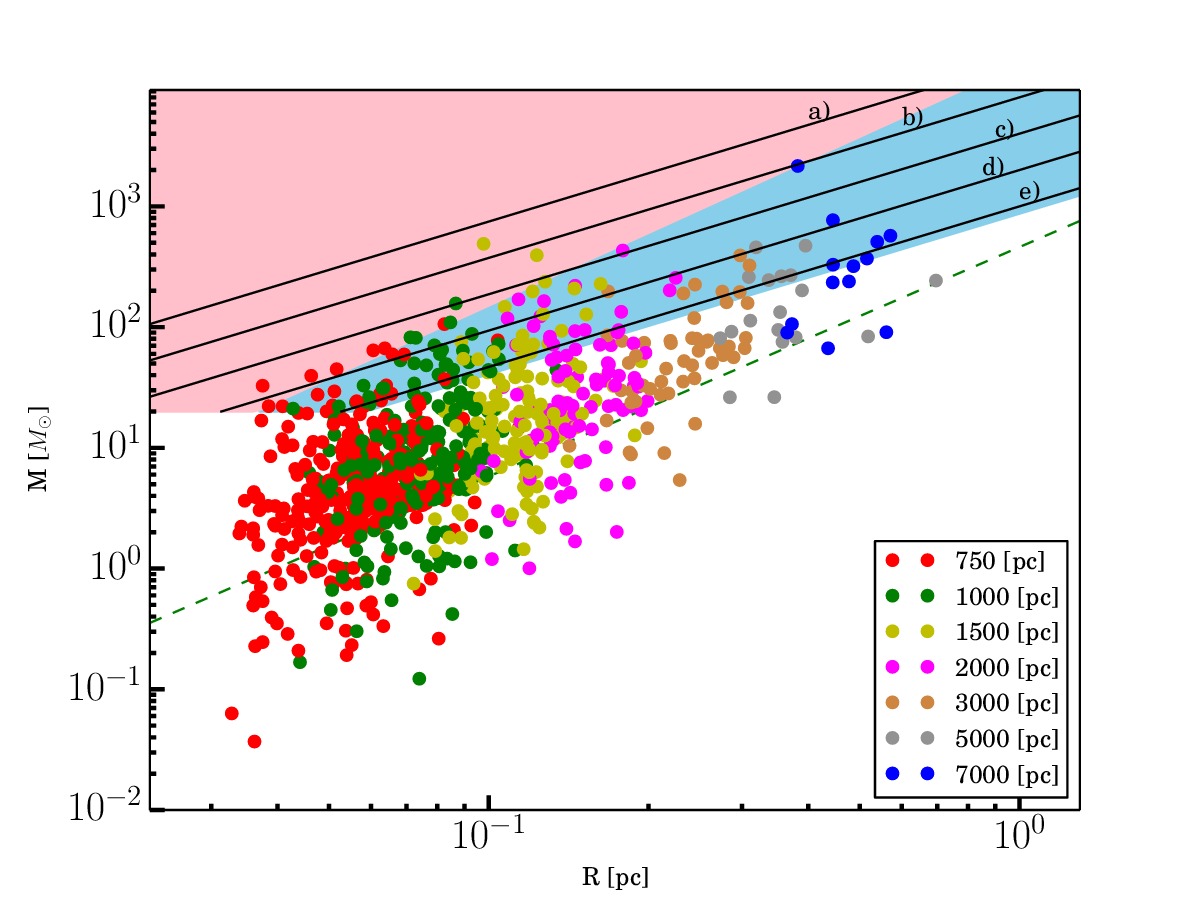}
\caption{Mass vs radius plot containing all the sources (solid circles) detected 
 at each virtual distance being associated with  sources detected
in the original map that are below the MSF zone (false positive and true negative). Black solid lines are lines parallel to the KP relation
 $M=k \ r^{1.33}$. The 
a) line is for $k=16000$ and delimits the probability of getting a false positive above this line,
b) is for $k=8000$, c) to $k=4000$, d) to $k=2000$ and e) $k=1000$. The $k=870$ line coincides with the KP relation. 
The probability values $p(k)$ (equation~\ref{eq:p(k)}) are shown in Table~\ref{tab:table_p_k}.
Dashed green line is the so called Larson's third law (see equation~\ref{eq:Larson}).
Pink and sky blue zones correspond to thresholds of \citet{Krumholz2008} and \citet{Kauffmann2010}, respectively.
}
\label{fig:massive_prescription}
\end{figure*}

 As one can see in Fig.\ref{fig:fraction_above_kauff_dist}, $f_{\mathrm{kauff}}$ increases between 750 and 1000 pc, remains almost constant between 1000
and 3000 pc, and then shows fluctuations above 3000 pc.
Such fluctuations are  due to the lack of statistics starting from that distance (indeed the error bars, estimates from the
Poisson statistics, are very large). The gap 
between 750 and 1000 pc is simply due to the fact that many sources found at 750 pc lie at $r<r_{\mathrm{break}}$ (see above). 
From 1000 up to 3000 pc $f_{\mathrm{kauff}}$ is of the order of $20\%$ while at larger distance it is larger but  the error bars 
are very large in turn.
Therefore we can reasonably say that   above 1000 pc 
 $f_{\mathrm{kauff}}$  slightly  increases with distance 
although the trend is  hard to quantify above 3000 pc.
 We can
 plot some lines, that are parallel to the KP relation (see Fig. \ref{fig:massive_prescription}), whose analytical form can be expressed as
\begin{equation}\label{eq:larsonk}
M_{\mathrm{k}}>k \ \left(\frac{r}{ [\mathrm{pc}]}\right)^{1.33} M_{\odot} ,
\end{equation}
and we can count the number of FP that lie above these lines.
Calling $n_{\mathrm{M_{k}}}$ the number of FP fulfilling  equation~\ref{eq:larsonk}, the relative fraction of FP
will be
\begin{equation}\label{eq:p(k)}
 p(k)=n_{M_{\mathrm{k}}}/(n_{\mathrm{FP}}+n_{\mathrm{TN}}).
\end{equation}
We have to make a distinction between the objects detected at distances lower and larger than 4000 pc respectively, since as we have seen in  
Fig.~\ref{fig:fraction_above_kauff_dist} for distance above 4000 pc $f_{\mathrm{kauff}}$ tends to to be larger.
 A list  of several values of $p(k)$ is reported in Table~\ref{tab:table_p_k} for the two different cases ($d<4000$ and $d>4000$ pc).
From these values we can infer that the fraction of misclassified objects $p(k)$ decreases with increasing $k$,  hence for a clump,
classified as MSF,  with an high 
value of $k$ indicates a lower probability of dealing with a FP and the other way around for a lower value of k.
The values reported in  Table~\ref{tab:table_p_k} may be very useful to the \emph{Herschel} astronomer. 
Suppose, for example, we find an Hi-GAL clump having  mass $M_{0}$ and radius $r_{0}$ and within the MSF zone, the coefficient $k$ can be found as 
$k=M_{0}/R_{0}^{1.33}$.
 Looking at Table~\ref{tab:table_p_k} one can get the corresponding value of $p(k)$ and
 hence estimate how likely it is that the clump   
 is a FP.
In section~\ref{correction} we provide a practical example of the use  of Table~\ref{tab:table_p_k} for a more 
correct interpretation of Hi-GAL data.



\begin{table}
  \centering
  \caption{ Some values of $p(k)$ (equation~\ref{eq:p(k)}), where $k=M/r^{1.33}$ and $p(k)$ is the fraction of FP
 fulfilling equation~\ref{eq:larsonk}. 
The mass is measured in $M_{\odot}$ and $r$ in pc. The values of $k$ are chosen such that $p(k)$, in the second
column, decreases in steps of 0.01. The values of the third and fifth columns are the error bars associated with the values of the
 second and fourth column respectively.
The second and the third column are valid for $d< 4 \, \mathrm{kpc}$, while the fourth and fifth ones are valid for  $d> 4 \, \mathrm{kpc}$. 
  }
  \label{tab:table_p_k}
  \begin{tabular}{c c c c c} 
\hline         
     &         &   $d< 4$ kpc \ \ \ \ \ \ \ \ \ \  & \ \ \ \ \ \ \ \ \ \  $d>4$ kpc  &     \\
\hline
                 $k$ &        $p(k)$    &  $\sigma_{p(k)}$    &    $p(k)$   &  $\sigma_{p(k)}$   \\
\hline       
                870     &   0.13    &      0.02          &      0.39         &     0.18                              \\
                922     &   0.12    &      0.02          &      0.35         &     0.17                          \\
                1000    &   0.11    &      0.01          &      0.32          &    0.16                           \\
                1096    &   0.10    &      0.01          &      0.22         &     0.12                         \\
                1122    &   0.09    &      0.01          &      0.22          &    0.12                           \\               
                1188    &   0.08    &      0.01         &       0.19          &    0.11                           \\
                1245    &   0.07    &      0.01         &       0.13         &     0.09                               \\
                1332    &   0.06    &      0.01         &       0.13          &    0.09                                  \\
                1508    &   0.05    &      0.01          &      0.13         &     0.09                                  \\
                1638    &   0.04    &      0.01         &       0.10         &     0.07                              \\
                1968    &   0.03    &      0.01         &       0.10          &    0.07                                 \\
                2263    &   0.02    &      0.01         &       0.03          &    0.04                                 \\                                               
\hline
  \end{tabular} 
\end{table}



\clearpage

\clearpage

\subsection{A new prescription for high-mass star formation}
\label{aaa_o}

\begin{figure}
\includegraphics[scale=0.45]{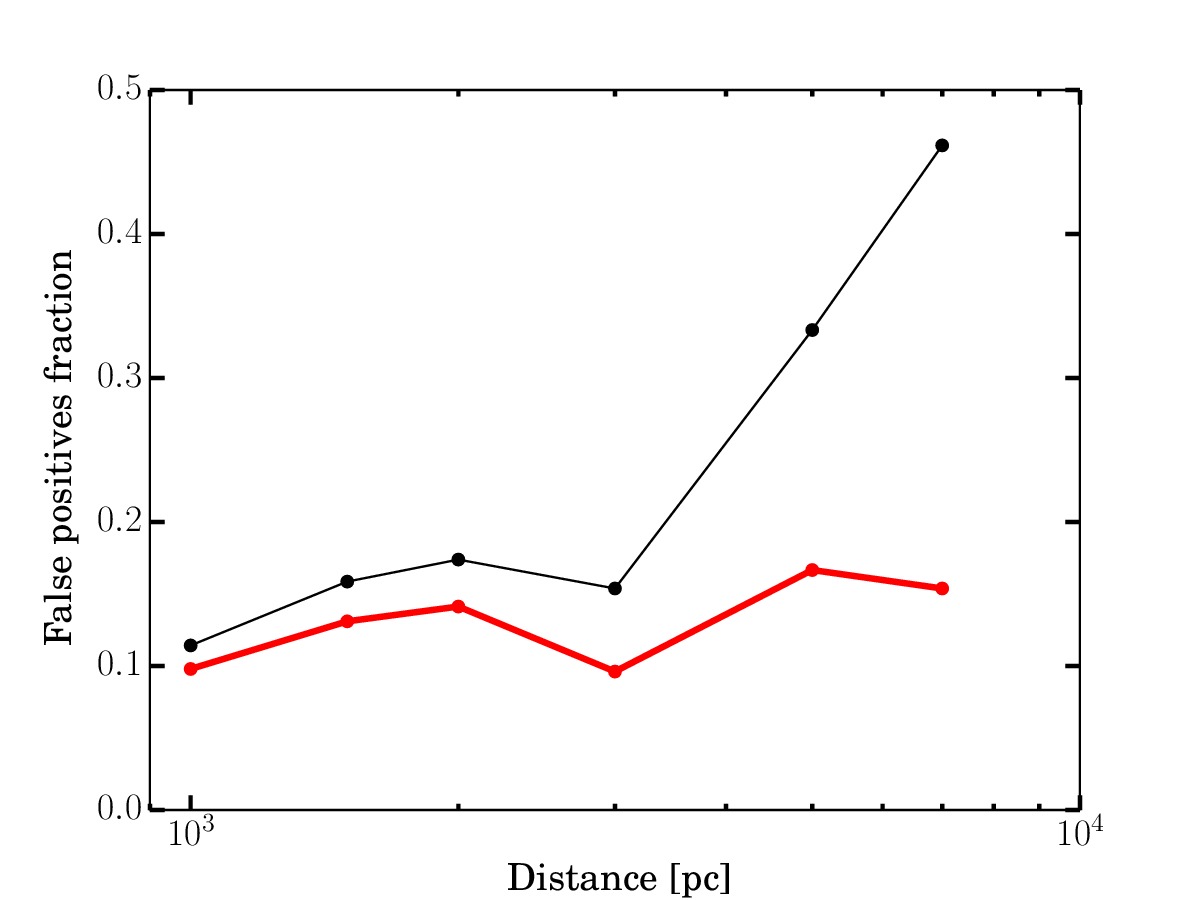}
\caption{Black line: fraction of objects $f_{\mathrm{kauff}}=n_{\mathrm{FP}}/(n_{\mathrm{FP}}+n_{\mathrm{TN}})$ that are classified as MSF
at the moved distances,
 according to the KP relation,
whereas   they are actually false positives (FP). 
Red line: same as the black one but for our prescription for MSF (equation~\ref{eq:ooprescription}).
}
\label{fig:fraction_above_oo}
\end{figure}

As pointed out in section~\ref{sec:mass}, the exponent of the $\left<M\right>_{\mathrm{d}}$ vs $\left<r\right>_{\mathrm{d}}$ power-law relation
is smaller than 2.0 (KM) and larger than 1.33 (KP), so that distance effects can bias the fraction of sources fulfilling the two 
aforementioned criteria and the character itself of star formation in the considered region.
A new prescription for identifying compatibility with high-mass star formation is therefore needed.
The prescription  we provide here is: 
\begin{equation}\label{eq:ooprescription}
M>1282 \ \left(\frac{r}{ [\mathrm{pc}]}\right)^{1.42} M_{\odot}
\end{equation} 
(see  Appendix~\ref{sec:oop} for the derivation of the way it was derived).
It is interesting to make a comparison of our formula with that of KP in terms of FP and TP above 1 kpc,
since about $99\%$ of the Hi-GAL sources lie in that range
of distances.
Figure~\ref{fig:fraction_above_oo}  display the fraction of FP  
as a function of distance for both thresholds.
The fraction of FP (Fig.~\ref{fig:fraction_above_oo}) generated by the KP prescription is  larger than ours at any distance; in particular 
it is larger by few percent between 1000 and 3000 pc, 
while it gets larger by $20-30\%$ at larger distances.
The number of TP and FN is comparable for both prescriptions\footnote{ 24 TP and 10 FN for our prescription and 26 
TP and 8 FN for the KP one}.
Notice that equation~\ref{eq:ooprescription} keeps  the fraction of FP  almost constant,  unlike the KP prescription.
 The main differences between the two are found in the 
 fraction of FP  at the largest probed distances namely 5000 and 7000 pc.
Notice also that using the KM prescription, one would get a smaller fraction of FP, but would also get a very low number 
of TP and a higher number of FN  (4  and 30, respectively).

Therefore, to minimize the occurrence of FP at all distances, we suggest to use the prescription presented here.
In particular this choice is critical at $d \geq 5000$ pc, where the majority of Hi-GAL sources ($73\%$) provided with a distance 
estimate are found to lie \citep{Elia2016}.

\section{Comparison with previous literature}
\label{correction}

The results obtained in this paper can be used to discuss some results found in literature on clump properties.
For example \citet{Wienen2015} discuss the MR diagram for the sources detected with the  APEX Telescope Large Area Survey of the whole inner
Galactic plane at $870 \, \mu \mathrm{m}$ (ATLASGAL).
These authors found that a relevant  fraction (namely $92  \%$) of the clumps are potentially forming massive stars
because they  fulfil the KP relation.
This fraction  is notably larger than the one found for
 the Hi-GAL sources \citep{Elia2016}, namely $71~\%$ and $65~\%$ for the protostellar and prestellar, respectively.
This relevant discrepancy may not be only explained with the Hi-GAL better sensitivity, but most likely it is implicit in the KP relation itself.
We have seen, indeed, at large distances the KP relation overestimates  the number 
of high-mass star forming candidates due to a shallow slope of 1.33.
In the MR plot   of \citet[][their Fig.~23]{Wienen2015} they found that for large values of the radius  ($\sim 1$ pc),
and hence at large distances, all the sources fulfil the KP relation while for smaller values of the radius ($\sim 0.1$ pc)
this fraction is smaller than 1. This effect is likely to be due to the distance bias that we discuss in our paper.
Even in the Hi-GAL catalogue we observe the same effect at large distances in the MR plot, but it is less evident
because the sources are more  scattered due to the
spread in temperature of the Hi-GAL sources.
The sources of \citet{Wienen2015} are less scattered in the MR plot, because the masses were derived using only two
 fixed values of the temperature. \\
It is noteworthy  that masses in ATLASGAL literature are also derived through  a different  dust opacity; this leads to rescale the value
 of the coefficient 
in the KP relation, to 580.
Therefore to properly rescale  equation~\ref{eq:ooprescription} to make it comparable  with ATLASGAL data we obtain:
\begin{equation}\label{eq:ooprescription_atlasgal}
M_{\mathrm{A}}>855 \ \left(\frac{r}{ [\mathrm{pc}]}\right)^{1.42} M_{\odot}.
\end{equation}

We apply equation~\ref{eq:ooprescription_atlasgal} to the data of Table~3 of \citet{Wienen2015}  
 to get the fraction of high-mass star forming candidates.
We find that this fraction  is $89\%$ while if we use the 
KP relation we get a $94\%$ fraction\footnote{Notice that source radii quoted by \citet{Wienen2015} are not beam-deconvolved.
Using, instead, deconvolved radii would increase, in principle, both the fractions reported here.}.
Note that \citet{Wienen2015} consider as a high mass star forming candidate a clump with a mass larger than 650 $M_{\odot}$
that fulfils the KP relation. By using such a more demanding prescription they found that  $71\%$ of the sources of the catalogue entries
 are massive star forming candidates.

We can also make the same test on the data of \citet{Ellsworth2015} from the Bolocam Galactic
Plane Survey \citep[BGPS][]{Ginsburg2013}, again using equation~\ref{eq:ooprescription_atlasgal}.
  since they used the same dust opacity of the ATLASGAL collaboration.
 We get  $25\%$ of the sources classified as  potentially forming massive protostar against  $36\%$ found using  the KP relation.

Similarly, we can also 
 apply the prescription of Table~\ref{tab:table_p_k} to the Hi-GAL data.
As already mentioned, the most comprehensive catalogue of compact objects, at present, is provided in \citet{Elia2016}. 
In that catalogue we find that, in the inner Galaxy, there are 62438 objects, with 35029 
of them having a known distance and classified as prestellar and protostellar.
We find  23733 objects  that fulfil the KP prescription.
This means that  $68 \%$ of the sources might be able to form a massive star according to the KP prescription.
One can apply now the values of Table~\ref{tab:table_p_k} to give an estimate of the fraction of FP. Clearly 
for each source we get a different value of $p(k)$ (equation~\ref{eq:p(k)}); taking the average of the values of $p(k)$ for all the sources 
we get $\left<p(k)\right>\sim 7 \%$.
Note however that the values of Table~\ref{tab:table_p_k} were derived for $d< 7 \, \mathrm{kpc}$  while in the  \citet{Elia2016} 
catalogue the $63\%$ of the  objects are found to lie  at
 larger distances

Finally, we  apply equation~\ref{eq:ooprescription} to the Hi-GAL catalogue to discriminate the fraction 
of high-mass star forming candidates.
We find that according to our prescription the fraction of such objects is $52 \%$ and $64 \%$ for the prestellar
and protostellar clumps, respectively.

\section{Conclusions}
\label{sec:conc}

Distance bias increasingly affects estimates of the physical parameters (mass, temperature and radius) of far-infrared sources.
This is particularly critical for the Hi-GAL survey, which observed a large area of the sky underlying
a wide range of
heliocentric distances. 
In this paper, using the information taken from nearby star forming
regions, we have shown how this bias influences the estimation of these quantities.
The main results of our work are:

\begin{enumerate}
\item We present an original pipeline to virtually ``move'' the maps at larger distances.
\item The number of sources detected with CuTEx decreases with  distance, 
for each band, as a power law with exponent between 1.1 and 1.9.
The smallest values for these exponent are found in the 70 $\mu \mathrm{m}$ band.
\item The protostellar fraction $n_{\mathrm{pro}}/n$ increases with distance until it reaches a plateau (above 1 kpc).
\item  The effect of the confusion  is to increase with distance the physical radius of the compact sources;
 we  show that the sources are classified, on average, as cores up to
1 kpc and as clumps at larger distances.
\item The contribution of the diffuse (inter-core) emission to the flux of a source increases, with respect to the
original population, with distance. This is due to both the increasing physical area of the source, and to 
the background   which gets lower at larger distances. The smallest effect is systematically found at 70 $\mu m$.
\item We found that the average core formation efficiency, for distances above 1500 pc,
 depends on the region and can go from few percent up to $20\%$.    
\item The average temperature derived from SED fits increases quite weakly with distance for the prestellar objects, whereas it decreases slowly for 
the protostellar objects.
\item  In the mass vs radius diagram the fraction of sources classified as compatible with high-mass star formation increases with
 distance if one considers  the Kauffmann-Pillai (KP) prescription,
whereas it decreases for the Krumholz-McKee (KM) prescription. This happens because 
the mean value of the mass is found to be related to the mean value of the radius with a power law with an exponent larger than 1.33 (KP criterion)
 and smaller than 2 (KM criterion) for all the investigated regions.
Therefore,  adopting the KM prescription to check the presence of MSF clumps one tends to lose genuine candidates  with distance, while  
adopting the KP criterion one tends to gain false positives at increasing distances.
\item  We show that the fraction of false positive (FP),
 defined as the number of FP (clumps that  are classified as MSF at a moved distance, according  the 
KP prescription, 
but do not have associations 
with MSF objects at the original distance) over the total number sources,  is  $13 \pm 2 \%$, keeping almost constant
between 1000 and 4000 pc. Above this distance the fraction of ``false'' high-mass star forming clumps according to the KP criterion climbs
up to almost $40\%$, but has large associated uncertainties due to small statistics. 
\item We  estimated how likely it is that    
a clump classified as MSF is actually a false positive, as a function of its position with respect to the parametrized area 
$M_{\mathrm{k}}>k \ \left(\frac{r}{ [\mathrm{pc}]}\right)^{1.33} M_{\odot}$
in the mass vs radius plot.
A dichotomy is found for distances shorter and larger of 4000 pc.
As an example, a probability of $10\%$ is achieved for $k=1096$ for $d<4000$ pc, and at $k=1969$ for $d>4000$ pc.
\item We derive a new prescription to discriminate possible MSF clumps:
 $M>1282.4 \ \left(\frac{r}{ [\mathrm{pc}]}\right)^{1.422} M_{\odot}$ that appears to produce a smaller amount of false positives  than the
KP relation and preserved the same rates of TP. 
\item We applied our prescription to discriminate
 high-mass star forming candidates in the  Hi-GAL dataset: we found that the fraction of high-mass star 
forming objects is $52\%$ and $64\%$, for prestellar and protostellar sources, respectively. 
\item Taking into account a recently derived distance estimate for the Serpens region of 436 pc \citep{Leon2016} instead of the one used
throughout this paper ($d_{0}=230$ pc), we notice (see appendix~\ref{sec:serp})  that  the mass  vs radius relation for a few cores changes so that they get
classifiable as high mass star forming candidates.
This might lead, in turn, to classify this region as compatible with  episodes of MSF.

\end{enumerate} 

\section*{Acknowledgments}
We are grateful to the referee for very useful comments
that helped to greatly improve the paper.
This research has made use of data from the Herschel Gould Belt survey project (http://gouldbelt-herschel.cea.fr). 
The HGBS is a Herschel Key Project jointly carried out by SPIRE Specialist Astronomy Group 3 (SAG3), scientists of several institutes 
in the PACS Consortium (CEA Saclay, INAF-IAPS Rome and INAF-Arcetri, KU Leuven, MPIA Heidelberg), and scientists of the Herschel Science Center (HSC).
AB, DE, SM, SP, ES, MB, AD, SL, MM's research activity is supported by the VIALACTEA Project, a
Collaborative Project under
Framework Programme 7 of the European Union funded un-der Contract \symbol{35}607380,
that is hereby acknowledged.

\newpage

\appendix
\section {\\Plot of the Perseus region}
\label{sec:app1}
In this Appendix we show the moved maps at 70, 160, 350, 500 $\mu \mathrm{m}$ for the Perseus region described in Section \ref{sec:MAM} and
the maps for the other regions at 250 $\mu \mathrm{m}$ only.
\newpage

\begin{figure*}
\includegraphics[scale=0.26]{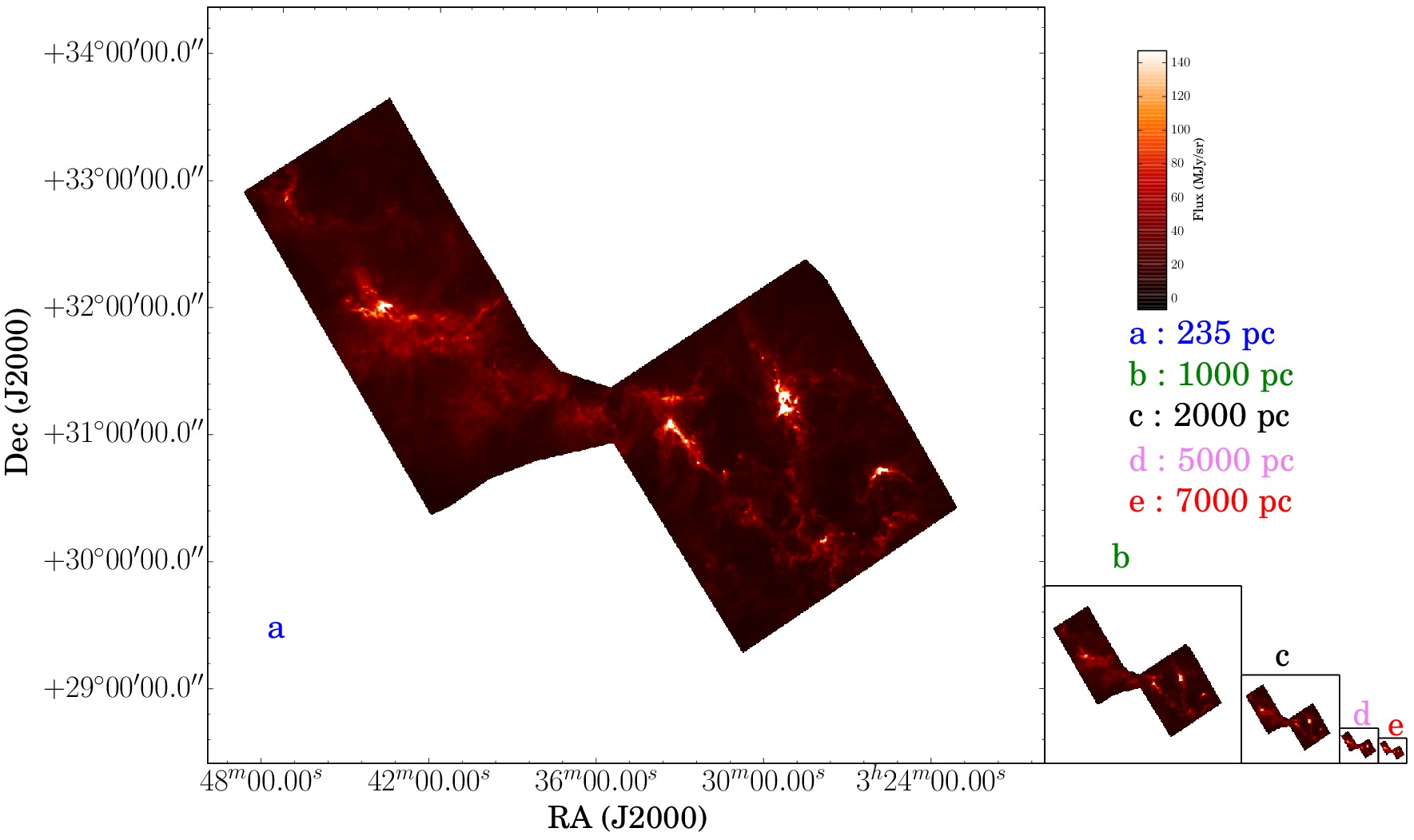}
\caption{Original and moved  Perseus maps at 500 $\mu \mathrm{m}$.
}
\label{fig:PesreusPLW}
\end{figure*}

\clearpage

\begin{figure*}
\includegraphics[scale=0.26]{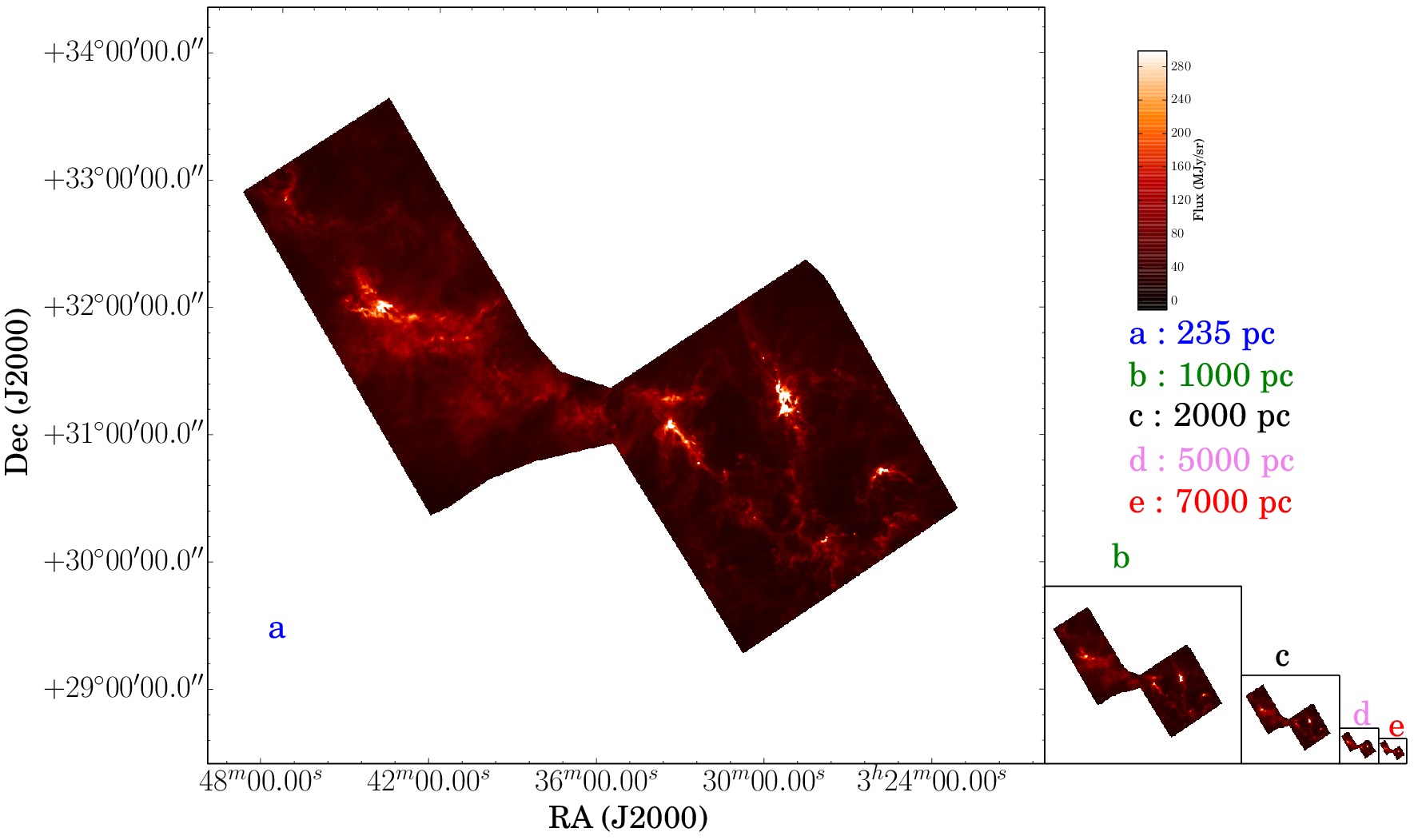}
\caption{Original and moved  Perseus maps at 350 $\mu \mathrm{m}$.
}
\label{fig:PerseusPMW}
\end{figure*}

\newpage

\begin{figure*}
\includegraphics[scale=0.26]{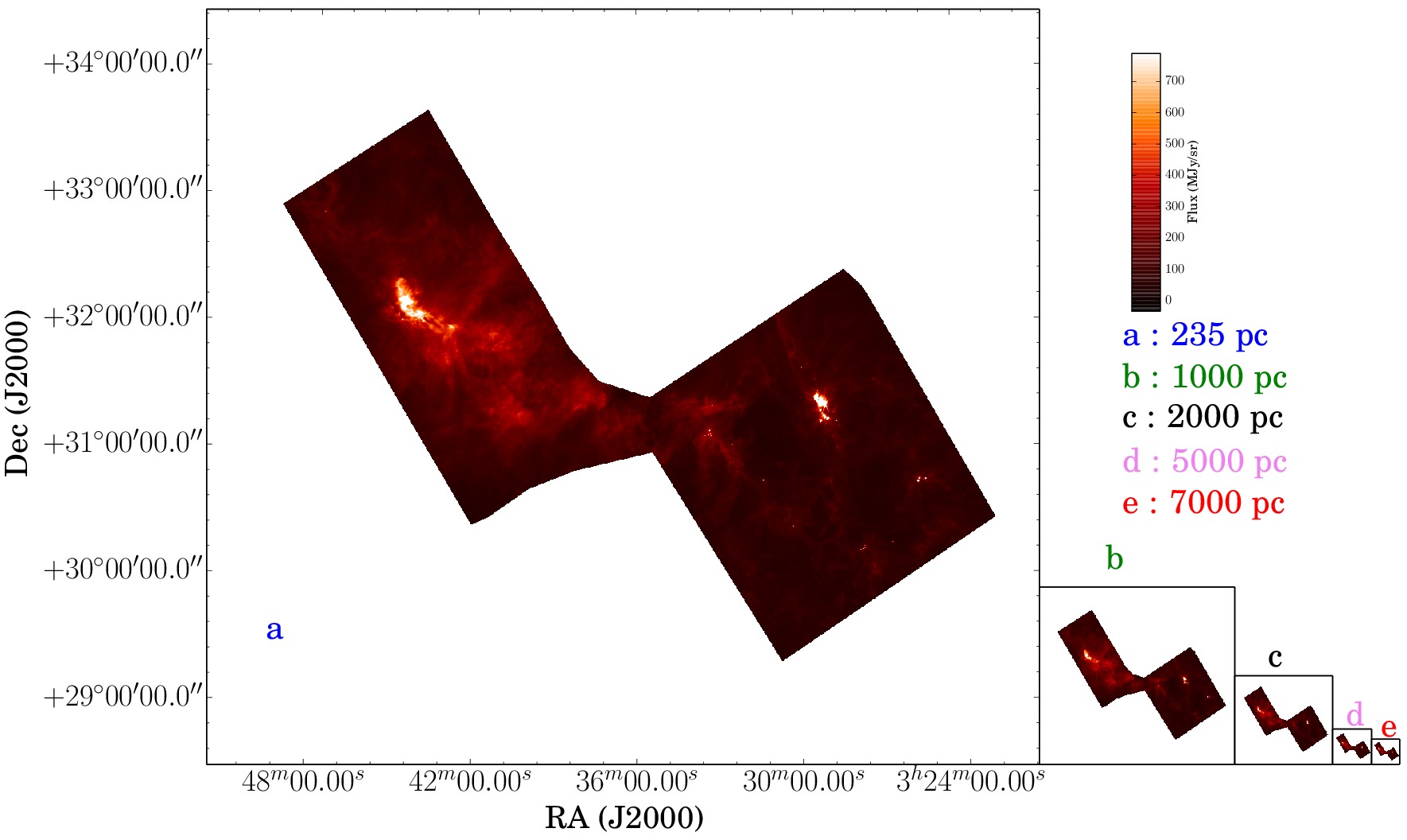}
\caption{Original and moved  Perseus maps at 160 $\mu \mathrm{m}$.
}
\label{fig:Pesreus_red}.
\end{figure*}

\newpage

\begin{figure*}
\includegraphics[scale=0.26]{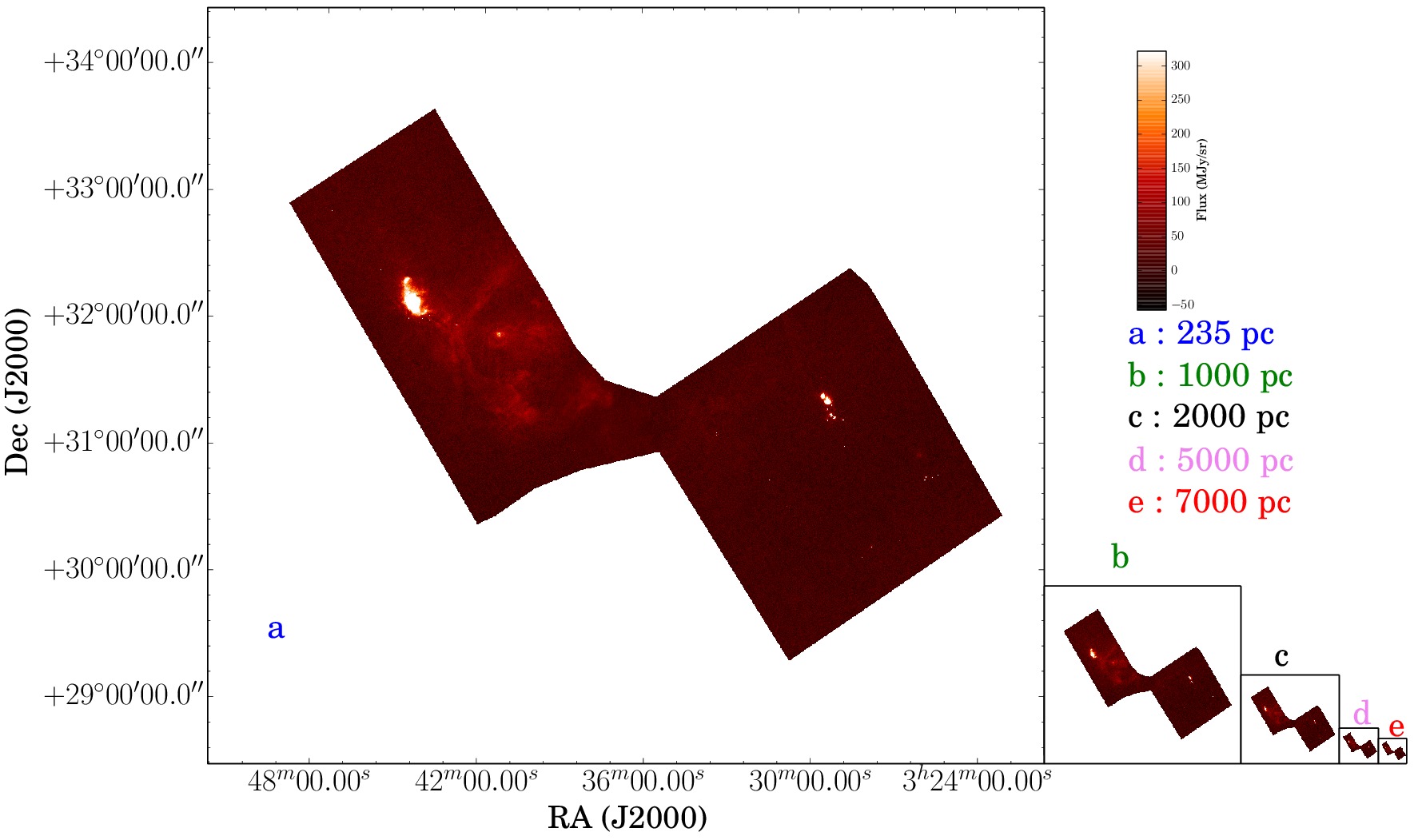}
\caption{Original and moved  Perseus maps at 70 $\mu \mathrm{m}$.
}
\label{fig:Pesreus_blue}
\end{figure*}

\begin{figure*}
\includegraphics[scale=0.26]{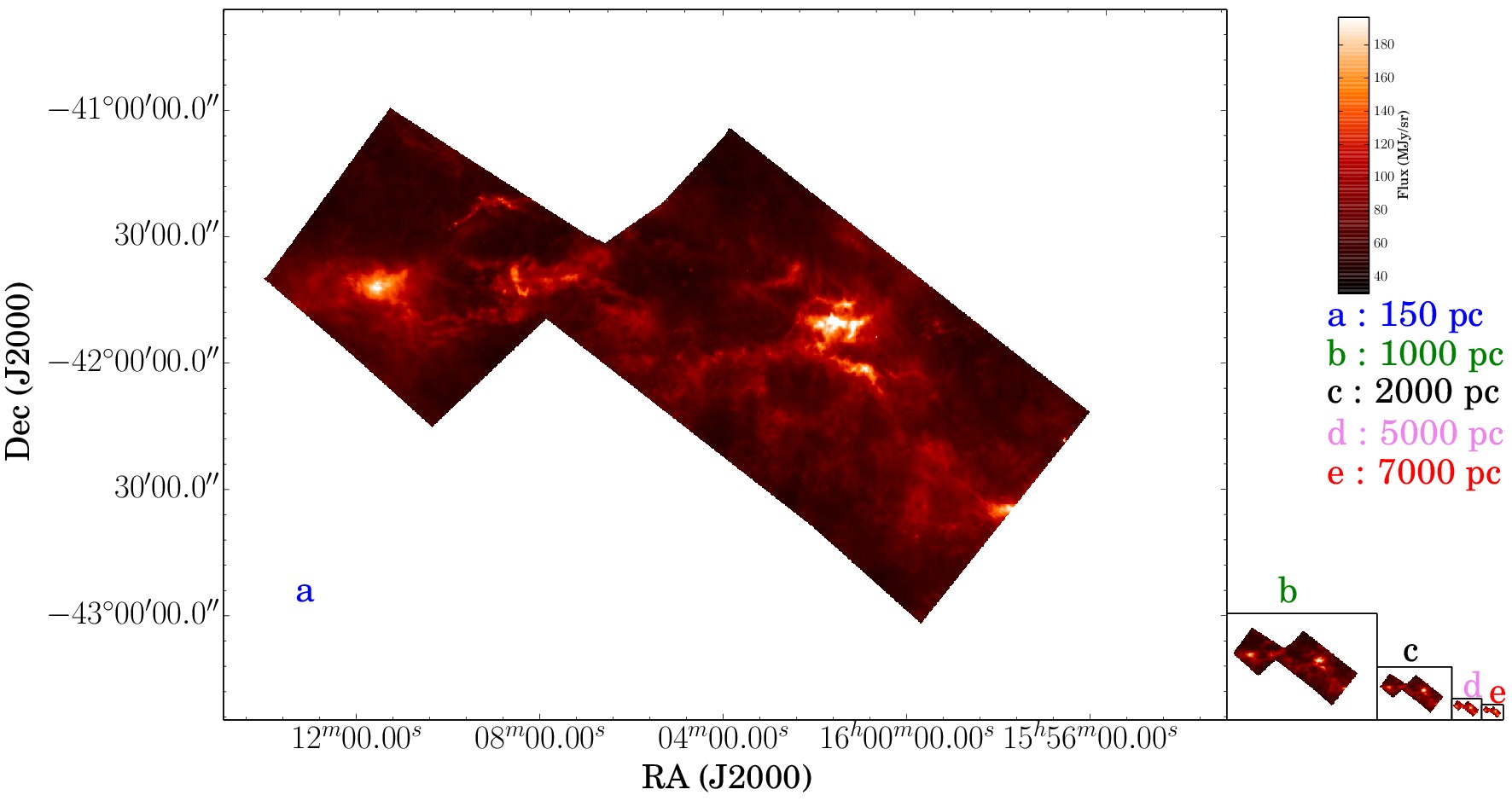}
\caption{Original and moved  Lupus IV maps at 250 $\mu \mathrm{m}$.
}
\label{fig:lupusIVPSW}
\end{figure*}

\begin{figure*}
\includegraphics[scale=0.26]{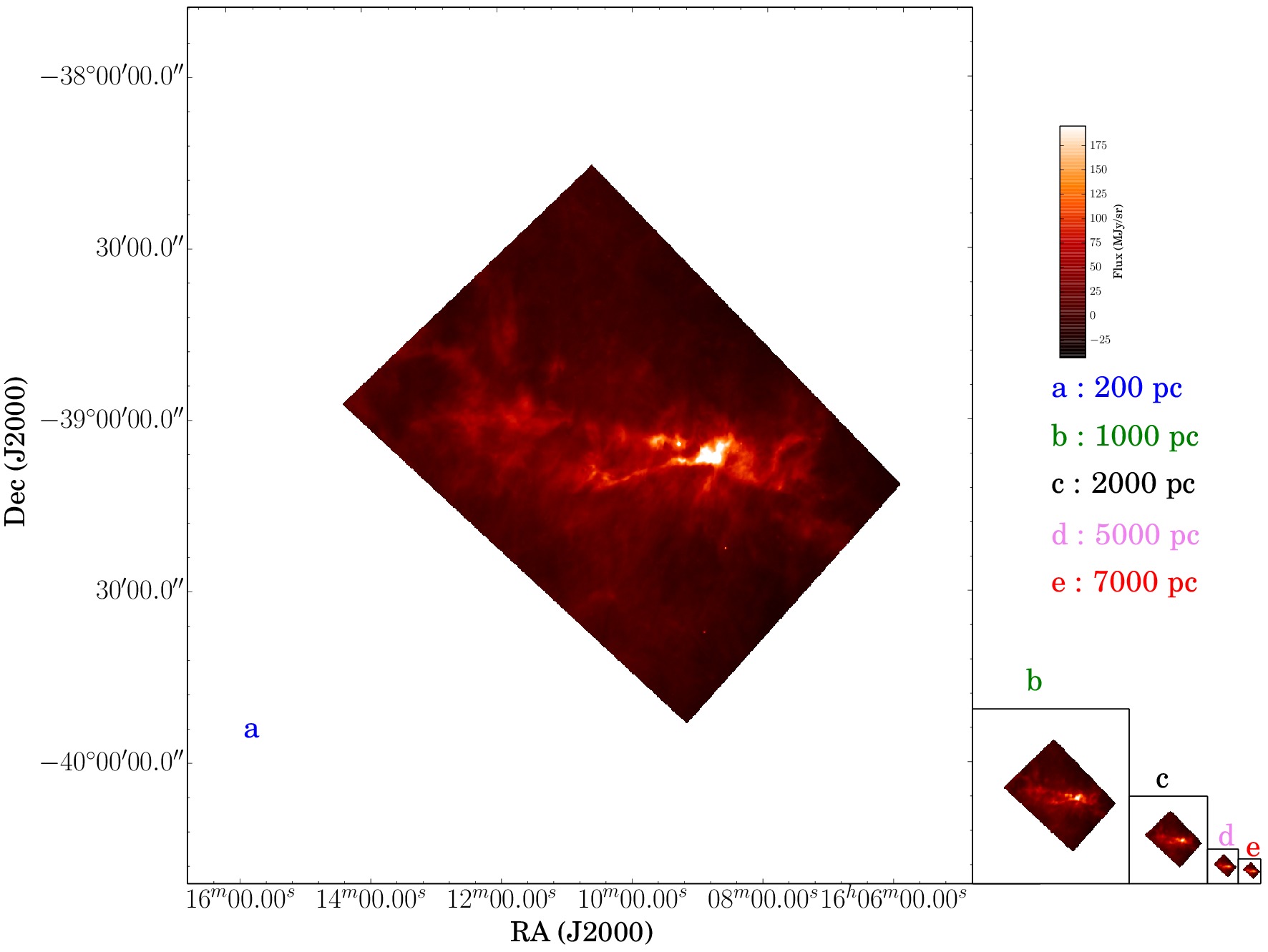}
\caption{Original and moved  Lupus III maps at 250 $\mu \mathrm{m}$.
}
\label{fig:LupusIIIPSW}
\end{figure*}

\begin{figure*}
\includegraphics[scale=0.26]{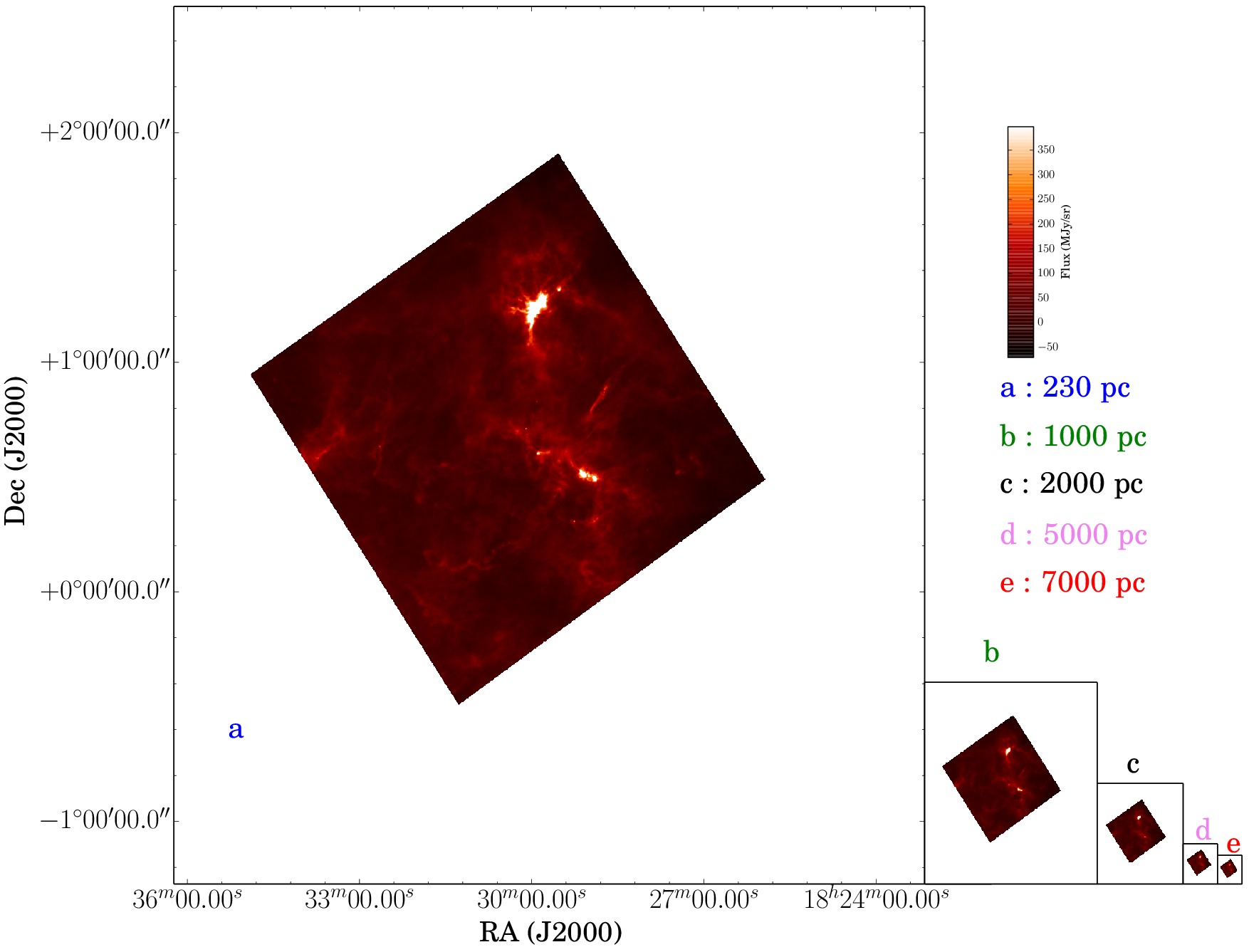}
\caption{Original and moved  Serpens  maps at 250 $\mu \mathrm{m}$.
}
\label{fig:SerpensPSW}
\end{figure*}

\begin{figure*}
\includegraphics[scale=0.26]{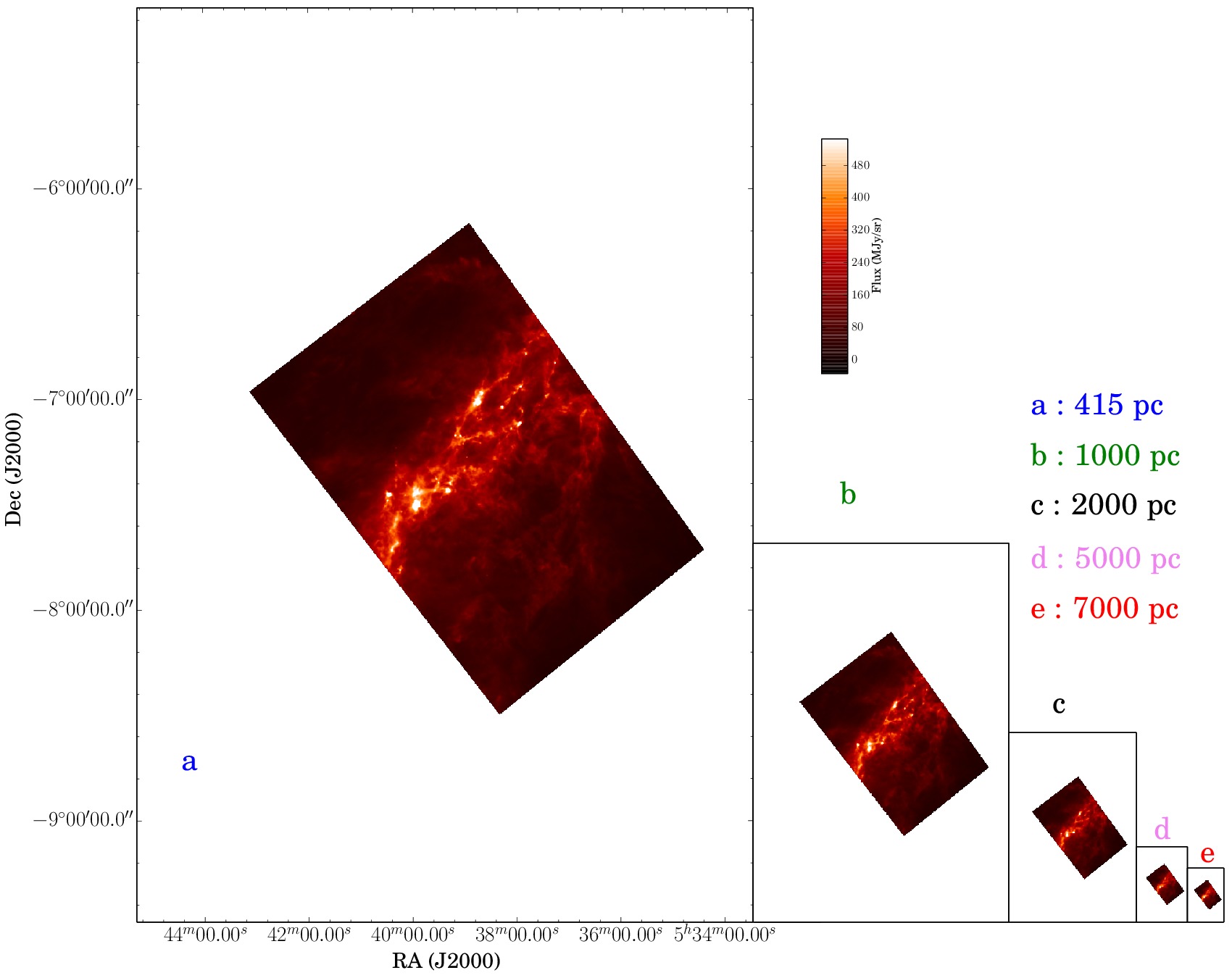}
\caption{Original and moved  Orion A center  maps at 250 $\mu \mathrm{m}$.
}
\label{fig:OrionCPSW}
\end{figure*}

\clearpage
\newpage
\begin{figure*}
\includegraphics[scale=0.26]{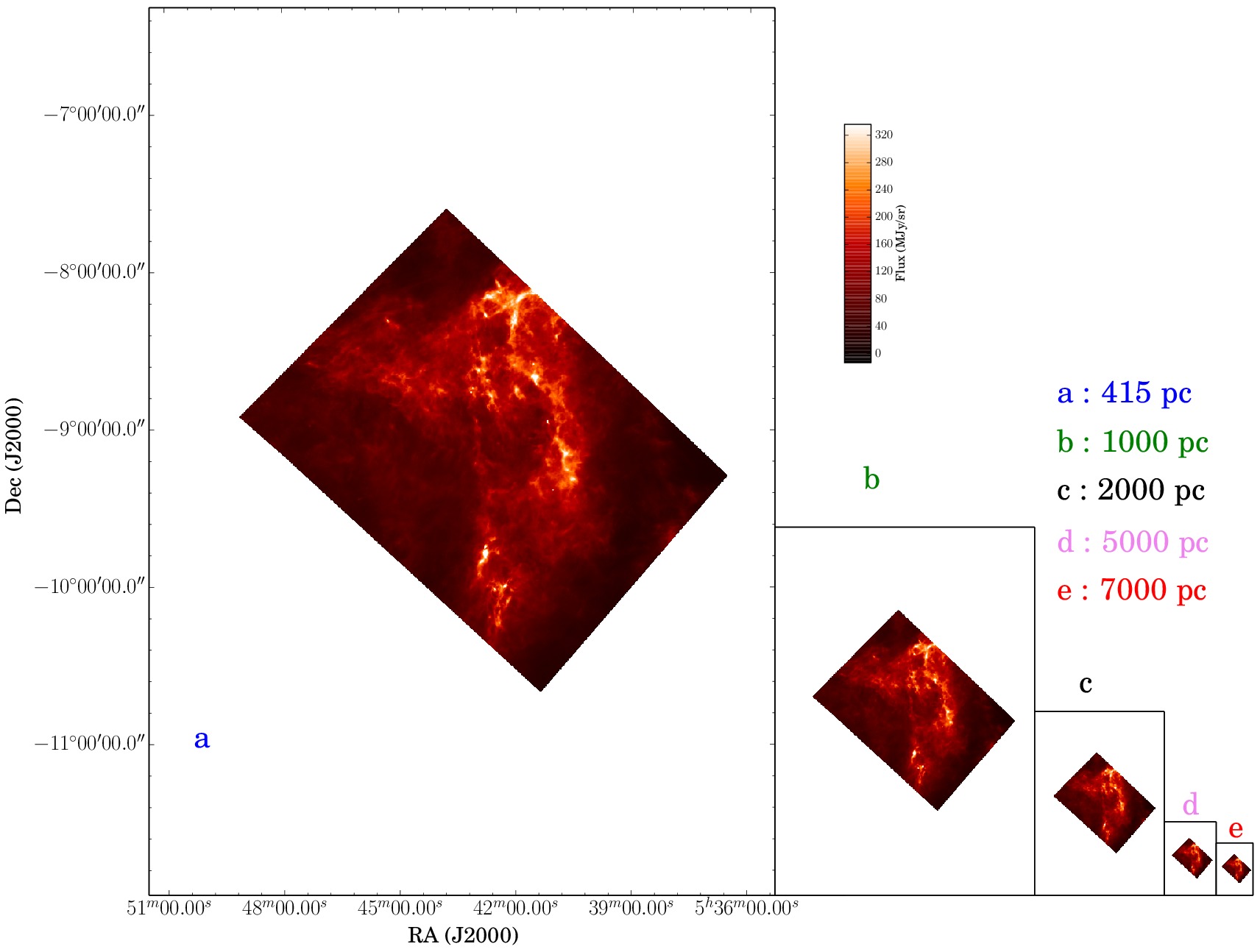}
\caption{Original and moved  Orion A south  maps at 250 $\mu \mathrm{m}$.
}
\label{fig:OrionSPSW}
\end{figure*}

\begin{figure*}
\includegraphics[scale=0.26]{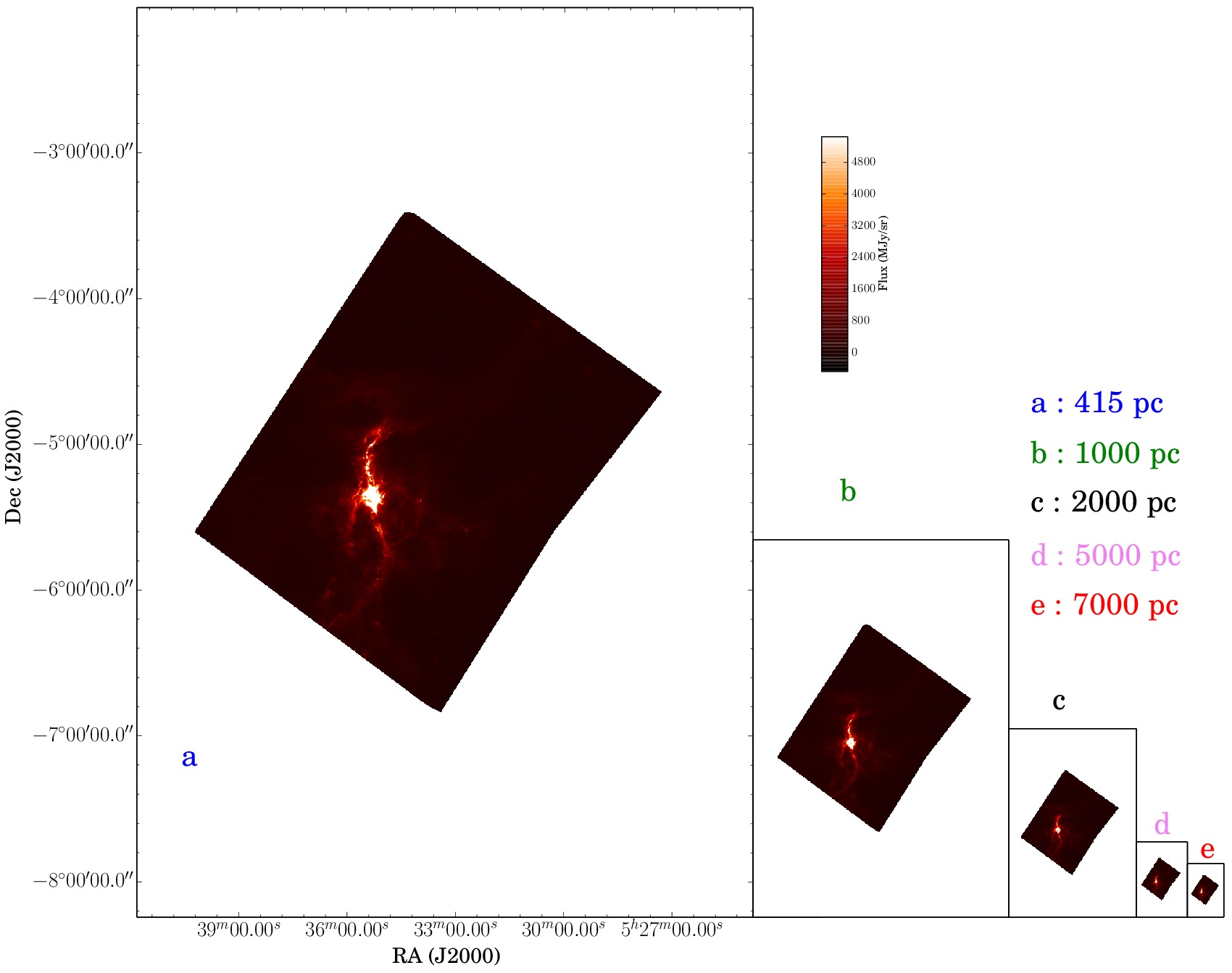}
\caption{Original and moved  Orion A north  maps at 250 $\mu \mathrm{m}$.
}
\label{fig:OrionNPSW}
\end{figure*}

\section{Derivation of a new high-mass star forming prescription}
\label{sec:oop}

In this Appendix we show how  the  prescription, expressed by equation~\ref{eq:ooprescription},
to discriminate high-mass star forming candidates is derived.

We consider all the objects (both prestellar and protostellar) detected in  all the regions  moved at 
 $d>1000$ pc.
We define a potentially high-mass star forming candidate $O_{\mathrm{hm}}$ (as defined in section~\ref{sec:assosiation}) as 
an object detected at a moved distance which is associated with, at least, a high-mass star forming core (according to the KP relation)
 detected at the original distance.
Similarly, we define a low-mass star forming candidate $O_{\mathrm{lm}}$ as 
an object detected at a moved distance which is associated  with  low-mass star
 forming cores detected at the original distance.

We search for  a prescription in the form $M>A \, r^{a}$. For a given $(A,a)$ pair,
we define as a true positive (TP) a $O_{\mathrm{hm}}$ whose mass is larger than $M$, a false negative (FN) a $O_{\mathrm{hm}}$
whose mass is smaller than $M$, a true negative (TN) a  $O_{\mathrm{lm}}$ whose mass is less than $M$ and a false positive (FP)
a  $O_{\mathrm{lm}}$ whose mass is larger than $M$.
We search  $(A,a)$ pairs that maximize the ratio $f=(N_{\mathrm{TP}}+N_{\mathrm{TN}})/(N_{\mathrm{FP}}+N_{\mathrm{FN}})$,
where $N_{\mathrm{TP}}$, $N_{\mathrm{TN}}$, $N_{\mathrm{FP}}$, $N_{\mathrm{FN}}$ are the numbers of TP, TN, FP and FN, respectively.
To do that we build a grid of values of $a$ and $A$ and then take those maximizing $f$.
The grid was built exploring values of $a$ between 1.33 and 2, in 30 steps, and values of $A$ between 870 and 1800, in 200 steps.
Then we impose the further constraint that
the fraction of FP  should be roughly constant with distance.
It turns out that the best values are $A=1282$ and $a=1.42$.
It is also possible to provide an uncertainty on the value of $A$:
we keep $a$ fixed and we vary $A$ as long as the total fraction of FP changes by $5\%$.
For $A=1282$ the fraction of FP is $\sim 9\%$ while $\sim4 \%$ and $\sim14 \%$ are achieved 
for $A=2000$ and $A=940$, respectively.

\section{Another determination of the distance of the Serpens cloud}
\label{sec:serp}

The distance of the Serpens molecular cloud has been a matter of debate.
In literature  a broad spectrum of distances can be found; for example \citet{Straizys2003} and \citet{Eiroa2008} estimate a distance of
225 and 230 pc, respectively, while recently \citet{Leon2016} place the Serpens cloud at 436 pc.

\begin{figure}
\includegraphics[scale=0.5]{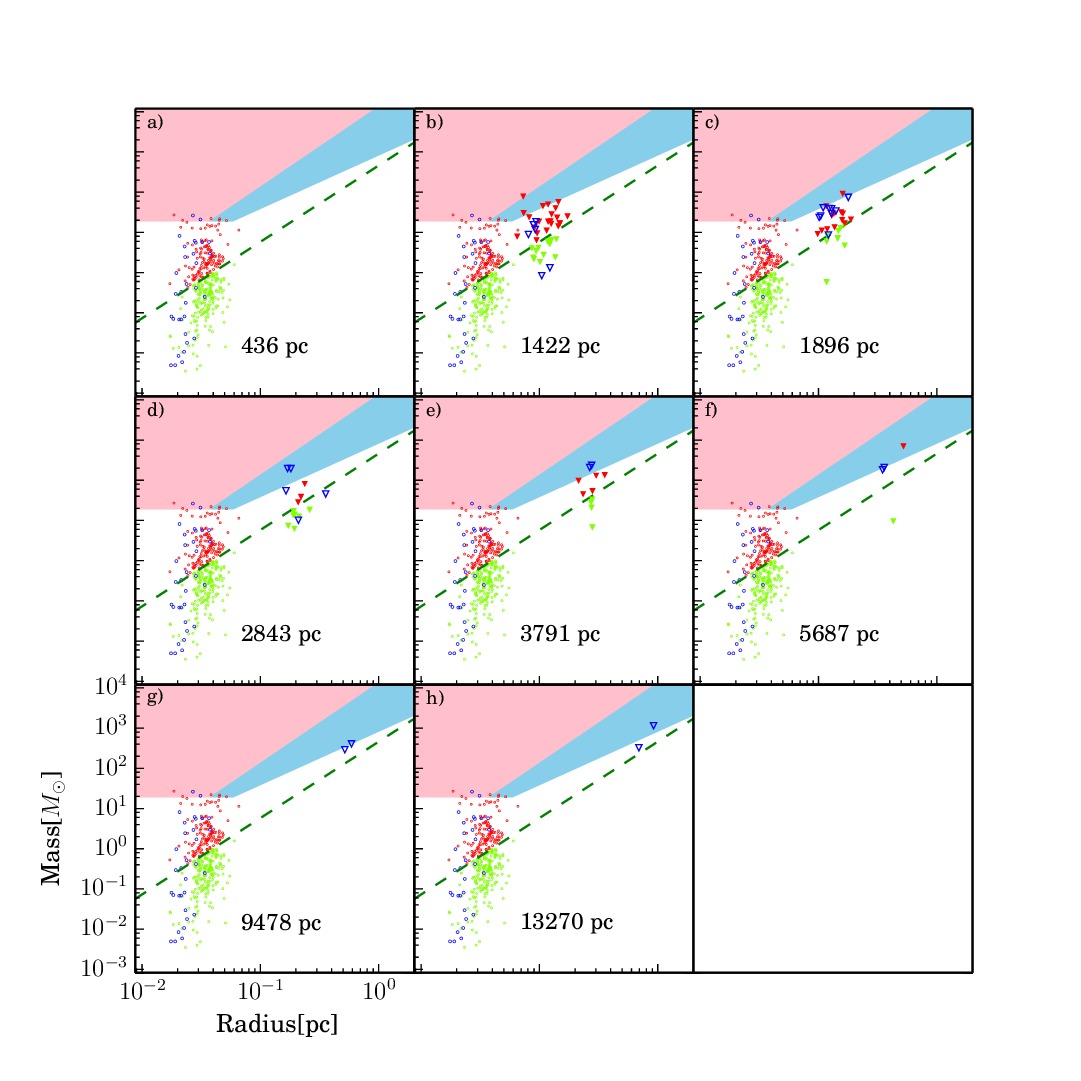}
\caption{The same as Fig.\ref{fig:mass_radius_Serpens00} but starting from $d_{0}=436$~pc instead of 230 pc, (simulated distances are scaled
accordingly, see text).}
\label{fig:mass_radius_Serpens00_new_dist}
\end{figure}

Here we want to check how the MR diagram, that we have shown in section~\ref{sec:mass}, is affected 
if we adopt a distance of 436 pc instead of
230 pc
(Figure~\ref{fig:mass_radius_Serpens00_new_dist}). 
Since the radius and the mass of the sources scale linearly and quadratically with distance, respectively, to get the new values of $r$ and $M$
we simply multiply these parameters by 436/230 and $(436/230)^2$, respectively.
Obviously, also the previously  simulated distances must be multiplied by 436/230 leading to new simulated distances of
1422,  1896, 2843, 3791, 5687, 9478 and  13270 pc.

Note that the Serpens cloud, with this new $d_{0}$ might be considered as a potentially MSF region, since in this case 
at the original distance
there are six sources fulfilling the KP relation. 
Note that also at the moved distances this cloud remains classifiable  as a MSF region.
Therefore the behaviour of this cloud in our analysis, if we assume $d_{0}=436$ pc, is very 
similar to the Orion A molecular cloud which
is a genuine high-mass star forming region \citep[e.g.][]{Genzel1989}.

\begin{figure}
\includegraphics[scale=0.5]{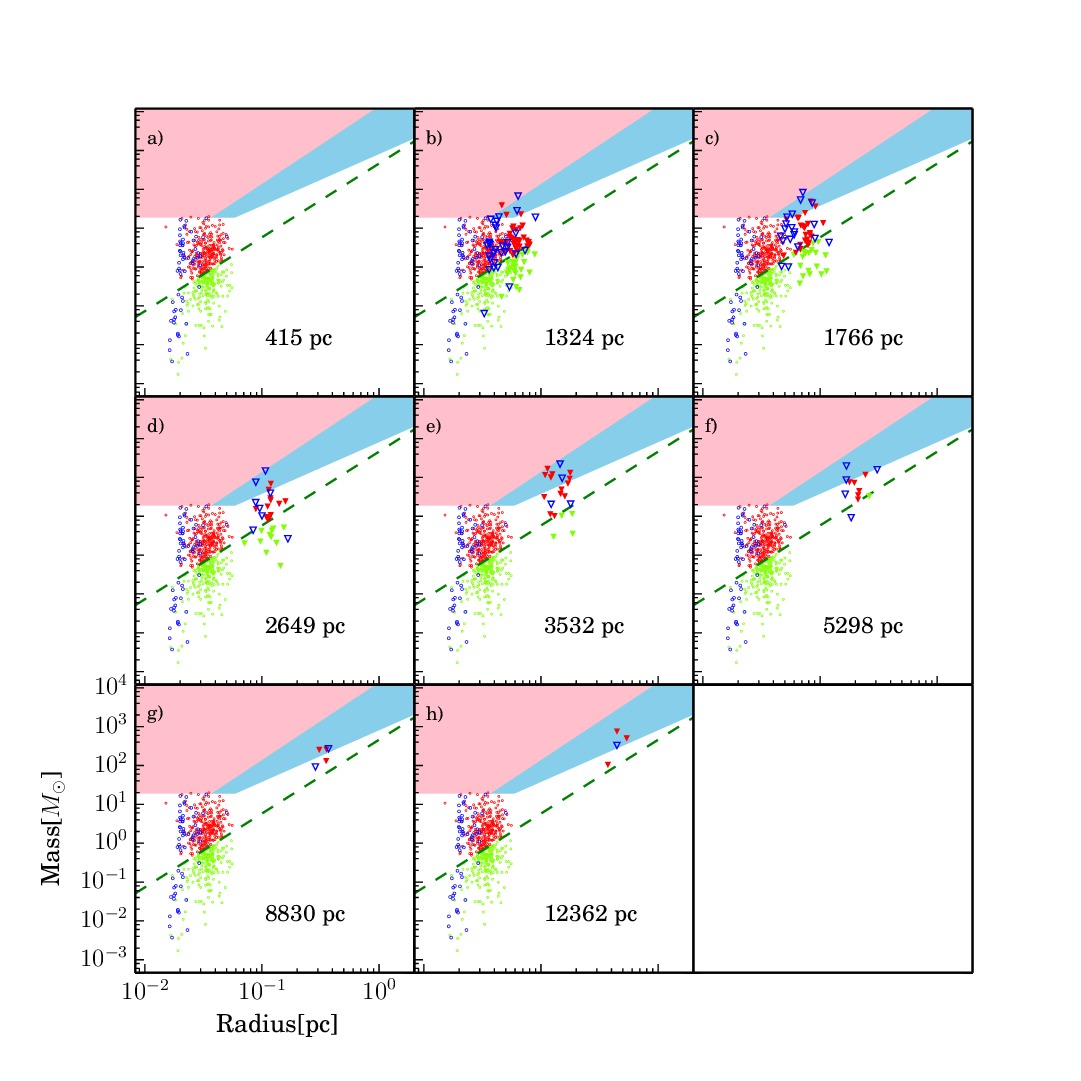}
\caption{The same as Fig.\ref{fig:mass_radius_perseus} but imposing $d_{0}=415$~pc instead of 235 pc for Perseus, to make a direct
comparison with Orion A (Fig. \ref{fig:mass_radius_Orione_unito}).
 Simulated distances are scaled
accordingly, see text.}
\label{fig:mass_radius_Perseus_new_dist}
\end{figure}

It is also interesting to apply the same technique developed in this section to make a further virtual comparative check  between 
the two regions of our sample with the  richest source statistics.
In practice we want
to understand how the properties of the Perseus cloud would change if it was located at the same distance
of Orion A (415 pc) instead of 235 pc.
We rescale the mass and the radius with the same procedure that we used above for Serpens data.
Figure~\ref{fig:mass_radius_Perseus_new_dist} displays the mass vs radius plot for the Perseus nebula assuming $d_{0}=415$ pc.
As one can see from Figs \ref{fig:mass_radius_Perseus_new_dist} and  \ref{radius_distribution_all_regions}
the radius of the sources at the original distance for Orion A and Perseus would be quite similar, while the masses in Orion A would be still 
larger than in  Perseus.
This suggest that, while the size distributions tend to be similar when reported at the same distance (this is somehow 
trivial since compact sources span the same range of angular sizes), the behaviour of the corresponding masses seems to
be more related to the intrinsic nature of the specific region.

\clearpage






\bibliographystyle{mnras}   

 \newcommand{\noop}[1]{}

\bsp
\label{lastpage}

\end{document}